%% file: tmi.tex
\documentclass[journal,twoside,web]{ieeecolor}
\usepackage{tmi}
\usepackage{cite}
\usepackage{amsmath,amssymb,amsfonts}
\usepackage{algorithmic}
\usepackage{graphicx}
\usepackage{textcomp}
\usepackage{graphicx}
\usepackage{booktabs}
\usepackage{multirow}
\usepackage{amsmath}

\usepackage{bm}
\usepackage[utf8]{inputenc}
\usepackage{amssymb, latexsym}
\usepackage{tikz}
\usetikzlibrary{shapes.geometric}
\usetikzlibrary{calc, decorations.pathreplacing}
\usetikzlibrary{fadings}
\usepackage{color, xcolor}
\usepackage{pgfplots}
\usepackage{pgf-umlsd}
\usepackage{ifthen}
\usetikzlibrary{fadings}
\usetikzlibrary{arrows,decorations.markings}
\usepackage{mathrsfs}

\usepackage{sidecap}

\usepackage{floatrow}
\usepackage{caption} 
\usepackage{subcaption}
\usepackage[framemethod=tikz]{mdframed}

\usepackage{makecell}

\usepackage{hyperref}

\usepackage{cleveref}
\Crefname{equation}{Eq.}{Eqs.}

\newcolumntype{P}[1]{>{\centering\arraybackslash}p{#1}}

\usepackage[linesnumbered,ruled,vlined]{algorithm2e}

\SetCommentSty{mycommfont}
\SetKwInput{Input}{Input}                
\SetKwInput{Output}{Output}         
\SetKwInput{Dataset}{Dataset}     
\SetKwInput{Settings}{Settings}      
\SetKwInput{Initialization}{Initialization}      
\SetAlCapNameFnt{\small}
\SetAlCapFnt{\small}
\SetAlgorithmName{Alg.}{Alg.}{List of Algorithms} 
\makeatletter
\newcommand{\algorithmfootnote}[2][\footnotesize]{%
  \let\old@algocf@finish\@algocf@finish% Store algorithm finish macro
  \def\@algocf@finish{\old@algocf@finish% Update finish macro to insert "footnote"
    \leavevmode\rlap{\begin{minipage}{\linewidth}
    #1#2
    \end{minipage}}%
  }%
}
\makeatother

\def\BibTeX{{\rm B\kern-.05em{\sc i\kern-.025em b}\kern-.08em
    T\kern-.1667em\lower.7ex\hbox{E}\kern-.125emX}}
\begin{document}
\title{Perfusion Imaging:\\A Data Assimilation Approach}
\author{Peirong Liu, Yueh Z. Lee, Stephen R. Aylward, and Marc Niethammer
\thanks{Submitted for review on September 4, 2020. This
work was supported by the NIH Grant 2R42NS086295-02A1. (Principal investigator: Yueh Z. Lee.)}
\thanks{Peirong Liu is with the Department of Computer Science, University of North Carolina at Chapel Hill, Chapel Hill, NC 27599, USA (e-mail: peirong@cs.unc.edu). }
\thanks{Yueh Z. Lee is with Department of Radiology, University of North Carolina at Chapel Hill, Chapel Hill, NC 27599, USA (e-mail: yueh\_lee@med.unc.edu).}
\thanks{Stephen R. Aylward is with Kitware, Inc., Carrboro, NC 27510, USA (e-mail: stephen.aylward@kitware.com).}
\thanks{Marc Niethammer is with the Department of Computer Science and the
Biomedical Research Imaging Center, University of North Carolina at Chapel Hill, Chapel Hill, NC 27599, USA (e-mail: mn@cs.unc.edu). }}

\maketitle

\begin{abstract}
Perfusion imaging (PI) is clinically used to assess strokes and brain tumors. Commonly used PI approaches based on magnetic resonance imaging (MRI) or computed tomography (CT) measure the effect of a contrast agent moving through blood vessels and into tissue. Contrast-agent free approaches, for example, based on intravoxel incoherent motion, also exist, but are so far not routinely used clinically. These methods rely on estimating on the arterial input function (AIF) to approximately model tissue perfusion, neglecting spatial dependencies, and reliably estimating the AIF is also non-trivial, leading to difficulties with standardizing perfusion measures. In this work we therefore propose a data-assimilation approach (PIANO) which estimates the velocity and diffusion fields of an advection-diffusion model that best explains the contrast dynamics. PIANO accounts for spatial dependencies and neither requires estimating the AIF nor relies on a particular contrast agent bolus shape. Specifically, we propose a convenient parameterization of the estimation problem, a numerical estimation approach, and extensively evaluate PIANO. We demonstrate that PIANO can successfully resolve velocity and diffusion field ambiguities and results in sensitive measures for the assessment of stroke, comparing favorably to conventional measures of perfusion. 
\end{abstract}


\input{sub/intro}
\input{sub/method}

\input{sub/exp_results}
\input{sub/exp_peclet}

\input{sub/exp_effect_robust}

\input{sub/exp_ident}
\input{sub/con}

%%\clearpage
\bibliographystyle{IEEEtran}
 \bibliography{tmi}
\end{document}

%% file: sub/intro.tex
\section{Introduction}
\label{intro}

Perfusion imaging (PI) allows for quantifying blood flow through the brain parenchyma by using an intravascular tracer and serial imaging. The resulting quantitative measures help clinical diagnosis and decision-making for cerebrovascular disease, particularly for acute stroke, and facilitate individualized treatment of stroke patients based on brain tissue status~\cite{demeestere2020stroke}. Despite its benefits, the widespread use of PI still faces many challenges. In fact, the postprocessing of PI is far from standardized. At present, the mainstream approach for postprocessing PI source data, a time series of 3D volumetric images, is done using tracer kinetic models to estimate hemodynamic parameters for each voxel, obtaining corresponding perfusion parameter maps in 3D~\cite{fieselmann2011sig2ctc}. Specifically, an arterial input function (AIF) is selected to approximate the delivery of intravascular tracer to tissue. Perfusion parameter maps are then computed based on the AIF and the observed concentration of contrast agents (CA) at each voxel by a deconvolution algorithm~\cite{mouridsen2006aif}.
However, there exist substantial differences in perfusion parameter maps generated across institutions, mainly caused by different AIF selection procedures, deconvolution techniques and interpretations of perfusion parameters \cite{mouridsen2006aif,schmainda2019a,schmainda2019b}. 

Moreover, postprocessing approaches for PI are performed on individual voxels, thereby disregarding spatial dependencies of contrast dynamics. Some efforts exist to fit CA transport via partial differential equations (PDEs)~\cite{cookson2014spatial,harabis2013dilution,strouthos2010dilution}, though these approaches ultimately reduce to voxel-based analyses {\textendash} parameters of a closed-form solution of the associated PDEs are estimated to fit the concentration time-curve voxel-by-voxel. The work by Cookson et al.~\cite{cookson2014spatial} is the most closely related work to our proposed approach, where advection-diffusion PDEs are used to model CA transport within cerebral blood vessels and brain tissue. However, that work assumes that the velocity and the diffusion are {\it constant} over the entire domain, which is unrealistic in real tissue. In fact, the spatially varying nature of perfusion is, for example, precisely the critical aspect of stroke assessment. As a result of the constancy assumption only simulations are considered in~\cite{cookson2014spatial}, but estimations based on real data are not explored.

{\bf Contributions:} We therefore propose a data-assimilation approach {\textendash} Perfusion Imaging via AdvectioN-diffusiOn (PIANO) {\textendash} which models CA transport by variable-coefficient advection-diffusion PDEs. To the best of our knowledge, PIANO is the first work taking into account the spatial relations between voxels in PI. Specifically, given a time series of CA concentration 3D images, PIANO estimates spatially-varying velocity and diffusion fields of the advection-diffusion model that best explain CA passage. By physically modeling CA transport via advection and diffusion, PIANO does not require AIF selection or deconvolution algorithms to compute perfusion parameter maps, which are required in conventional PI postprocessing approaches and may yield differences in parameter map estimations. We extensively assess the estimation behavior of PIANO. In particular, we assess PIANO's ability to disentangle velocity from diffusion estimates and its robustness to noise. Quantitative comparisons further demonstrate the advantage of feature maps from PIANO over conventional perfusion parameter maps. We describe and test PIANO in the context of brain PI. The approach, however, is general and could conceivably be applied to PI of other organs.

This manuscript is a significant extension of our work: Peirong Liu, Yueh Z. Lee, Stephen R. Aylward and Marc Niethammer, ``PIANO: Perfusion Imaging via Advection-diffusion'': In \textit{$23$rd International Conference on Medical Image Computing and Computer Assisted Intervention (MICCAI)}, 2020. Specifically, in this paper, we provide detailed experimental evaluations from multiple aspects: (1) We show that the velocity and diffusion fields estimated by PIANO fall within reasonable value ranges that are consistent with value ranges reported in literature; (2) We demonstrate the effectiveness and robustness of PIANO, by exploring its robustness to noise; (3) We further verify the capability of PIANO to disentangle the estimation of advection velocities from the estimation of the diffusion process.

%% file: sub/method.tex
\section{Perfusion Imaging via AdvectioN-diffusiOn (PIANO)}
\label{sec:piano}

First, Sec~\ref{sec:governing_equations} describes how we model CA transport as a combination of advection and diffusion. Sec.~\ref{sec:estimating_advection_diffusion} then discusses how PIANO estimates the velocity and the diffusion fields that best explain the contrast dynamics.

%%%%%%%%%%%%%%%%%%%%%%%%%%%%%%%%%
\input{sub/exp_results/2d_demo}
%%%%%%%%%%%%%%%%%%%%%%%%%%%%%%%%%

\subsection{Governing Equations}
\label{sec:governing_equations}

After the injected CA has fully flowed into the brain, the observed local changes of CA concentration (which we refer to as concentration in what follows) in the brain can generally be explained by two dominating macroscopic effects: advection and diffusion. Advection mainly describes the transport of CA driven by the blood flow within the blood vessels, while diffusion captures the movements of freely-diffusive CA within the extracellular space as well as aspects of capillary transport. Note that because voxel sizes in PI ($\approx 1~mm$) are orders of magnitude larger than capillary radii~\cite{marin2012}, capillary blood transport may also manifests as diffusion macroscopically. In this work, we refer to diffusion as the effective diffusion observable at voxel scale combining these effects.

Let $C({\bf{x}}, t)$ denote the concentration at location ${\bf{x}}$ in the brain $\Omega \subset \mathbb{R}^3$, at time $t$. Local concentration may be modeled as an advection-diffusion equation:
\begin{align}
\label{eq: full}
& \frac{\partial C({\bf{x}}, t)}{\partial t} = - \nabla \cdot \left({\bf{V}}({\bf{x}})\, C({\bf{x}}, t)\right) + \nabla \cdot \left({\bf{D}}({\bf{x}})\, \nabla C({\bf{x}}, t)\right),  % \\
%\label{eq: dirichlet}
%& s.t. \quad C({\bf{x}},\, t) = f({\bf{x}},\, t), \quad \forall {\bf{x}} \in \partial \Omega,\, \forall t \in \mathbb{R}
\end{align}
%\Cref{eq: dirichlet} imposes the Dirichlet boundary condition, where $f({\bf{x}},\, t)$ specifies the concentration values along the boundary $\partial \Omega$. %By \Cref{eq: neumann} we assume that there is no CA passing through the brain boundary. 
where ${\bf{V}}({\bf{x}}) = (V^x({\bf{x}}), V^y({\bf{x}}), V^z({\bf{x}}))^T$ is the spatially-varying velocity, with each component referring to the blood flow velocity in directions $x,\, y,\, z$ respectively. ${\bf{D}}$ is a spatially-varying diffusion tensor field governing CA diffusion, where each ${\bf{D}}(x)$ is assumed to be a $3\times 3$ symmetric positive semi-definite (PSD) matrix~\cite{niethammer2006dti}. We assume ${\bf{V}}$ and ${\bf{D}}$ to be constant in time to simplify our estimation problem. Further, assuming the blood flow is incompressible everywhere, i.e., ${\bf{V}}$ is divergence-free ($\nabla\cdot {\bf{V}}({\bf{x}}) = 0,\,\forall {\bf{x}} \in \Omega$), \Cref{eq: full} can be rewritten as:
\begin{equation}
\frac{\partial C({\bf{x}}, t)}{\partial t} = -  {\bf{V}}({\bf{x}})\cdot\nabla C({\bf{x}}, t) + \nabla \cdot \left({\bf{D}}({\bf{x}})\, \nabla C({\bf{x}}, t)\right).
\label{eq: div_free}
\end{equation}

%\begin{equation}
%\frac{\partial C}{\partial t} = -  {\bf{V}} \cdot \nabla C + \nabla \cdot \left({\bf{D}}\, \nabla C\right)
%\label{eq: div_free}
%\end{equation}

%%%%%%%%%%%%%%%%%%%%%%%%%%%%%

\subsection{Estimating Advection and Diffusion}
\label{sec:estimating_advection_diffusion}

%%%%%%%%%%%%%%%%%%%%%%%%%%%%%

\input{sub/fig_framework_tmi}

\input{sub/pseudocode}

%%%%%%%%%%%%%%%%%%%%%%%%%%%%%

Sec.~\ref{sec:governing_equations} described PIANO's advection-diffusion model for CA transport. Here, we focus on a particular approach to estimate divergence-free vector fields ${\bf{V}}$ and PSD diffusion tensor fields ${\bf{D}}$ from time series of measured 3D volumetric concentration images, $\{\left(C^{t_i} \right)_{N_x \times N_y \times N_z }\in \mathbb{R}(\Omega) \vert\, i = 0,\, 1,\, \ldots,\, T\}$, with temporal resolution $\Delta t$.

%%%%%%%%%%%%%%%%%%%%%%%%%%%%%%

\subsubsection{Parametrization of Velocity and Diffusion Fields}
\label{sec:parametrization}
To ensure that the vector field ${\bf{V}}$ is divergence-free, we represent it by two scalar fields $\Gamma_1, \, \Gamma_2$~\cite{barbarosie2011divfree}: 
\begin{equation}
{\bf{V}}({\bf{x}}) = \nabla \Gamma_1({\bf{x}}) \wedge \nabla \Gamma_2 ({\bf{x}}), \quad \Gamma_1, \, \Gamma_2 \in \mathbb{R}(\Omega),\, \forall {\bf{x}} \in \Omega,
\label{eq: v_div_free}
\end{equation}
where $\wedge$ denotes the exterior product between vectors in $\mathbb{R}^3$. To construct a PSD tensor field, we parametrize ${\bf{D}}$ by its Cholesky factorization: %(Appendix \ref{app: cholesky_decomp})
\begin{equation}
{\bf{D}}({\bf{x}}) = {\bf{L}}({\bf{x}})^T{\bf{L}}({\bf{x}}), \quad {\bf{L}} \in \mathbb{R}^{3 \times 3}(\Omega), \, \forall {\bf{x}} \in \Omega,
\label{eq: d_psd}
\end{equation}
%${\bf{L}}({\bf{x}})$ is an upper triangular matrix with non-negative diagonals. Assuming the isotropic diffusion of CA, i.e., ${\bf{D}} = D\,I$ ($I$ is the identity matrix), \Cref{eq: d_psd} simplifies to 
where ${\bf{L}}({\bf{x}})$ is an upper triangular matrix with non-negative diagonals. Assuming the diffusion of CA is isotropic, \Cref{eq: d_psd} simplifies to ($I$ is the identity matrix)
\begin{equation}
{\bf{D}}({\bf{x}}) = D({\bf{x}})\, I = L^2({\bf{x}})\, I, \quad L \in \mathbb{R}(\Omega), \, \forall {\bf{x}} \in \Omega.
\label{eq: d_scalar}
\end{equation}

%Replace ${\bf{V}}$ and ${\bf{D}}$ with the above parametrizations, \Cref{eq: div_free} equals
%\begin{equation}
%\frac{\partial C}{\partial t} = -  \left( \nabla \Gamma_1 \wedge \nabla \Gamma_2 \right) \cdot \nabla C + \nabla \cdot \left(L^2\, \nabla C\right)
%\label{eq: sim_param}
%\end{equation}

%%%%%%%%%%%%%%%%%%%%%%%%

\subsubsection{Numerical Flow}
\label{sec:numerical_flow}
The voxel spacings $\Delta x, \Delta y, \Delta z$ of the given 3D volumetric concentration images naturally introduce corresponding grid sizes in axial, coronal and sagittal directions. We use a first-order upwind scheme \cite{leveque2002} to approximate the partial differential operators of the advection term in \Cref{eq: div_free}, and nested forward-backward differences for the diffusion term: forward differences for $\nabla\cdot$ and backward differences for $\nabla C$ in \Cref{eq: div_free}. Discretizing all spatial derivatives on the right hand side of \Cref{eq: div_free} results in a system of ordinary differential equations, which we solve by numerical integration. Specifically, we impose a mixed boundary condition (BC) for the system: Dirichlet BCs are applied on the first and last axial slices\footnote{Our dataset is acquired axially, but BCs could be modified for different acquisition formats as needed. This BC essentially replaces determining the AIF.} which simply impose the {\it measured} concentrations. We impose homogeneous Neumann BCs on the outer brain contours in the remaining axial slices, assuming no contrast agent passes through these boundaries. We use a Runge-Kutta-Fehlberg method to advance in time ($\delta t$) to predict $\widehat{C}^{t+\delta t}$. Note that the chosen $\delta t$ is typically smaller than the temporal resolution of the given concentration time series images ($\Delta t$), to satisfy the Courant-Friedrichs-Lewy (CFL) condition \cite{leveque2002} and thereby to ensure stable numerical integration.

%%%%%%%%%%%%%%%%%%%%%%%%%%%

\subsubsection{Estimation}
\label{sec:estimation}
Given an initial state $C^t$, PIANO applies the current estimate of ${\bf{V}},\, {\bf{D}}$ to $C^t$ by \Cref{eq: div_free} and predicts subsequent concentration images with time step $\delta t$. Instead of starting from a specific concentration image, we randomly pick an image from the given concentration time series as the initial condition for each estimating iteration. We then integrate the PIANO model forward to time frame $T_{\text{pd}}$ (Fig. \ref{fw}). This reduces the sensitivity of the estimated ${\bf{V}}$ and ${\bf{D}}$ to varying initial conditions. We define our estimation losses as follows.

\paragraph{Collocation Concentration Loss.} Given a sample $\{C^{t_i} \in \mathbb{R}(\Omega) \vert\, i = 0,\, 1,\, \ldots,\, T_{\text{pd}}\}$, with $t_0,\, t_1,\, \ldots,\, T_{\text{pd}}$ as collocation points, we define the collocation concentration loss ($\mathcal{L}_{CC}$) as the mean squared error of the predicted concentrations at $t_1,\, \ldots,\, T_{\text{pd}}$. This encourages estimates to be close to the measurements:

\begin{equation}
\mathcal{L}_{CC} = \frac{1}{T_{\text{pd}}} \sum_{i = 1}^{T_{\text{pd}}}  \frac{1}{\vert\Omega\vert} \int_{\Omega}(C^{t_i}({\bf{x}}) - \widehat{C}^{t_i}({\bf{x}}))^2 d{\bf{x}}. 
\label{eq: loss_cc}
\end{equation}

\paragraph{Anisotropic Smoothness Regularizations.} Assuming the estimated fields are spatially smooth, we impose regularization terms on $\nabla {\bf{V}},\, \nabla D$ as
\begin{equation}
\left\{
\begin{aligned}
 \mathcal{L}_{AS_{\bf{V}}} & =\sum_{ax\in\{x,y,z\}}  \frac{1}{\vert\Omega\vert} \int_{\Omega} \alpha_{\bf{V}}\, \|\nabla V^{ax} \|_2^2\,d{\bf{x}},\\
 \mathcal{L}_{AS_D} & = \frac{1}{\vert\Omega\vert}\,\int_{\Omega} \alpha_D \, \|\nabla D \|_2^2\,d{\bf{x}},   
 \label{eq: ani_loss}
\end{aligned}
\right.
\end{equation}
%\begin{equation}
 %\mathcal{L}_{AS_{\bf{V}}} =\sum_{ax\in\{x,y,z\}}  \frac{1}{\vert\Omega\vert} \int_{\Omega} \alpha_{\bf{V}}\, \|\nabla V^{ax} \|_2^2\,d{\bf{x}},\quad
 %\mathcal{L}_{AS_D} = \frac{1}{\vert\Omega\vert}\,\int_{\Omega} \alpha_D \, \|\nabla D \,d{\bf{x}},   
 %\label{eq: ani_loss}
%\end{equation}
where the associated coefficients $\alpha_{\bf{V}},\, \alpha_D$ are computed as
\begin{equation}
\left\{
\begin{aligned}
\alpha_{\bf{V}} &= \sum_{ax\in\{x,y,z\}} \frac{g(\| \nabla (K_{\sigma} \ast V^{ax})\|_2^2)}{3},\\
\alpha_D &= g(\| \nabla (K_{\sigma} \ast D)\|_2^2), \quad \sigma > 0,   
 \label{eq: coef}
\end{aligned}
\right.
\end{equation}
%\begin{equation}
%\alpha_{\bf{V}} = \sum_{ax\in\{x,y,z\}} \frac{g(\| \nabla (K_{\sigma} \ast V^{ax})\|_2^2)}{3},\quad \alpha_D = g(\| \nabla (K_{\sigma} \ast D)\|_2^2), \quad \sigma > 0,
%\label{eq: coef}
%\end{equation}
with $g(s) = exp(- {s} / {k})$ ($k > 0$). The decreasing function $g$ is added to reduce the gradient penalty on those regions which have a large likelihood to be edges \cite{pm1990pm}. %This likelihood is measured by $\vert \nabla V_{\sigma}^{ax}\vert^2$. 
To make the estimation relatively insensitive to noise, Gaussian smoothing ($K_{\sigma}$) is applied to the parameter fields first. To avoid the undesirable effect that edges might be formed at different locations for different velocity channels, we average over axes to obtain a {\it common} coefficient $\alpha_{\bf{V}}$ at each location \cite{weickert1998anidiff}.

Overall, PIANO estimates ${\bf{V}},\, D$ by minimizing the following sum of losses:%, $\mathcal{L}$:
\begin{equation}
\min\limits_{{\bf{V}},\, D}\, \mathcal{L} = \mathcal{L}_{CC} + \lambda_{\bf{V}}\,\mathcal{L}_{AS_{\bf{V}}} +  \lambda_D\,\mathcal{L}_{AS_D},\quad \lambda_{\bf{V}},\, \lambda_D > 0.
\label{eq: loss}
\end{equation}
%where $\lambda_{\bf{V}},\, \lambda_D > 0$ are weighting factors for $\mathcal{L}_{AS_{\bf{V}}}$ and $\mathcal{L}_{AS_D}$.

%% file: sub/exp_results/2d_demo.tex
\begin{figure}
	\noindent\resizebox{\textwidth}{!}{
	\begin{tikzpicture}
	\node at (0, 0) {\includegraphics[width=1.9cm]{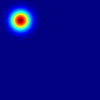}};
	\node at (2, 0) {\includegraphics[width=1.9cm]{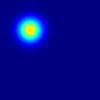}};
	\node at (4, 0) {\includegraphics[width=1.9cm]{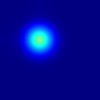}};
	\node at (6, 0) {\includegraphics[width=1.9cm]{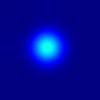}};
	\node at (8, 0) {\includegraphics[width=1.9cm]{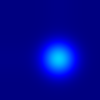}};
	\node at (10, 0) {\includegraphics[width=1.9cm]{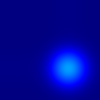}};
	
	\node at (0, -2) {\includegraphics[width=1.9cm]{fig/2d_demo/gt_1.png}};
	\node at (2, -2) {\includegraphics[width=1.9cm]{fig/2d_demo/gt_400.png}};
	\node at (4, -2) {\includegraphics[width=1.9cm]{fig/2d_demo/gt_800.png}};
	\node at (6, -2) {\includegraphics[width=1.9cm]{fig/2d_demo/gt_1200.png}};
	\node at (8, -2) {\includegraphics[width=1.9cm]{fig/2d_demo/gt_1600.png}};
	\node at (10, -2) {\includegraphics[width=1.9cm]{fig/2d_demo/gt_2000.png}};
	\node at (0,-3.2)  {{\footnotesize{$t = 0$}}};
	\node at (2,-3.2)  {{\footnotesize{$t = 4$}}};
	\node at (4,-3.2)  {{\footnotesize{$t = 8$}}};
	\node at (6,-3.2)  {{\footnotesize{$t = 12$}}};
	\node at (8,-3.2)  {{\footnotesize{$t = 16$}}};
	\node at (10,-3.2)  {{\footnotesize{$t = 20$}}};
	
	%\node[label=below:\rotatebox{90}{\footnotesize Simulation}] at (-1.2,1) {};
	%\node[label=below:\rotatebox{90}{\footnotesize Prediction}] at (-1.2,-1) {};
	\node at (-1.25, 0) {\footnotesize (a)};
	\node at (-1.25, -2) {\footnotesize (b)};
	\node at (11.2, -1) {\includegraphics[width=0.4cm]{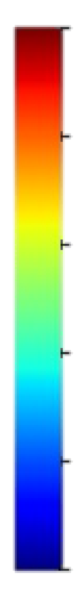}};
	\node at (11.575, 0.45) {\tiny{$1.0$}};
	\node at (11.575, -0.125) {\tiny{$0.8$}};
	\node at (11.575, -0.7) {\tiny{$0.6$}};
	\node at (11.575, -1.275) {\tiny{$0.4$}};
	\node at (11.575, -1.85) {\tiny{$0.2$}};
	\node at (11.575, -2.425) {\tiny{$0.0$}};

	\end{tikzpicture}
	}
	\caption{Toy example of 2D PIANO estimation. (a) Simulated advection-diffusion process with constant velocity and diffusivity; (b) Estimated advection-diffusion process from $t=0$. PIANO successfully captures the advection-diffusion process.} % Settings: D = 0.0004, V = (Vx, Vy) = (0.012, 0.012), Voxel spacing = (0.5, 0.5, dt = 0.01); Ratio of mean absolute error: 7e-5%\caption{Toy illustration of PIANO estimation in 2D. (a) Simulated advection-diffusion process with constant diffusivity and velocity; (b) Predicted advection-diffusion process starting from $t=0$.} % Settings: D = 0.0004, V = (Vx, Vy) = (0.012, 0.012), Voxel spacing = (0.5, 0.5, dt = 0.01); Ratio of mean absolute error: 7e-5%
	\label{2d_demo}
\end{figure}

%% file: sub/fig_framework_tmi.tex
\begin{figure}[t]
	\noindent\resizebox{\textwidth}{!}{
	\begin{tikzpicture}
		% Define the macro.
		% 1st argument: Height and width of the layer rectangle slice.
		% 2nd argument: Depth of the layer slice
		% 3rd argument: X Offset --> use it to offset layers from previously drawn layers.
		% 4th argument: Options for filldraw.
		% 5th argument: Text to be placed below this layer.
		% 6th argument: Y Offset --> Use it when an output needs to be fed to multiple layers that are on the same X offset.
		\newcommand{\networkLayer}[6]{
			\def\a{#1} % Used to distinguish input resolution for current layer.
			\def\b{0.02}
			\def\c{#2} % Width of the cube to distinguish number of input channels for current layer.
			\def\t{#3} % X offset for current layer.
			\def\d{#4} % Y offset for current layer.

			% Draw the layer body.
			\draw[line width=0.3mm](\c+\t,0,\d) -- (\c+\t,\a,\d) -- (\t,\a,\d);                                                      % back plane
			\draw[line width=0.3mm](\t,0,\a+\d) -- (\c+\t,0,\a+\d) node[midway,below] {#6} -- (\c+\t,\a,\a+\d) -- (\t,\a,\a+\d) -- (\t,0,\a+\d); % front plane
			\draw[line width=0.3mm](\c+\t,0,\d) -- (\c+\t,0,\a+\d);
			\draw[line width=0.3mm](\c+\t,\a,\d) -- (\c+\t,\a,\a+\d);
			\draw[line width=0.3mm](\t,\a,\d) -- (\t,\a,\a+\d);

			% Recolor visible surfaces
			\filldraw[#5] (\t+\b,\b,\a+\d) -- (\c+\t-\b,\b,\a+\d) -- (\c+\t-\b,\a-\b,\a+\d) -- (\t+\b,\a-\b,\a+\d) -- (\t+\b,\b,\a+\d); % front plane
			\filldraw[#5] (\t+\b,\a,\a-\b+\d) -- (\c+\t-\b,\a,\a-\b+\d) -- (\c+\t-\b,\a,\b+\d) -- (\t+\b,\a,\b+\d);
			% Colored slice.
			\ifthenelse {\equal{#5} {}}
			{} % Do not draw colored slice if #4 is blank.
			{\filldraw[#5] (\c+\t,\b,\a-\b+\d) -- (\c+\t,\b,\b+\d) -- (\c+\t,\a-\b,\b+\d) -- (\c+\t,\a-\b,\a-\b+\d);} % Else, draw a colored slice.
		}
		
		% Denotations
		%\draw [dash dot] (5.85,1.05) -- (11.5,1.05) -- (11.5,-0.5) -- (5.85,-0.5) -- (5.85,1.05);
		\filldraw[draw=black,fill=blue!15] (6.2,0.73) rectangle (6.95,0.63);
		\node at (8.925,0.67){\begin{tabular}{c}{Measured concentration}\end{tabular}};
		\filldraw[draw=black,fill=red!15] (6.2,0.33) rectangle (6.95,0.23);
		\node at (8.93,0.28){\begin{tabular}{c}{Predicted concentration}\end{tabular}};
		\filldraw[draw=black,fill=orange!30] (6.2,-0.07) rectangle (6.95,-0.17);
		\node at (8.8,-0.12){\begin{tabular}{c}{Estimated parameters}\end{tabular}};

		% Time Series Dataset
		% Box gap: 0.6
		% Dot gap: 0.25
		\draw [dash dot] (-1.4,1.8) -- (4.85,1.8) -- (4.85,-1) -- (-1.4,-1) -- (-1.4,1.8);
		\node at (1.725,1.5){\begin{tabular}{c}{{\bf{Dataset: }}CAs Concentration Time-Series}\end{tabular}};
		\networkLayer{1.0}{0.1}{-0.5}{0.0}{color=gray!20}{}
		\node at (-0.75,-0.7) {$C^{t_0}$};
		\networkLayer{1.0}{0.1}{0.1}{0.0}{color=gray!20}{}
		\node at (-0.15,-0.7) {$C^{t_1}$};
		\networkLayer{1.0}{0.1}{0.7}{0}{color=gray!20}{}
		\node at (0.45,-0.7) {$C^{t_2}$};
		\draw[gray,fill=gray] (0.75,0.28) circle (1pt);
		\draw[gray,fill=gray]  (1,0.28) circle (1pt);
		\draw[gray,fill=gray]  (1.25,0.28) circle (1pt);
		\networkLayer{1.0}{0.1}{1.8}{0}{color=blue!15}{}
		\node at (1.55,-0.7) {\color{blue!70}$C^{t_i}$};
		\draw[blue!70,fill=blue!70] (1.85,0.28) circle (1pt);
		\draw[blue!70,fill=blue!70] (2.1,0.28) circle (1pt);
		\draw[blue!70,fill=blue!70] (2.35,0.28) circle (1pt);
		\networkLayer{1.0}{0.1}{2.9}{0}{color=blue!15}{}
		\node at (2.9,-0.74) {\color{blue!70}$C^{t_{i + T_\text{pd}}}$};
		\draw[gray,fill=gray]  (2.95,0.28) circle (1pt);
		\draw[gray,fill=gray]  (3.2,0.28) circle (1pt);
		\draw[gray,fill=gray]  (3.45,0.28) circle (1pt);
		\networkLayer{1.0}{0.1}{4}{0}{color=gray!20}{}
		\node at (3.9,-0.7) {$C^{t_T}$};
		
		\draw [decorate,decoration={brace,amplitude=3pt,mirror,raise=4ex},line width=1.25pt,color = blue!50]
  (1.55,-0.5) -- (2.9,-0.5);
		\tikzstyle{myarrows}=[line width=0.7mm,draw=blue!50,-triangle 45,postaction={draw, line width=0.1mm, shorten >=0.1mm, -}]
		\draw [myarrows](2.225,-1.2)--(2.225,-2.2);
		\node at (4.7,-1.45){\begin{tabular}{c}{\color{blue!70}\emph{Input sample $S_i$, starting from}}\end{tabular}};
		\node at (5.325,-1.78){\begin{tabular}{c}{\color{blue!70}\emph{randomly selected initial condition $C^{t_i}$}}\end{tabular}};

		% Forward-in-time Block
		\node at (-0.05,-2){\begin{tabular}{c}{\bf{Forward in Time}}\end{tabular}};
		% Dot gap: 0.25
		\draw [dash dot] (-1.4,-2.25) -- (8,-2.25) -- (8,-7) -- (-1.4,-7) -- (-1.4,-2.25);
		% Ground Truth
		\networkLayer{1.0}{0.1}{2.9765}{9}{color=blue!15}{}
		\node at (-0.75,-4.2) {\color{blue!70}$C^{t_i}$};
		\networkLayer{1.0}{0.1}{6.11}{9}{color=blue!15}{}
		\node at (2.6,-4.2) {\color{blue!70}$C^{t_{i+1}}$};
		\draw[blue!70,fill=blue!70] (4,-3.25) circle (1pt);
		\draw[blue!70,fill=blue!70] (4.75,-3.25) circle (1pt);
		\draw[blue!70,fill=blue!70] (5.5,-3.25) circle (1pt);
		\networkLayer{1.0}{0.1}{10.7}{9}{color=blue!15}{}
		\node at (7.25,-4.2) {\color{blue!70}$C^{t_{i + T_\text{pd}}}$};

		% Prediction
		% Box gap: 1.2
		% Arrow: 0.85
		\networkLayer{1.0}{0.1}{5.865}{16.5}{color=blue!15}{}
		\node at (-0.75,-5.1) {\color{blue!70}$C^{t_i}$};
		\draw [->, line width = 0.3mm, color = red!60](-0.6,-6)--(0.25,-6);
		\networkLayer{1.0}{0.1}{7.065}{16.5}{color=red!15}{}
		\node at (0.7,-5.05) {\color{red!60}$\widehat{C}^{t_i+\delta t}$};
		\draw [->, line width = 0.3mm, color = red!60](0.6,-6)--(1.45,-6);
		\draw[red!60,fill=red!60] (1.6,-6) circle (1pt);
		\draw[red!60,fill=red!60] (1.85,-6) circle (1pt);
		\draw[red!60,fill=red!60] (2.1,-6) circle (1pt);
		\networkLayer{1.0}{0.1}{9}{16.5}{color=red!15}{}
		\node at (2.6,-5.05) {\color{red!60}$\widehat{C}^{t_{i+1}}$};
		\draw [->, line width = 0.3mm, color = red!60](2.55,-6)--(3.4,-6);
		\networkLayer{1.0}{0.1}{10.2}{16.5}{color=red!15}{}
		\node at (4,-5.05) {\color{red!60}$\widehat{C}^{t_{i+1}+\delta t}$};
		\draw [->, line width = 0.3mm, color = red!60](3.75,-6)--(4.6,-6);
		\draw[red!60,fill=red!60] (4.75,-6) circle (1pt);
		\draw[red!60,fill=red!60] (5,-6) circle (1pt);
		\draw[red!60,fill=red!60] (5.25,-6) circle (1pt);
		\draw[red!60,fill=red!60] (5.5,-6) circle (1pt);
		\draw[red!60,fill=red!60] (5.75,-6) circle (1pt);
		\draw [->, line width = 0.3mm, color = red!60](5.85,-6)--(6.7,-6);
		\networkLayer{1.0}{0.1}{13.59}{16.5}{color=red!15}{}
		\node at (7.25,-5.05) {\color{red!60}$\widehat{C}^{t_{i + T_\text{pd}}}$};
		%\node at (4.8,-7.25) {\color{red!60}\bf{\Cref{eq: div_free}} and BC};
		\node at (0.6,-7.75) {\color{red!60}\bf{\Cref{eq: div_free}} and BC};
		
		% V-D Block
		\node at (-0.5,-8.45){\begin{tabular}{c}{\bf{Parameters}}\end{tabular}};
		\draw [dash dot] (-1.4,-8.7) -- (6,-8.7) -- (6,-11) -- (-1.4,-11) -- (-1.4,-8.7);
		\draw [thick, color=red!60] (-0.6,-6) -- (-0.6,-7.5) -- (0.6,-7.5) -- (0.6,-6);
		\draw [thick, color=red!60] (0.6,-7.5) -- (2.55,-7.5) -- (2.55,-6);
		\draw [thick, color=red!60] (2.55,-7.5) -- (3.75,-7.5) -- (3.75,-6);
		\draw [thick, color=red!60] (3.75,-7.5) -- (5.85,-7.5) -- (5.85,-6);
		\node at (3.95,-8.25){\begin{tabular}{c}{\color{orange}\emph{Apply updated $\bf{D},\,\bf{V}$}}\end{tabular}};
		\tikzstyle{myarrows}=[line width=0.7mm,draw=orange!85,-triangle 45,postaction={draw, line width=0.1mm, shorten >=0.1mm, -}]
		\draw [myarrows](2.225,-8.5) -- (2.225,-7.52);
		\draw [decorate,decoration={brace,amplitude=3pt,raise=4ex},line width=1.25pt,color = orange!85]
  (1.55,-9.2) -- (2.9,-9.2);
		\networkLayer{1.0}{0.1}{9.8}{26}{color=orange!30}{}
		\node at (0.725,-9.35) {\color{orange}\bf{\Cref{eq: d_scalar}}};
		\draw [->, line width = 0.3mm, color = orange](0.15,-9.7)--(1.2,-9.7);
		\node at (-0.55,-10.7) {\color{orange}$L$};
		\networkLayer{1.0}{0.1}{11.82}{26}{color=orange!30}{}
		\node at (1.5,-10.7) {\color{orange}$\bf{D}$};
		\networkLayer{1.0}{0.1}{12.92}{26}{color=orange!30}{}
		\node at (2.6,-10.7) {\color{orange}$\bf{V}$};
		\node at (3.775,-9.35) {\color{orange}\bf{\Cref{eq: v_div_free}}};
		\draw [->, line width = 0.3mm, color = orange](4.35,-9.7)--(3.2,-9.7);
		\networkLayer{1.0}{0.1}{15.3}{25.5}{color=orange!30}{}
		\node at (5.71,-10.1) {\color{orange}$\Gamma_1$};
		\networkLayer{1.0}{0.1}{15.3}{27.2}{color=orange!30}{}
		\node at (5.05,-10.75) {\color{orange}$\Gamma_2$};
		
		% Loss Block
		\node at (10.35,-2){\begin{tabular}{c}{\bf{Losses}}\end{tabular}};
		\draw [dash dot] (9.875,-2.25) -- (11.6,-2.25) -- (11.6,-11) -- (9.875,-11) -- (9.875,-2.25);
		\draw [decorate,decoration={brace,amplitude=5pt,raise=5ex},line width=1.75pt,color = gray]
  (7.5,-3.25) -- (7.5,-6);
  		\draw [line width = 1.75pt, color = gray, ->] (8.5,-4.62) -- (9.8,-4.62);
		\node at (9.1,-4.3) {\color{gray}\bf{\Cref{eq: loss_cc}}};
		\node at (10.4,-4.62) {\color{gray}$\bf{\mathcal{L}_{CC}}$};
		\draw [decorate,decoration={brace,amplitude=5pt,raise=5ex},line width=1.75pt,color = gray]
  (5.5,-8.7) -- (5.5,-11);
		\draw [line width = 1.75pt, color = gray, ->] (6.5,-9.85) -- (9.8,-9.85);
		\node at (8.1,-9.5) {\color{gray}\bf{\Cref{eq: ani_loss,eq: coef}}};
		\node at (10.4,-9.85) {\color{gray}$\bf{\mathcal{L}_{AS}}$};
		\draw [decorate,decoration={brace,amplitude=5pt,raise=5ex},line width=1.75pt,color = gray]
  (10.1,-4.65) -- (10.1,-9.85);
		%\node at (10.75,-10.5) {\color{gray}\bf{\Cref{eq: loss}}};
		\node at (10.75,-3.5) {\color{gray}\bf{\Cref{eq: loss}}};
		\node at (11.325,-7.25) {\color{gray}$\bf{\mathcal{L}}$};
		
	\end{tikzpicture}
	}
	\caption{Estimation framework of PIANO for one iteration (See Alg.~\ref{algorithm} for the entire estimation approach), given training sample $S^i = \{ C^{t_j} \vert j = i,\, i+1,\, \ldots,\, i + T_\text{pd} \}$.}

	\label{fw}
\end{figure}
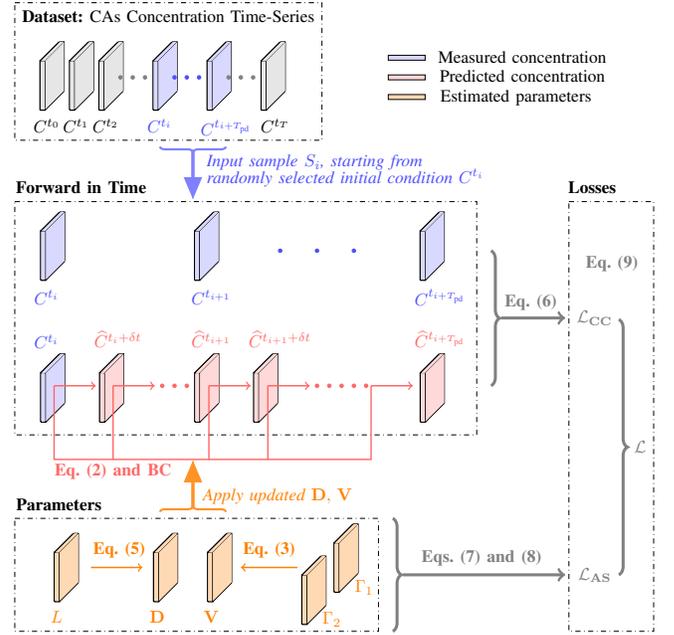

%% file: sub/pseudocode.tex
\begin{algorithm}[t]
\caption{Pseudo-code for PIANO}
\label{algorithm}
{\small
  \Input{Time series of CA concentration images $\{C^{t_i} \in \mathbb{R}(\Omega) \vert\, i = 0,\, 1,\, \ldots,\, T\}$}
    \Output{Estimated ${\bf{V}}$ and ${\bf{D}}$, predicted CA concentrations $\{\widehat{C}^{t_i} \in \mathbb{R}(\Omega) \vert\, i = 0,\, 1,\, \ldots,\, T\}$}
  %\Output{(a) Estimated velocity field ${\bf{V}}$ and diffusion field ${\bf{D}}$ \\ ~~~~~~~~~~~~ (b) Predicted CA concentration maps after $t_0$ with time interval $\delta t$}
  \Settings{$\lambda_{\bf{V}},\, \sigma_{\bf{V}},\, \lambda_{\bf{D}},\, \sigma_{\bf{D}},\, k,\, \sigma$ in \Cref{eq: loss,eq: loss_cc,eq: ani_loss,eq: coef}, $\delta t$, $T_\text{pd}$, $lr$}
 % \Dataset{$S_i = \{ C^{t_j} \vert j = i,\, i+1,\, \ldots,\, i + T_\text{pd} \},\quad i = 0,\, 1,\, \ldots,\, T - T_\text{pd}$}
  \Initialization{$\Gamma_1({\bf{x}}),\, \Gamma_2({\bf{x}}),\, L({\bf{x}}) \sim 0.001\times\mathcal{N}(0,\,1),\quad \forall {\bf{x}} \in \Omega$}

%\tcc{More details of one PIANO estimation iteration shown in Fig. \ref{fw}}
\While{$\mathcal{L}$ not converged}   
   {
   Randomly select sample $S_i = \{ C^{t_j} \vert j = i,\, i+1,\, \ldots,\, i + T_\text{pd} \}$ from $\{C^{t_i}\}$\\  % \tcp*{Add comment} ,\quad i = 0,\, 1,\, \ldots,\, T - T_\text{pd}
   \For{$t = t_i + \delta t,\, \ldots,\, t_{i+1},\, t_{i+1} + \delta t,\, \ldots,\, t_{i+T_{\text{pd}}} $}
   { \label{time step}
   Discretize in space and compute advection-diffusion PDE via \Cref{eq: div_free}\\
   Impose the mixed boundary condition and integrate in time to obtain $\widehat{C}^{t+\delta t}$\\
   }
   %Predict concentration time-series: $\widehat{S}_{i} = \{ \widehat{C}^{t_j} \vert j = i,\, i+1,\, \ldots,\, i + T_\text{pd} \}$\\
   Compute $\mathcal{L}$ (\Cref{eq: loss}) and propagate backward (SGD with momentum)\\
   Update $\Gamma_1,\, \Gamma_2,\, L$ by learning rate $lr$ and update ${\bf{V}},\,{\bf{D}}$ via \Cref{eq: v_div_free,eq: d_scalar}
   }
Predict the entire concentration time-series $\{ \widehat{C}^{t_i} \vert i = 0,\, 1,\, \ldots,\, T \}$ starting from $C^{t_0}$}
\algorithmfootnote{\emph{{$T_{\text{pd}}$}: the number of consecutive time points in one training sample;} \\ 
{\emph{Convergence criterion: ${\vert\mathcal{L}\text{ of current iteration} - \mathcal{L} \text{ of last iteration}\vert} / {\mathcal{L} \text{ of}}$\\${\text{last iteration}}< 0.001$ for 10 subsequent iterations.}}}
\end{algorithm}

%% file: sub/exp_results.tex
\section{Experimental Results}
\label{sec:experimental_results}
We tested PIANO on the Ischemic Stroke Lesion Segmentation (ISLES) 2017~\cite{isles2015b,isles2015a} dataset. %The %training
The dataset includes images for 43 ischemic stroke patients. Each patient has the following images: an apparent diffusion coefficient (ADC) map, a 4D dynamic susceptibility contrast (DSC) MR perfusion image (from 40 to 80 available time points; temporal resolution $\approx 1~s$)~\cite{essig2013mrp}, and a segmented lesion map %established by T2w or FLAIR scans 
viewed as the gold-standard lesion. For each patient the dataset also includes five perfusion summary maps: (1) Cerebral blood flow (CBF); (2) Cerebral blood volume (CBV); (3) Mean transit time (MTT); (4) Time to peak (TTP); and (5) Time to peak for the deconvolved residue function (Tmax). In this work, we focus on the ADC map and perfusion parameter maps which correspond to \emph{physical} measures, i.e., CBF and CBV, MTT for further quantitative comparison. \footnote{We do not compare with TTP and Tmax. Specifically, TTP refers to the CA concentration peak time; Tmax is the time need at which the residue function reaches its maximum  which is a relative rather than an absolute measure~\cite{essig2013mrp}. Hence, both measures depend on the onset of perfusion measurements and do not correspond to direct physical tissue measures.}
  %  which are not defined on cerebral structures. TTP refers to the CA concentration peak time, which is neither related to AIF nor the deconvolution; Tmax is the time need at which the residue function reaches maximum, which is more a relative than absolute value~\cite{essig2013mrp}.}}

We first convert DSC MR perfusion images to concentration images using the relation between the MR signal and CA concentration~\cite{fieselmann2011sig2ctc}. Specifically, the concentration can be determined as follows:
\begin{equation}
C(\mathbf{x}, \, t_i) = - \frac{k_{\text{mr}}}{\text{TE}}\, \ln\bigg( \frac{S(\mathbf{x}, \, t_i)}{S_0} \bigg),\quad i = 1,\,...,\, nT,
\label{eq: mrp2ctc}
\end{equation}
where $C(\mathbf{x}, \, t_i), \, S(\mathbf{x}, \, t_i)$ denote the CA concentration value and the received MR signal at voxel position $\mathbf{x}$ and time $t_i$, respectively. $\frac{k_{\text{mr}}}{\text{TE}}$ is a constant of proportionality related to the image acquisition process, which is usually set to $1$ for the sake of simplicity~\cite{fieselmann2011sig2ctc}. The baseline value $S_0$ is obtained by the mean of $S(\mathbf{x},\, t_j)$ during the $B$ acquired time frames before the CA bolus arrival:
\begin{equation}
S_0 = \frac{1}{B} \,\sum_{j= 1}^B\, S(\mathbf{x},\, t_j).
\end{equation}
The original perfusion images are typically anisotropic, with a much larger voxel size along the axial ($6.5~mm$) than in the other two directions ($1.2~mm$). To obtain a more uniform computational grid for the model, we upsample each concentration image along the axial direction (to $1.3~mm$ grid size) using the Lanczos Windowed Sinc method~\cite{Meijering1999lancoz}. Then we create a concentration time-series dataset for each patient $N$: $\{C^{t_i} \in \mathbb{R}(\Omega) \vert\, i = 0,\, 1,\, \ldots,\, T_N\}$, starting from the time when the total concentration over the entire brain reaches its maximum, at which we assume the CA has been fully transported into the brain, till the last available time point. We test PIANO on all patients with identical model settings. Specifically, we set $\lambda_{\bf{V}} = \lambda_D = 0.1$ (\Cref{eq: loss}). In \Cref{eq: coef}, $\sigma = 0.6$; $k$ was treated as a `noise estimator' \cite{pm1990pm}, where a histogram of the absolute values of the gradient throughout the current image was computed, and $k$ was set as 90\% of the histogram's integral at every estimating iteration. Throughout the estimation, the prediction temporal resolution is $\delta t = 0.02~s$, and $T_\text{pd} = \lfloor{\frac{T_k}{3}}\rfloor$. (See Alg.~\ref{algorithm}.)%The convergence and stopping criterion for PIANO are described in Algorithm \ref{algorithm}.

\subsection{PIANO Feature Maps}
\label{sec:piano_feature_maps}
%%%%%%%%%%%%%%%%%%%%%%%%%%%%%%%%%
\input{sub/exp_results/param_14}
\input{sub/exp_results/param_16}
\input{sub/exp_results/ctc_14}
\input{sub/exp_results/ctc_16}
 
%%%%%%%%%%%%%%%%%%%%%%%%%%%%%%%%%

For a better insight into an estimated velocity field ${\bf{V}}$ and diffusion field ${\bf{D}}$, we compute the following maps: (1) ${\bf{V}}_{rgb}$: Color-coded orientation map of ${\bf{V}} = (V^x, V^y, V^z)^T$, obtained by normalizing ${\bf{V}}$ to unit length and mapping its 3 components to red, green, blue respectively; (2) $\| {\bf{V}} \|_2$: $2$ norm of ${\bf{V}}$; (3) $D$: scalar field in \Cref{eq: d_scalar}.

Fig. \ref{M14_param} and Fig. \ref{M16_param} show the PIANO feature maps estimated from two ISLES 2017 patients: all are highly consistent with the lesion in both cases. Details of the blood flow trajectories are revealed in ${\bf{V}}_{rgb}$ by the ridged patterns and the sharp changes of colors in the unaffected (right) hemisphere, while the flat patterns appearing within the lesion provide little directional information about the velocity and indicate low velocity magnitudes. Velocity magnitudes are more directly visualized via $\|{\bf{V}} \|_2$, from which one can easily locate the lesion where $\|{\bf{V}} \|_2$ is low. $D$ also indicates lower diffusion values in the lesion, though with less contrast potentially due to the fact that it captures the accumulated effect of CA diffusion at the voxel-level. %Based on extensive experiments (not shown due to page limits) PIANO feature maps are resilient to noise (i.e., estimates also get noisy, but generally retain their structure) and can disambiguate ${\bf V}$ and ${D}$.

%%%%%%%%%%%%%%%%%%%%%%%%%%%%%%%%%

\subsection{Predicted CA Concentration}
\label{sec:predicted_concentration}
To better illustrate the prediction accuracy, and therefore the estimation accuracy of ${\bf V}$ and ${D}$, of PIANO, we provide the corresponding predicted time-series of CA concentration images in Fig.~\ref{M14_ctc} and Fig.~\ref{M16_ctc} for the same patients in Fig.~\ref{M14_param} and Fig.~\ref{M16_param}, respectively. We see that PIANO is capable of predicting the CA concentration given their initial state, indicating its ability to successfully capture ${\bf V}$ and ${D}$. Note that although the concentration values for these two patients differ considerably, caused by the different total volume of injected CA, PIANO is still able to provide plausible estimates.
 
%%%%%%%%%%%%%%%%%%%%%%%%%%%%%%%%%

\subsection{Quantitative Comparison}
\label{sec:quantiative_comparison}

To quantitatively compare PIANO feature maps with the maps provided by ISLES 2017 in their ability to detect the lesion, we compare feature values in the lesion with the values in the contralateral region of the lesion (c-lesion). The c-lesion region is determined by mirroring the lesion to the unaffected side via the midline of the cerebral hemispheres. Values in the c-lesion provide a reference for the normal values. We consider the following three metrics for comparison between the different maps: (1) Relative mean value between lesion and c-lesion ($\mu^r \in [0, 1]$): 
\begin{equation}
\mu^r = min\{\frac{\text{mean in lesion}}{\text{mean in c-lesion}}, \frac{\text{mean in c-lesion}}{\text{mean in lesion}}\};
\label{eq: rel_mean}
\end{equation}
(2) Relative standard deviation (STD) between lesion and c-lesion ($\sigma^r \in [0, 1]$):
\begin{equation}
\sigma^r = min\{\frac{\text{STD in lesion}}{\text{STD in c-lesion}}, \frac{\text{STD in c-lesion}}{\text{STD in lesion}}\};
\label{eq: rel_std}
\end{equation}
(3) Absolute t-value\footnote{PIANO feature maps, ADC, CBF and CBV typically have smaller values in the lesion than c-lesion, and therefore a negative t-statistic between the values of the lesion and the c-lesion. While the case for MTT is opposite: values in the lesion are typically larger than c-lesion due to its definition, resulting in a positive t-statistic between values in the lesion and c-lesion. For more explicit measurements of the differences between lesion and c-lesion, we take the minimum of fractions in Eq.~(\ref{eq: rel_mean}-\ref{eq: rel_std}) and absolute value of the t-statistic.}: the absolute value of unpaired t-statistic between the values in the lesion and the c-lesion\footnote{While a paired test between corresponding voxels is possible and results in similar measures, we opt for the unpaired test to avoid voxel-level correspondence issues.}

%%%%%%%%%%%%%%%%%%%%%%%%%%%%%%%%%
%\input{sub/exp_results/box_plots_tmi}
\input{sub/exp_results/box_plots_tmi2}
\input{sub/exp_results/table_tmi2_4maps}
Fig. \ref{box_plots} compares the PIANO and ISLES 2017 maps based on the above three metrics computed from 43 patients, where $\mu^r$ of $\| {\bf{V}} \|_2$ achieves the lowest value, meaning more significant differences between lesion and c-lesion.
Moreover, Fig. \ref{box_plots} (b) shows $\| {\bf{V}} \|_2$ reveals much stronger differences between a lesion and its c-lesion compared to all other maps. Tab.~\ref{table} summarizes results over all patients. The most distinguishing results are obtained from PIANO feature maps.

%% file: sub/exp_results/param_14.tex
\begin{figure}[t]
	\noindent\resizebox{\textwidth}{!}{
	\begin{tikzpicture}
	% Lesion
	\node at (-1.5,2.25) {\footnotesize{Lesion}};
	%\node[label=below:\rotatebox{90}{\footnotesize Lesion}] at (-1.3,2.9) {};
	\node at (0, 2.25) {\includegraphics[width=1.9cm]{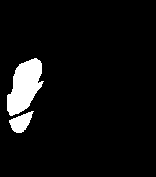}};
	\node at (2, 2.25) {\includegraphics[width=1.9cm]{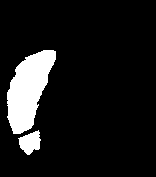}};
	\node at (4, 2.25) {\includegraphics[width=1.9cm]{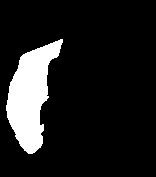}};
	\node at (6, 2.25) {\includegraphics[width=1.9cm]{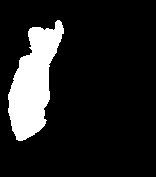}};
	\node at (8, 2.25) {\includegraphics[width=1.9cm]{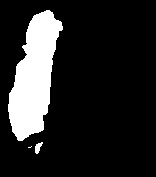}};
	\node at (10, 2.25) {\includegraphics[width=1.9cm]{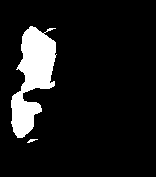}};

	% V color-by-orientation
	\node at (-1.5,0) {\footnotesize{${\bf{V}}_{rgb}$}};
	%\node[label=below:\rotatebox{90}{\footnotesize ${\bf{V}}_{rgb}$}] at (-1.3,0.65) {};
	\node at (0, 0) {\includegraphics[width=1.9cm]{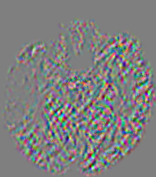}};
	\node at (2, 0) {\includegraphics[width=1.9cm]{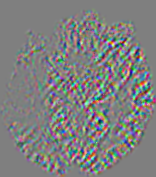}};
	\node at (4, 0) {\includegraphics[width=1.9cm]{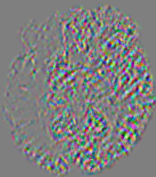}};
	\node at (6, 0) {\includegraphics[width=1.9cm]{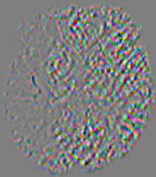}};
	\node at (8, 0) {\includegraphics[width=1.9cm]{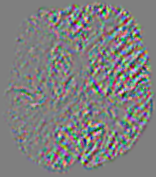}};
	\node at (10, 0) {\includegraphics[width=1.9cm]{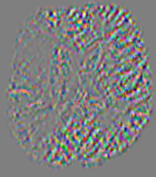}};
	
	% Norm V
	\node at (-1.5, -2.25) {\footnotesize{${\|\bf{V}} \|_2$}};
	%\node[label=below:\rotatebox{90}{\footnotesize ${\Vert \bf{V}} \Vert$}] at (-1.3,-1.6) {};
	\node at (0, -2.25) {\includegraphics[width=1.9cm]{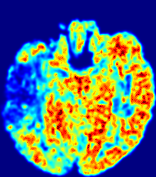}};
	\node at (2, -2.25) {\includegraphics[width=1.9cm]{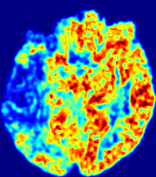}};
	\node at (4, -2.25) {\includegraphics[width=1.9cm]{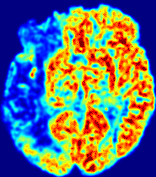}};
	\node at (6, -2.25) {\includegraphics[width=1.9cm]{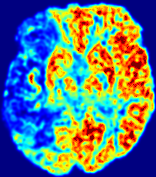}};
	\node at (8, -2.25) {\includegraphics[width=1.9cm]{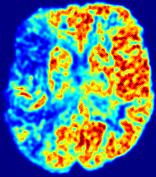}};
	\node at (10, -2.25) {\includegraphics[width=1.9cm]{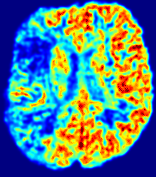}};
	% Color bar shift: +0.1; Text shift: +1; Text Gap: -0.36
	\node at (11.12, -2.15) {\includegraphics[width=0.25cm]{fig/cb.png}};
	\node at (11.56, -1.25) {\tiny{$3.5$}};
	\node at (11.56, -1.61) {\tiny{$2.8$}};
	\node at (11.56, -1.97) {\tiny{$2.1$}};
	\node at (11.56, -2.33) {\tiny{$1.4$}};
	\node at (11.56, -2.69) {\tiny{$0.7$}};
	\node at (11.56, -3.025) {\tiny{$0.0$}};
	\node at (11.4, -3.235) {\tiny{$(mm/s)$}};
	
	% D
	\node at (-1.5, -4.5) {\footnotesize{$D$}};
	%\node[label=below:\rotatebox{90}{\footnotesize $D$}] at (-1.3,-3.85) {};
	\node at (0, -4.5) {\includegraphics[width=1.9cm]{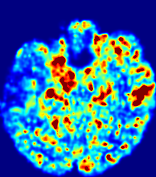}};
	\node at (2, -4.5) {\includegraphics[width=1.9cm]{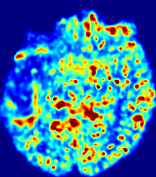}};
	\node at (4, -4.5) {\includegraphics[width=1.9cm]{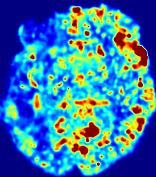}};
	\node at (6, -4.5) {\includegraphics[width=1.9cm]{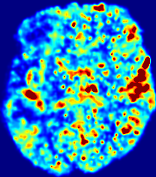}};11.56
	\node at (8, -4.5) {\includegraphics[width=1.9cm]{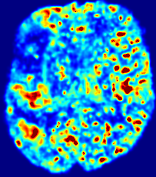}};
	\node at (10, -4.5) {\includegraphics[width=1.9cm]{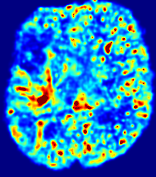}};
	% Color bar shift: +0.1; Unit shift: -0.985;  Text shift: +1; Text Gap: -0.36
	\node at (11.12, -4.35) {\includegraphics[width=0.25cm]{fig/cb.png}};
	\node at (11.56, -3.45) {\tiny{$0.020$}};
	\node at (11.56, -3.81) {\tiny{$0.016$}};
	\node at (11.56, -4.17) {\tiny{$0.012$}};
	\node at (11.56, -4.53) {\tiny{$0.008$}};
	\node at (11.56, -4.89) {\tiny{$0.004$}};
	\node at (11.56, -5.235) {\tiny{$0.000$}};
	\node at (11.45, -5.48) {\tiny{$(mm^2/s)$}};

	% x axis legend
	\node at (0,-5.9)  {{\footnotesize{Slice \#1}}};
	\node at (2,-5.9)  {{\footnotesize{Slice \#2}}};
	\node at (4,-5.9)  {{\footnotesize{Slice \#3}}};
	\node at (6,-5.9)  {{\footnotesize{Slice \#4}}};
	\node at (8,-5.9)  {{\footnotesize{Slice \#5}}};
	\node at (10,-5.9)  {{\footnotesize{Slice \#6}}};

	\end{tikzpicture}
	}
	\caption{PIANO feature maps for one stroke patient, where the lesion is located in the left hemisphere. Top row: segmented stroke lesion region (white) on different slices, obtained from ISLES 2017. The corresponding slices for the PIANO feature maps are shown in the following rows.}
	\label{M14_param}
\end{figure}

%% file: sub/exp_results/param_16.tex
\begin{figure}[t]
	\noindent\resizebox{\textwidth}{!}{
	\begin{tikzpicture}
	% Lesion
	\node at (-1.5,2.25) {\footnotesize{Lesion}};
	%\node[label=below:\rotatebox{90}{\footnotesize Lesion}] at (-1.3,2.9) {};
	\node at (0, 2.25) {\includegraphics[width=1.9cm]{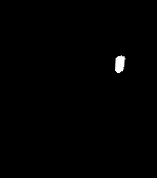}};
	\node at (2, 2.25) {\includegraphics[width=1.9cm]{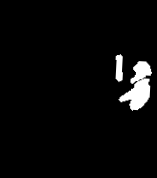}};
	\node at (4, 2.25) {\includegraphics[width=1.9cm]{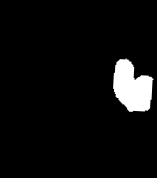}};
	\node at (6, 2.25) {\includegraphics[width=1.9cm]{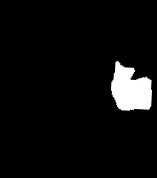}};
	\node at (8, 2.25) {\includegraphics[width=1.9cm]{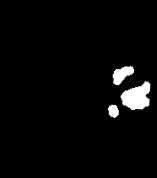}};
	\node at (10, 2.25) {\includegraphics[width=1.9cm]{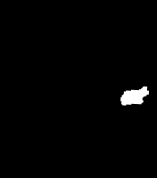}};

	% V color-by-orientation
	\node at (-1.5,0) {\footnotesize{${\bf{V}}_{rgb}$}};
	%\node[label=below:\rotatebox{90}{\footnotesize ${\bf{V}}_{rgb}$}] at (-1.3,0.65) {};
	\node at (0, 0) {\includegraphics[width=1.9cm]{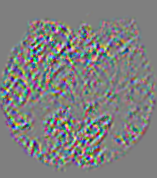}};
	\node at (2, 0) {\includegraphics[width=1.9cm]{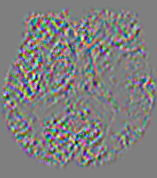}};
	\node at (4, 0) {\includegraphics[width=1.9cm]{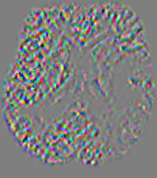}};
	\node at (6, 0) {\includegraphics[width=1.9cm]{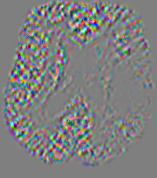}};
	\node at (8, 0) {\includegraphics[width=1.9cm]{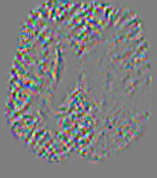}};
	\node at (10, 0) {\includegraphics[width=1.9cm]{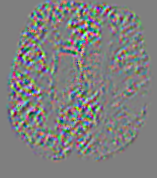}};
	
	% Norm V
	\node at (-1.5, -2.25) {\footnotesize{${\Vert \bf{V}} \Vert_2$}};
	%\node[label=below:\rotatebox{90}{\footnotesize ${\Vert \bf{V}} \Vert$}] at (-1.3,-1.6) {};
	\node at (0, -2.25) {\includegraphics[width=1.9cm]{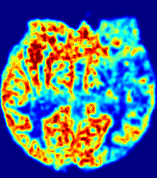}};
	\node at (2, -2.25) {\includegraphics[width=1.9cm]{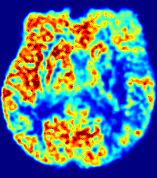}};
	\node at (4, -2.25) {\includegraphics[width=1.9cm]{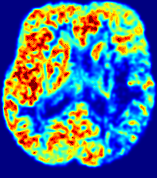}};
	\node at (6, -2.25) {\includegraphics[width=1.9cm]{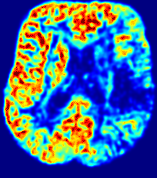}};
	\node at (8, -2.25) {\includegraphics[width=1.9cm]{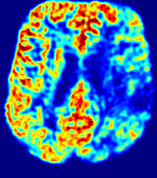}};
	\node at (10, -2.25) {\includegraphics[width=1.9cm]{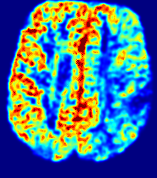}};
	% Color bar shift: +0.1; Text shift: +1; Text Gap: -0.36
	\node at (11.12, -2.15) {\includegraphics[width=0.25cm]{fig/cb.png}};
	\node at (11.56, -1.25) {\tiny{$3.5$}};
	\node at (11.56, -1.61) {\tiny{$2.8$}};
	\node at (11.56, -1.97) {\tiny{$2.1$}};
	\node at (11.56, -2.33) {\tiny{$1.4$}};
	\node at (11.56, -2.69) {\tiny{$0.7$}};
	\node at (11.56, -3.025) {\tiny{$0.0$}};
	\node at (11.4, -3.235) {\tiny{$(mm/s)$}};
	
	% D
	\node at (-1.5, -4.5) {\footnotesize{$D$}};
	%\node[label=below:\rotatebox{90}{\footnotesize $D$}] at (-1.3,-3.85) {};
	\node at (0, -4.5) {\includegraphics[width=1.9cm]{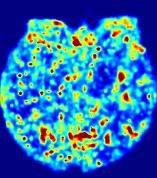}};
	\node at (2, -4.5) {\includegraphics[width=1.9cm]{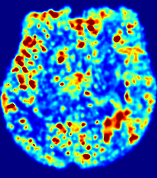}};
	\node at (4, -4.5) {\includegraphics[width=1.9cm]{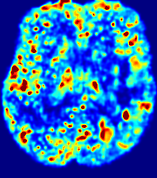}};
	\node at (6, -4.5) {\includegraphics[width=1.9cm]{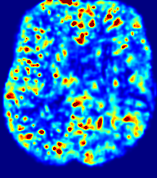}};
	\node at (8, -4.5) {\includegraphics[width=1.9cm]{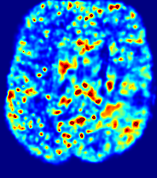}};
	\node at (10, -4.5) {\includegraphics[width=1.9cm]{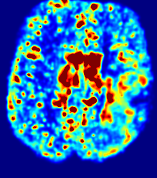}};
	% Color bar shift: +0.1; Unit shift: -0.985;  Text shift: +1; Text Gap: -0.36
	\node at (11.12, -4.35) {\includegraphics[width=0.25cm]{fig/cb.png}};
	\node at (11.56, -3.45) {\tiny{$0.020$}};
	\node at (11.56, -3.81) {\tiny{$0.016$}};
	\node at (11.56, -4.17) {\tiny{$0.012$}};
	\node at (11.56, -4.53) {\tiny{$0.008$}};
	\node at (11.56, -4.89) {\tiny{$0.004$}};
	\node at (11.56, -5.235) {\tiny{$0.000$}};
	\node at (11.45, -5.48) {\tiny{$(mm^2/s)$}};

	% x axis legend
	\node at (0,-5.9)  {{\footnotesize{Slice \#1}}};
	\node at (2,-5.9)  {{\footnotesize{Slice \#2}}};
	\node at (4,-5.9)  {{\footnotesize{Slice \#3}}};
	\node at (6,-5.9)  {{\footnotesize{Slice \#4}}};
	\node at (8,-5.9)  {{\footnotesize{Slice \#5}}};
	\node at (10,-5.9)  {{\footnotesize{Slice \#6}}};

	\end{tikzpicture}
	}
	\caption{PIANO feature maps for another patient in the ISLES 2017 training set, where the lesion is located in the right hemisphere. Top row: segmented stroke lesion region (white) on different slices. The corresponding slices for the PIANO feature maps are shown in the following rows.}
	\label{M16_param}
\end{figure}

%% file: sub/exp_results/ctc_14.tex
\begin{figure}[t]
	\noindent\resizebox{\textwidth}{!}{
	\begin{tikzpicture}
	
	% Slice Gap: 4.75
	
	%%%%%%%%%%%%%%%%%%%%%%%
	%%%%%%%% Denotations %%%%%%%%
	%%%%%%%%%%%%%%%%%%%%%%%
	
	\draw [->, line width = 0.4mm, color = black](-1.75, 23)--(17, 23);
	\node at (16.5, 22.75)  {\large Time};
	\draw [->, line width = 0.4mm, color = black](-1.75, 23)--(-1.75, -7);
	%\node[label=below:\rotatebox{90}{\large Slice}] at (-1.4, -5.95) {};
	\node at (-1.1, -6.8)  {\large Slice};
	
	\node at (-2.1, 20.125) {\footnotesize{(i)}}; 
	\node at (-2.1, 15.375) {\footnotesize{(ii)}}; 
	\node at (-2.1, 10.625) {\footnotesize{(iii)}}; 
	\node at (-2.1, 5.875) {\footnotesize{(iv)}}; 
	\node at (-2.1, 1.125) {\footnotesize{(v)}}; 
	\node at (-2.1, -3.625) {\footnotesize{(vi)}}; 
	
	%\node at (0,-6.15)  {{\footnotesize{$t_0$}}};
	%\node at (2,-6.15)  {{\footnotesize{$t_1$}}};
	%\node at (4,-6.15)  {{\footnotesize{$t_2$}}};
	%\node at (6,-6.15)  {{\footnotesize{$t_3$}}};
	%\node at (8,-6.15)  {{\footnotesize{$t_4$}}};
	%\node at (10,-6.15)  {{\footnotesize{$t_5$}}};
	%\node at (12,-6.15)  {{\footnotesize{$t_6$}}};
	%\node at (14,-6.15)  {{\footnotesize{$t_7$}}};
	\node at (8, -6.35) {\includegraphics[width=6cm]{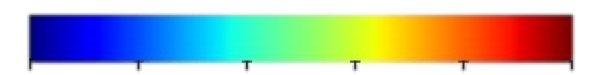}};
	\node at (5.3, -6.9)  {$0$};
	\node at (6.4, -6.9)  {$7$};
	\node at (7.4775, -6.9)  {$14$};
	\node at (8.55, -6.9)  {$21$};
	\node at (9.65, -6.9)  {$28$};
	\node at (10.75, -6.9)  {$35$};

	%\node[label=below:\rotatebox{90}{\footnotesize Measured}] at (-1.2, 22.2) {};
	%\node[label=below:\rotatebox{90}{\footnotesize Measured}] at (-1.2, 17.45) {};
	%\node[label=below:\rotatebox{90}{\footnotesize Measured}] at (-1.2, 12.7) {};
	%\node[label=below:\rotatebox{90}{\footnotesize Measured}] at (-1.2, 7.95) {};
	%\node[label=below:\rotatebox{90}{\footnotesize Measured}] at (-1.2, 3.2) {};
	%\node[label=below:\rotatebox{90}{\footnotesize Measured}] at (-1.2, -1.55) {};
	
	%\node[label=below:\rotatebox{90}{\footnotesize Prediction}] at (-1.2, 19.95) {};
	%\node[label=below:\rotatebox{90}{\footnotesize Prediction}] at (-1.2, 15.2) {};
	%\node[label=below:\rotatebox{90}{\footnotesize Prediction}] at (-1.2, 10.45) {};
	%\node[label=below:\rotatebox{90}{\footnotesize Prediction}] at (-1.2,5.7) {};
	%\node[label=below:\rotatebox{90}{\footnotesize Prediction}] at (-1.2,0.95) {};
	%\node[label=below:\rotatebox{90}{\footnotesize Prediction}] at (-1.2,-3.8) {};
	
	\node at (11.12, -4.35) {\includegraphics[width=0.25cm]{fig/cb.png}};
	
	%%%%%%%%  Slice (i)  %%%%%%%%
	
	\node at (-1.25, 21.25) {\footnotesize{(a)}}; 
	\node at (-1.25, 19) {\footnotesize{(b)}}; 
	
	\node at (0, 21.25) {\includegraphics[width=1.9cm]{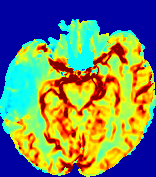}};
	\node at (2, 21.25) {\includegraphics[width=1.9cm]{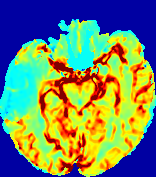}};
	\node at (4, 21.25) {\includegraphics[width=1.9cm]{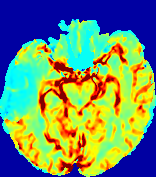}};
	\node at (6, 21.25) {\includegraphics[width=1.9cm]{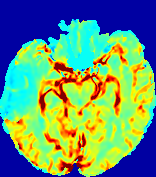}};
	\node at (8, 21.25) {\includegraphics[width=1.9cm]{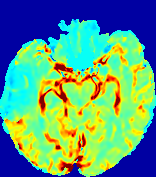}};
	\node at (10, 21.25) {\includegraphics[width=1.9cm]{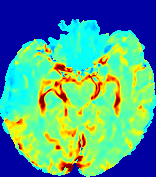}};
	\node at (12, 21.25) {\includegraphics[width=1.9cm]{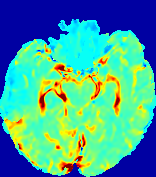}};
	\node at (14, 21.25) {\includegraphics[width=1.9cm]{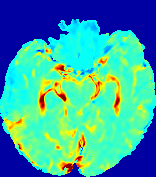}};
	\node at (16, 21.25) {\includegraphics[width=1.9cm]{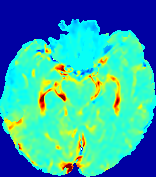}};
	
	\node at (0, 19) {\includegraphics[width=1.9cm]{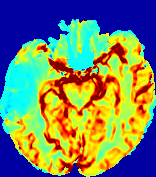}};
	\node at (2, 19) {\includegraphics[width=1.9cm]{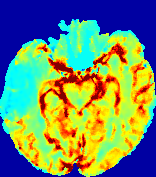}};
	\node at (4, 19) {\includegraphics[width=1.9cm]{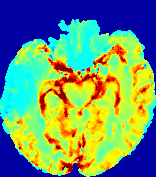}};
	\node at (6, 19) {\includegraphics[width=1.9cm]{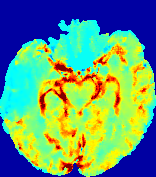}};
	\node at (8, 19) {\includegraphics[width=1.9cm]{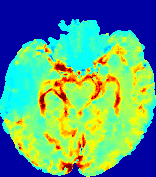}};
	\node at (10, 19) {\includegraphics[width=1.9cm]{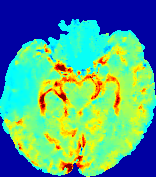}};
	\node at (12, 19) {\includegraphics[width=1.9cm]{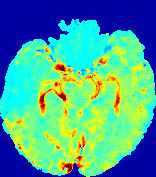}};
	\node at (14, 19) {\includegraphics[width=1.9cm]{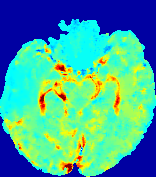}};
	\node at (16, 19) {\includegraphics[width=1.9cm]{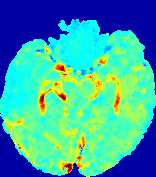}};
	
	%%%%%%%%  Slice (ii)  %%%%%%%%
	
	\node at (-1.25, 16.5) {\footnotesize{(a)}}; 
	\node at (-1.25, 14.25) {\footnotesize{(b)}}; 
	
	\node at (0, 16.5) {\includegraphics[width=1.9cm]{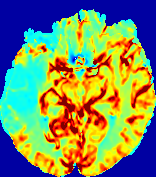}};
	\node at (2, 16.5) {\includegraphics[width=1.9cm]{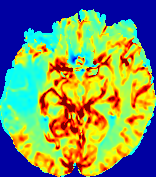}};
	\node at (4, 16.5) {\includegraphics[width=1.9cm]{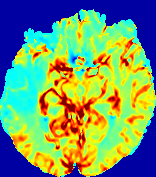}};
	\node at (6, 16.5) {\includegraphics[width=1.9cm]{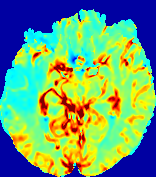}};
	\node at (8, 16.5) {\includegraphics[width=1.9cm]{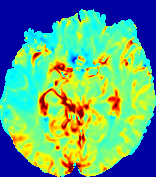}};
	\node at (10, 16.5) {\includegraphics[width=1.9cm]{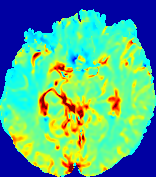}};
	\node at (12, 16.5) {\includegraphics[width=1.9cm]{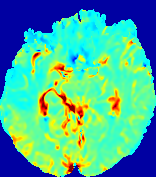}};
	\node at (14, 16.5) {\includegraphics[width=1.9cm]{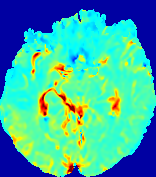}};
	\node at (16, 16.5) {\includegraphics[width=1.9cm]{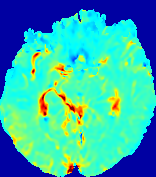}};
	
	\node at (0, 14.25) {\includegraphics[width=1.9cm]{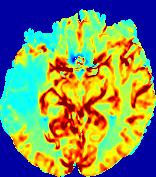}};
	\node at (2, 14.25) {\includegraphics[width=1.9cm]{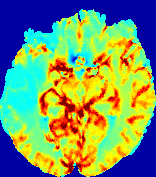}};
	\node at (4, 14.25) {\includegraphics[width=1.9cm]{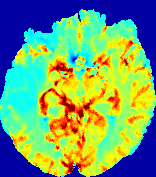}};
	\node at (6, 14.25) {\includegraphics[width=1.9cm]{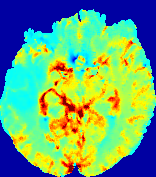}};
	\node at (8, 14.25) {\includegraphics[width=1.9cm]{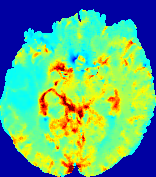}};
	\node at (10, 14.25) {\includegraphics[width=1.9cm]{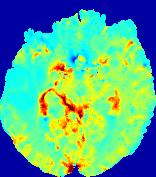}};
	\node at (12, 14.25) {\includegraphics[width=1.9cm]{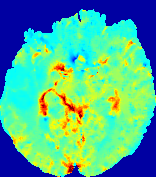}};
	\node at (14, 14.25) {\includegraphics[width=1.9cm]{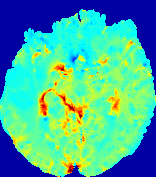}};
	\node at (16, 14.25) {\includegraphics[width=1.9cm]{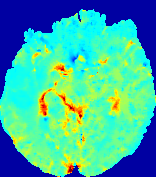}};
	
	%%%%%%%%  Slice (iii)  %%%%%%%%
	
	\node at (-1.25, 11.75) {\footnotesize{(a)}}; 
	\node at (-1.25, 9.5) {\footnotesize{(b)}}; 
	
	\node at (0, 11.75) {\includegraphics[width=1.9cm]{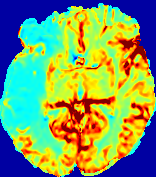}};
	\node at (2, 11.75) {\includegraphics[width=1.9cm]{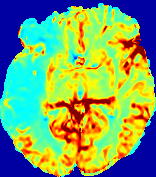}};
	\node at (4, 11.75) {\includegraphics[width=1.9cm]{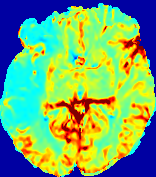}};
	\node at (6, 11.75) {\includegraphics[width=1.9cm]{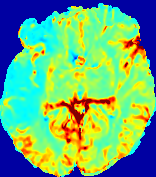}};
	\node at (8, 11.75) {\includegraphics[width=1.9cm]{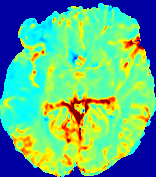}};
	\node at (10, 11.75) {\includegraphics[width=1.9cm]{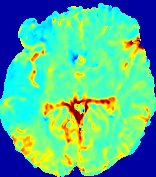}};
	\node at (12, 11.75) {\includegraphics[width=1.9cm]{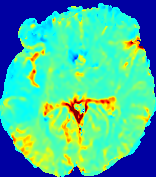}};
	\node at (14, 11.75) {\includegraphics[width=1.9cm]{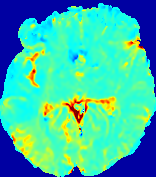}};
	\node at (16, 11.75) {\includegraphics[width=1.9cm]{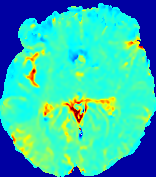}};
	
	\node at (0, 9.5) {\includegraphics[width=1.9cm]{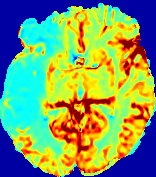}};
	\node at (2, 9.5) {\includegraphics[width=1.9cm]{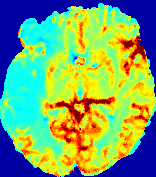}};
	\node at (4, 9.5) {\includegraphics[width=1.9cm]{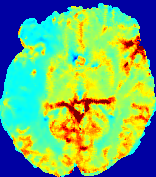}};
	\node at (6, 9.5) {\includegraphics[width=1.9cm]{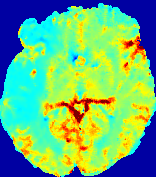}};
	\node at (8, 9.5) {\includegraphics[width=1.9cm]{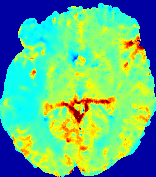}};
	\node at (10, 9.5) {\includegraphics[width=1.9cm]{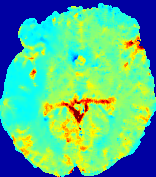}};
	\node at (12, 9.5) {\includegraphics[width=1.9cm]{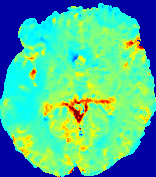}};
	\node at (14, 9.5) {\includegraphics[width=1.9cm]{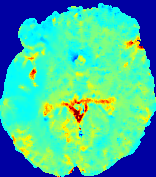}};
	\node at (16, 9.5) {\includegraphics[width=1.9cm]{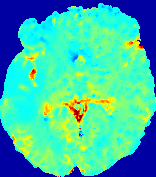}};
	
	%%%%%%%%  Slice (iv)  %%%%%%%%
	
	\node at (-1.25, 7) {\footnotesize{(a)}}; 
	\node at (-1.25, 4.75) {\footnotesize{(b)}}; 
	
	\node at (0, 7) {\includegraphics[width=1.9cm]{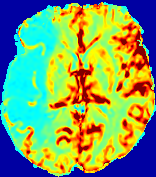}};
	\node at (2, 7) {\includegraphics[width=1.9cm]{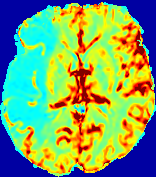}};
	\node at (4, 7) {\includegraphics[width=1.9cm]{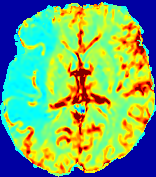}};
	\node at (6, 7) {\includegraphics[width=1.9cm]{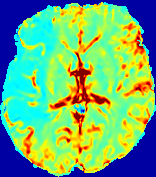}};
	\node at (8, 7) {\includegraphics[width=1.9cm]{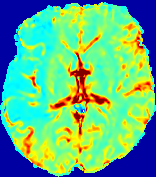}};
	\node at (10, 7) {\includegraphics[width=1.9cm]{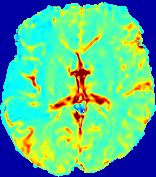}};
	\node at (12, 7) {\includegraphics[width=1.9cm]{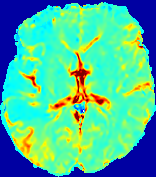}};
	\node at (14, 7) {\includegraphics[width=1.9cm]{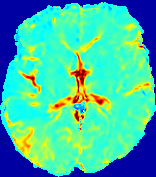}};
	\node at (16, 7) {\includegraphics[width=1.9cm]{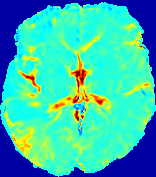}};
	
	\node at (0, 4.75) {\includegraphics[width=1.9cm]{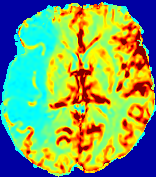}};
	\node at (2, 4.75) {\includegraphics[width=1.9cm]{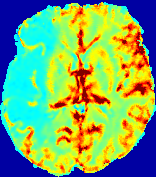}};
	\node at (4, 4.75) {\includegraphics[width=1.9cm]{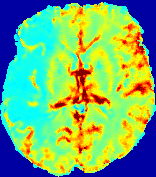}};
	\node at (6, 4.75) {\includegraphics[width=1.9cm]{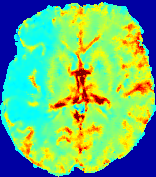}};
	\node at (8, 4.75) {\includegraphics[width=1.9cm]{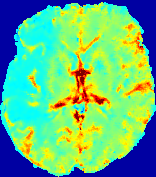}};
	\node at (10, 4.75) {\includegraphics[width=1.9cm]{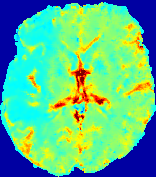}};
	\node at (12, 4.75) {\includegraphics[width=1.9cm]{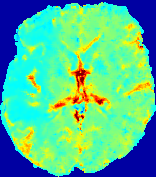}};
	\node at (14, 4.75) {\includegraphics[width=1.9cm]{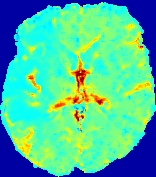}};
	\node at (16, 4.75) {\includegraphics[width=1.9cm]{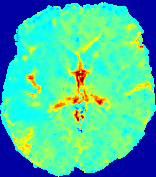}};

	%%%%%%%%  Slice (v)  %%%%%%%%
	
	\node at (-1.25, 2.25) {\footnotesize{(a)}}; 
	\node at (-1.25, 0) {\footnotesize{(b)}}; 
	
	\node at (0, 2.25) {\includegraphics[width=1.9cm]{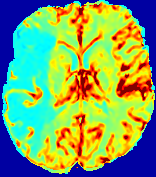}};
	\node at (2, 2.25) {\includegraphics[width=1.9cm]{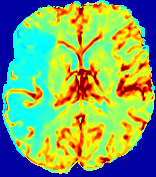}};
	\node at (4, 2.25) {\includegraphics[width=1.9cm]{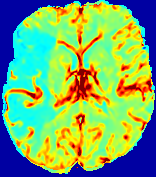}};
	\node at (6, 2.25) {\includegraphics[width=1.9cm]{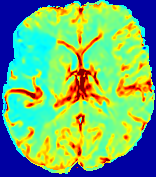}};
	\node at (8, 2.25) {\includegraphics[width=1.9cm]{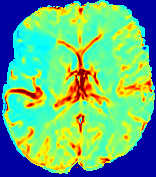}};
	\node at (10, 2.25) {\includegraphics[width=1.9cm]{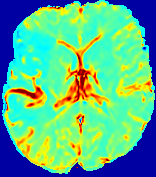}};
	\node at (12, 2.25) {\includegraphics[width=1.9cm]{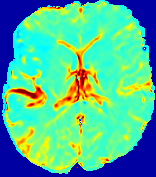}};
	\node at (14, 2.25) {\includegraphics[width=1.9cm]{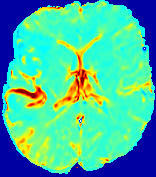}};
	\node at (16, 2.25) {\includegraphics[width=1.9cm]{fig/M14/GT/7/8.png}};
	
	\node at (0, 0) {\includegraphics[width=1.9cm]{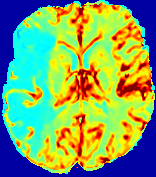}};
	\node at (2, 0) {\includegraphics[width=1.9cm]{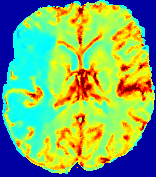}};
	\node at (4, 0) {\includegraphics[width=1.9cm]{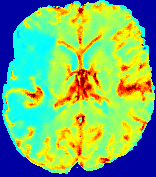}};
	\node at (6, 0) {\includegraphics[width=1.9cm]{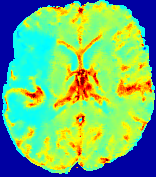}};
	\node at (8, 0) {\includegraphics[width=1.9cm]{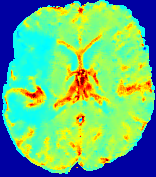}};
	\node at (10, 0) {\includegraphics[width=1.9cm]{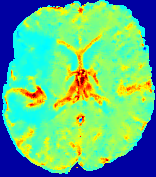}};
	\node at (12, 0) {\includegraphics[width=1.9cm]{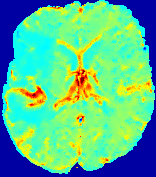}};
	\node at (14, 0) {\includegraphics[width=1.9cm]{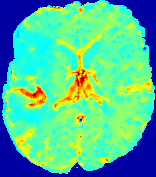}};
	\node at (16, 0) {\includegraphics[width=1.9cm]{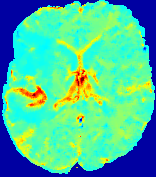}};
	
	%%%%%%%%  Slice (vi)  %%%%%%%%
	
	\node at (-1.25, -2.5) {\footnotesize{(a)}}; 
	\node at (-1.25, -4.75) {\footnotesize{(b)}}; 
	
	\node at (0, -2.5) {\includegraphics[width=1.9cm]{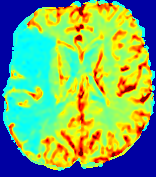}};
	\node at (2, -2.5) {\includegraphics[width=1.9cm]{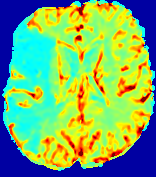}};
	\node at (4, -2.5) {\includegraphics[width=1.9cm]{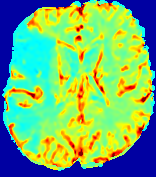}};
	\node at (6, -2.5) {\includegraphics[width=1.9cm]{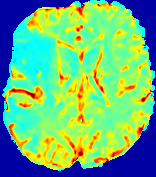}};
	\node at (8, -2.5) {\includegraphics[width=1.9cm]{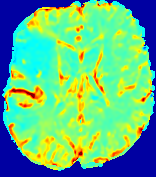}};
	\node at (10, -2.5) {\includegraphics[width=1.9cm]{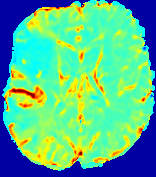}};
	\node at (12, -2.5) {\includegraphics[width=1.9cm]{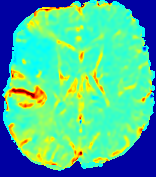}};
	\node at (14, -2.5) {\includegraphics[width=1.9cm]{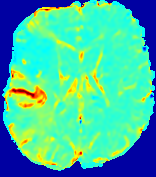}};
	\node at (16, -2.5) {\includegraphics[width=1.9cm]{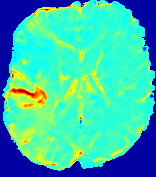}};
	
	\node at (0, -4.75) {\includegraphics[width=1.9cm]{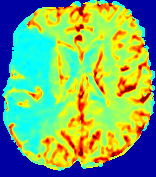}};
	\node at (2, -4.75) {\includegraphics[width=1.9cm]{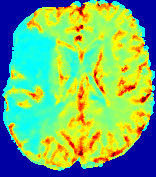}};
	\node at (4, -4.75) {\includegraphics[width=1.9cm]{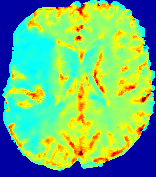}};
	\node at (6, -4.75) {\includegraphics[width=1.9cm]{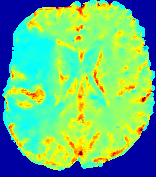}};
	\node at (8, -4.75) {\includegraphics[width=1.9cm]{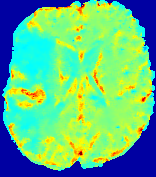}};
	\node at (10, -4.75) {\includegraphics[width=1.9cm]{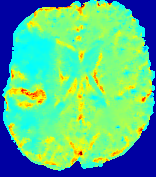}};
	\node at (12, -4.75) {\includegraphics[width=1.9cm]{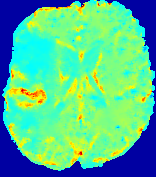}};
	\node at (14, -4.75) {\includegraphics[width=1.9cm]{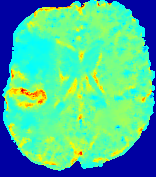}};
	\node at (16, -4.75) {\includegraphics[width=1.9cm]{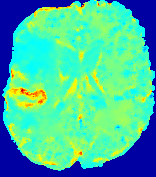}};
	
	\end{tikzpicture}
	} 
	\caption{Predicted concentration time series for the same patient shown in Fig. \ref{M14_param}, where (i)-(vi) correspond to slices \#1-6 respectively. Each grouped row displays (a) the measured concentration image sequences and (b) the predicted concentrations at corresponding time points.}
	\label{M14_ctc}
\end{figure}

%% file: sub/exp_results/ctc_16.tex
\begin{figure}[t]
	\noindent\resizebox{\textwidth}{!}{
	\begin{tikzpicture}
	
	% Slice Gap: 4.75
	
	%%%%%%%%%%%%%%%%%%%%%%%
	%%%%%%%% Denotations %%%%%%%%
	%%%%%%%%%%%%%%%%%%%%%%%
	
	\draw [->, line width = 0.4mm, color = black](-1.75, 23)--(17, 23);
	\node at (16.5, 22.75)  {\large Time};
	\draw [->, line width = 0.4mm, color = black](-1.75, 23)--(-1.75, -7);
	%\node[label=below:\rotatebox{90}{\large Slice}] at (-1.4, -5.95) {};
	\node at (-1.1, -6.8)  {\large Slice};
	
	\node at (-2.1, 20.125) {\footnotesize{(i)}}; 
	\node at (-2.1, 15.375) {\footnotesize{(ii)}}; 
	\node at (-2.1, 10.625) {\footnotesize{(iii)}}; 
	\node at (-2.1, 5.875) {\footnotesize{(iv)}}; 
	\node at (-2.1, 1.125) {\footnotesize{(v)}}; 
	\node at (-2.1, -3.625) {\footnotesize{(vi)}}; 
	
	%\node at (0,-6.15)  {{\footnotesize{$t_0$}}};
	%\node at (2,-6.15)  {{\footnotesize{$t_1$}}};
	%\node at (4,-6.15)  {{\footnotesize{$t_2$}}};
	%\node at (6,-6.15)  {{\footnotesize{$t_3$}}};
	%\node at (8,-6.15)  {{\footnotesize{$t_4$}}};
	%\node at (10,-6.15)  {{\footnotesize{$t_5$}}};
	%\node at (12,-6.15)  {{\footnotesize{$t_6$}}};
	%\node at (14,-6.15)  {{\footnotesize{$t_7$}}};
	\node at (8, -6.35) {\includegraphics[width=6cm]{fig/cb2.png}};
	\node at (5.3, -6.9)  {$0$};
	\node at (6.4, -6.9)  {$5$};
	\node at (7.4775, -6.9)  {$10$};
	\node at (8.55, -6.9)  {$15$};
	\node at (9.65, -6.9)  {$20$};
	\node at (10.75, -6.9)  {$25$};

	%\node[label=below:\rotatebox{90}{\footnotesize Measured}] at (-1.2, 22.2) {};
	%\node[label=below:\rotatebox{90}{\footnotesize Measured}] at (-1.2, 17.45) {};
	%\node[label=below:\rotatebox{90}{\footnotesize Measured}] at (-1.2, 12.7) {};
	%\node[label=below:\rotatebox{90}{\footnotesize Measured}] at (-1.2, 7.95) {};
	%\node[label=below:\rotatebox{90}{\footnotesize Measured}] at (-1.2, 3.2) {};
	%\node[label=below:\rotatebox{90}{\footnotesize Measured}] at (-1.2, -1.55) {};
	
	%\node[label=below:\rotatebox{90}{\footnotesize Prediction}] at (-1.2, 19.95) {};
	%\node[label=below:\rotatebox{90}{\footnotesize Prediction}] at (-1.2, 15.2) {};
	%\node[label=below:\rotatebox{90}{\footnotesize Prediction}] at (-1.2, 10.45) {};
	%\node[label=below:\rotatebox{90}{\footnotesize Prediction}] at (-1.2,5.7) {};
	%\node[label=below:\rotatebox{90}{\footnotesize Prediction}] at (-1.2,0.95) {};
	%\node[label=below:\rotatebox{90}{\footnotesize Prediction}] at (-1.2,-3.8) {};
	
	\node at (11.12, -4.35) {\includegraphics[width=0.25cm]{fig/cb.png}};
	
	%%%%%%%%  Slice (i)  %%%%%%%%
	
	\node at (-1.25, 21.25) {\footnotesize{(a)}}; 
	\node at (-1.25, 19) {\footnotesize{(b)}}; 
	
	\node at (0, 21.25) {\includegraphics[width=1.9cm]{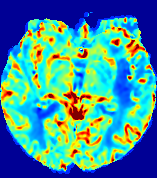}};
	\node at (2, 21.25) {\includegraphics[width=1.9cm]{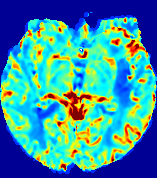}};
	\node at (4, 21.25) {\includegraphics[width=1.9cm]{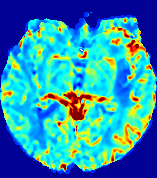}};
	\node at (6, 21.25) {\includegraphics[width=1.9cm]{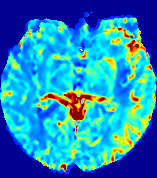}};
	\node at (8, 21.25) {\includegraphics[width=1.9cm]{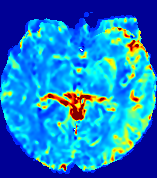}};
	\node at (10, 21.25) {\includegraphics[width=1.9cm]{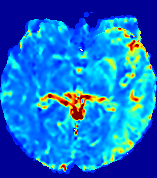}};
	\node at (12, 21.25) {\includegraphics[width=1.9cm]{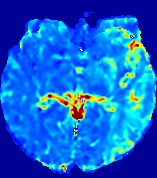}};
	\node at (14, 21.25) {\includegraphics[width=1.9cm]{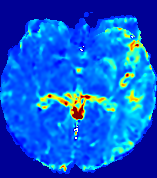}};
	\node at (16, 21.25) {\includegraphics[width=1.9cm]{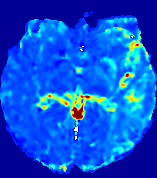}};
	
	\node at (0, 19) {\includegraphics[width=1.9cm]{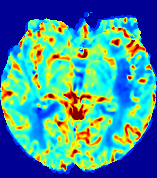}};
	\node at (2, 19) {\includegraphics[width=1.9cm]{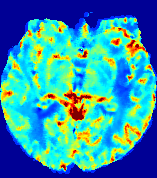}};
	\node at (4, 19) {\includegraphics[width=1.9cm]{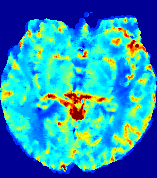}};
	\node at (6, 19) {\includegraphics[width=1.9cm]{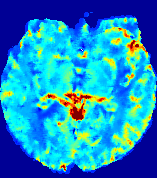}};
	\node at (8, 19) {\includegraphics[width=1.9cm]{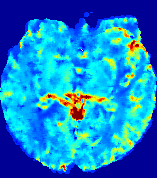}};
	\node at (10, 19) {\includegraphics[width=1.9cm]{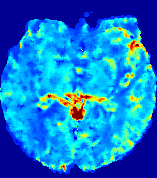}};
	\node at (12, 19) {\includegraphics[width=1.9cm]{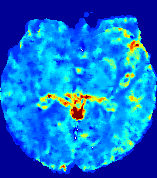}};
	\node at (14, 19) {\includegraphics[width=1.9cm]{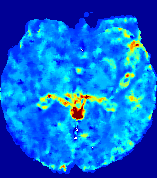}};
	\node at (16, 19) {\includegraphics[width=1.9cm]{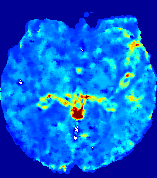}};
	
	%%%%%%%%  Slice (ii)  %%%%%%%%
	
	\node at (-1.25, 16.5) {\footnotesize{(a)}}; 
	\node at (-1.25, 14.25) {\footnotesize{(b)}}; 
	
	\node at (0, 16.5) {\includegraphics[width=1.9cm]{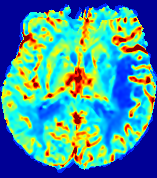}};
	\node at (2, 16.5) {\includegraphics[width=1.9cm]{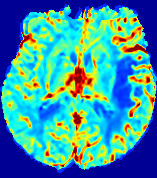}};
	\node at (4, 16.5) {\includegraphics[width=1.9cm]{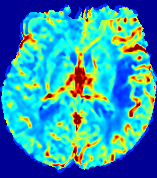}};
	\node at (6, 16.5) {\includegraphics[width=1.9cm]{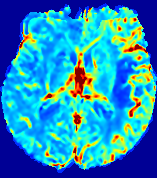}};
	\node at (8, 16.5) {\includegraphics[width=1.9cm]{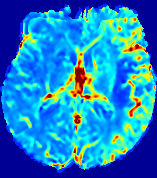}};
	\node at (10, 16.5) {\includegraphics[width=1.9cm]{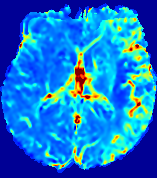}};
	\node at (12, 16.5) {\includegraphics[width=1.9cm]{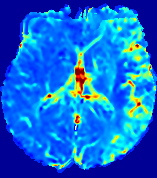}};
	\node at (14, 16.5) {\includegraphics[width=1.9cm]{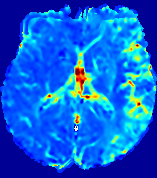}};
	\node at (16, 16.5) {\includegraphics[width=1.9cm]{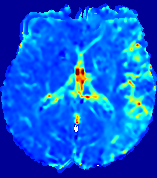}};
	
	\node at (0, 14.25) {\includegraphics[width=1.9cm]{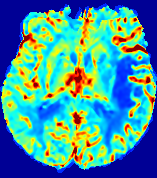}};
	\node at (2, 14.25) {\includegraphics[width=1.9cm]{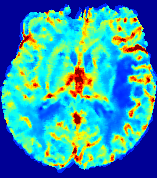}};
	\node at (4, 14.25) {\includegraphics[width=1.9cm]{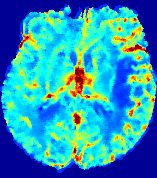}};
	\node at (6, 14.25) {\includegraphics[width=1.9cm]{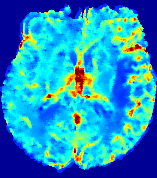}};
	\node at (8, 14.25) {\includegraphics[width=1.9cm]{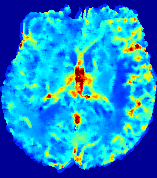}};
	\node at (10, 14.25) {\includegraphics[width=1.9cm]{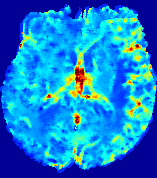}};
	\node at (12, 14.25) {\includegraphics[width=1.9cm]{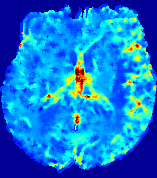}};
	\node at (14, 14.25) {\includegraphics[width=1.9cm]{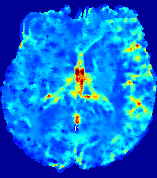}};
	\node at (16, 14.25) {\includegraphics[width=1.9cm]{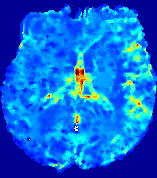}};
	
	%%%%%%%%  Slice (iii)  %%%%%%%%
	
	\node at (-1.25, 11.75) {\footnotesize{(a)}}; 
	\node at (-1.25, 9.5) {\footnotesize{(b)}}; 
	
	\node at (0, 11.75) {\includegraphics[width=1.9cm]{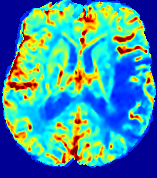}};
	\node at (2, 11.75) {\includegraphics[width=1.9cm]{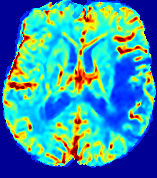}};
	\node at (4, 11.75) {\includegraphics[width=1.9cm]{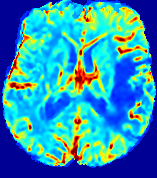}};
	\node at (6, 11.75) {\includegraphics[width=1.9cm]{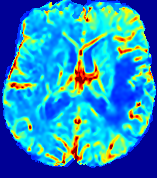}};
	\node at (8, 11.75) {\includegraphics[width=1.9cm]{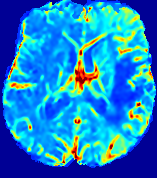}};
	\node at (10, 11.75) {\includegraphics[width=1.9cm]{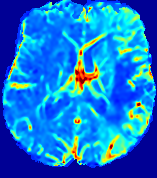}};
	\node at (12, 11.75) {\includegraphics[width=1.9cm]{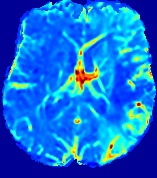}};
	\node at (14, 11.75) {\includegraphics[width=1.9cm]{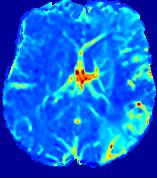}};
	\node at (16, 11.75) {\includegraphics[width=1.9cm]{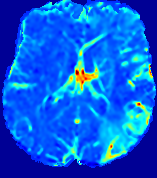}};
	
	\node at (0, 9.5) {\includegraphics[width=1.9cm]{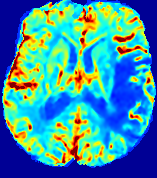}};
	\node at (2, 9.5) {\includegraphics[width=1.9cm]{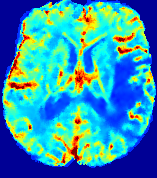}};
	\node at (4, 9.5) {\includegraphics[width=1.9cm]{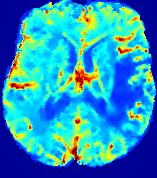}};
	\node at (6, 9.5) {\includegraphics[width=1.9cm]{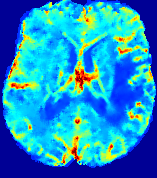}};
	\node at (8, 9.5) {\includegraphics[width=1.9cm]{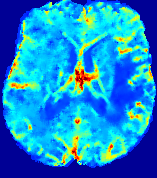}};
	\node at (10, 9.5) {\includegraphics[width=1.9cm]{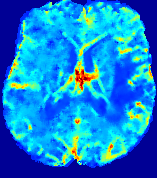}};
	\node at (12, 9.5) {\includegraphics[width=1.9cm]{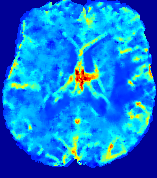}};
	\node at (14, 9.5) {\includegraphics[width=1.9cm]{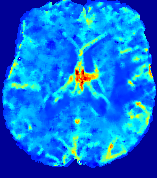}};
	\node at (16, 9.5) {\includegraphics[width=1.9cm]{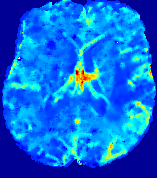}};
	
	%%%%%%%%  Slice (iv)  %%%%%%%%
	
	\node at (-1.25, 7) {\footnotesize{(a)}}; 
	\node at (-1.25, 4.75) {\footnotesize{(b)}}; 
	
	\node at (0, 7) {\includegraphics[width=1.9cm]{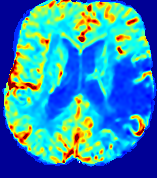}};
	\node at (2, 7) {\includegraphics[width=1.9cm]{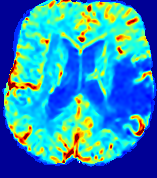}};
	\node at (4, 7) {\includegraphics[width=1.9cm]{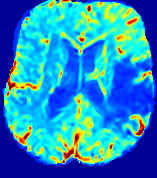}};
	\node at (6, 7) {\includegraphics[width=1.9cm]{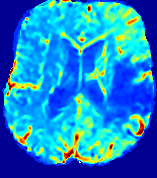}};
	\node at (8, 7) {\includegraphics[width=1.9cm]{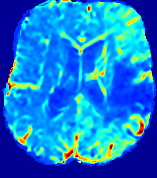}};
	\node at (10, 7) {\includegraphics[width=1.9cm]{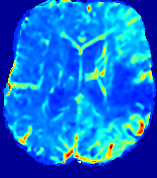}};
	\node at (12, 7) {\includegraphics[width=1.9cm]{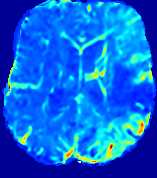}};
	\node at (14, 7) {\includegraphics[width=1.9cm]{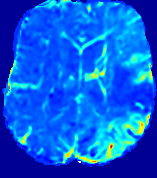}};
	\node at (16, 7) {\includegraphics[width=1.9cm]{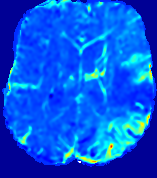}};
	
	\node at (0, 4.75) {\includegraphics[width=1.9cm]{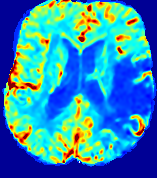}};
	\node at (2, 4.75) {\includegraphics[width=1.9cm]{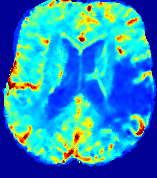}};
	\node at (4, 4.75) {\includegraphics[width=1.9cm]{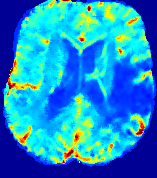}};
	\node at (6, 4.75) {\includegraphics[width=1.9cm]{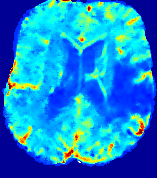}};
	\node at (8, 4.75) {\includegraphics[width=1.9cm]{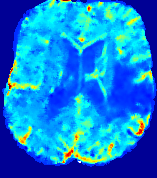}};
	\node at (10, 4.75) {\includegraphics[width=1.9cm]{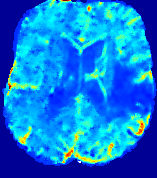}};
	\node at (12, 4.75) {\includegraphics[width=1.9cm]{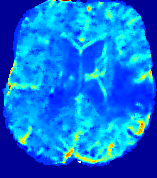}};
	\node at (14, 4.75) {\includegraphics[width=1.9cm]{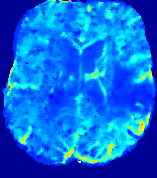}};
	\node at (16, 4.75) {\includegraphics[width=1.9cm]{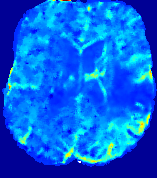}};

	%%%%%%%%  Slice (v)  %%%%%%%%
	
	\node at (-1.25, 2.25) {\footnotesize{(a)}}; 
	\node at (-1.25, 0) {\footnotesize{(b)}}; 
	
	\node at (0, 2.25) {\includegraphics[width=1.9cm]{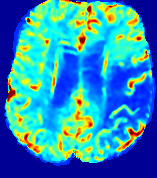}};
	\node at (2, 2.25) {\includegraphics[width=1.9cm]{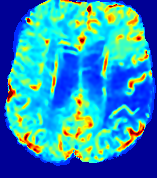}};
	\node at (4, 2.25) {\includegraphics[width=1.9cm]{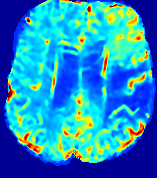}};
	\node at (6, 2.25) {\includegraphics[width=1.9cm]{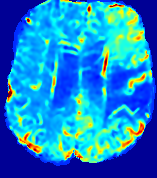}};
	\node at (8, 2.25) {\includegraphics[width=1.9cm]{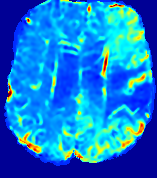}};
	\node at (10, 2.25) {\includegraphics[width=1.9cm]{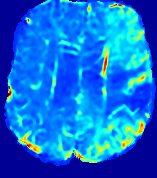}};
	\node at (12, 2.25) {\includegraphics[width=1.9cm]{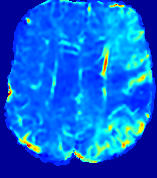}};
	\node at (14, 2.25) {\includegraphics[width=1.9cm]{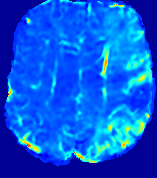}};
	\node at (16, 2.25) {\includegraphics[width=1.9cm]{fig/M16/GT/9/8.png}};
	
	\node at (0, 0) {\includegraphics[width=1.9cm]{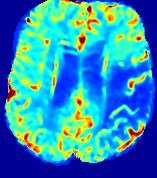}};
	\node at (2, 0) {\includegraphics[width=1.9cm]{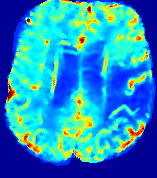}};
	\node at (4, 0) {\includegraphics[width=1.9cm]{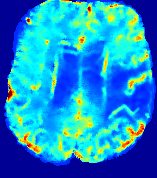}};
	\node at (6, 0) {\includegraphics[width=1.9cm]{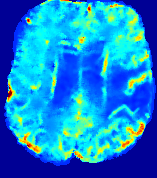}};
	\node at (8, 0) {\includegraphics[width=1.9cm]{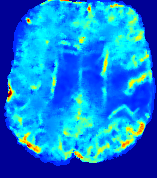}};
	\node at (10, 0) {\includegraphics[width=1.9cm]{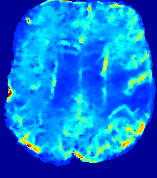}};
	\node at (12, 0) {\includegraphics[width=1.9cm]{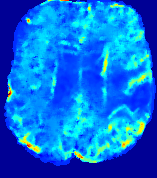}};
	\node at (14, 0) {\includegraphics[width=1.9cm]{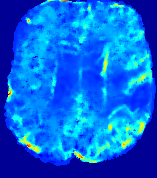}};
	\node at (16, 0) {\includegraphics[width=1.9cm]{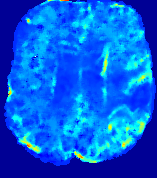}};
	
	%%%%%%%%  Slice (vi)  %%%%%%%%
	
	\node at (-1.25, -2.5) {\footnotesize{(a)}}; 
	\node at (-1.25, -4.75) {\footnotesize{(b)}}; 
	
	\node at (0, -2.5) {\includegraphics[width=1.9cm]{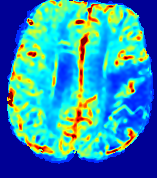}};
	\node at (2, -2.5) {\includegraphics[width=1.9cm]{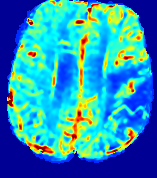}};
	\node at (4, -2.5) {\includegraphics[width=1.9cm]{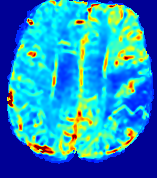}};
	\node at (6, -2.5) {\includegraphics[width=1.9cm]{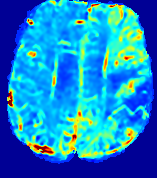}};
	\node at (8, -2.5) {\includegraphics[width=1.9cm]{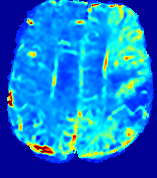}};
	\node at (10, -2.5) {\includegraphics[width=1.9cm]{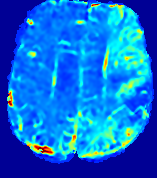}};
	\node at (12, -2.5) {\includegraphics[width=1.9cm]{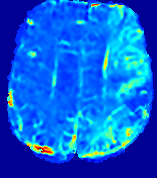}};
	\node at (14, -2.5) {\includegraphics[width=1.9cm]{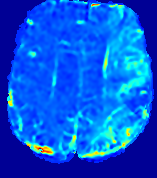}};
	\node at (16, -2.5) {\includegraphics[width=1.9cm]{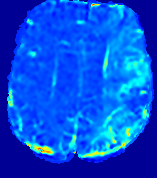}};
	
	\node at (0, -4.75) {\includegraphics[width=1.9cm]{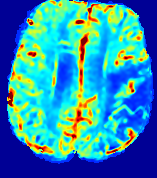}};
	\node at (2, -4.75) {\includegraphics[width=1.9cm]{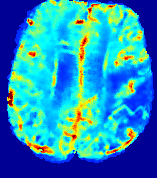}};
	\node at (4, -4.75) {\includegraphics[width=1.9cm]{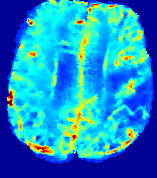}};
	\node at (6, -4.75) {\includegraphics[width=1.9cm]{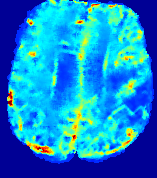}};
	\node at (8, -4.75) {\includegraphics[width=1.9cm]{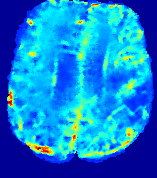}};
	\node at (10, -4.75) {\includegraphics[width=1.9cm]{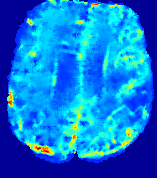}};
	\node at (12, -4.75) {\includegraphics[width=1.9cm]{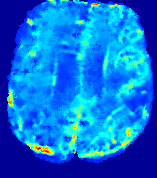}};
	\node at (14, -4.75) {\includegraphics[width=1.9cm]{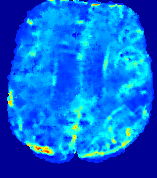}};
	\node at (16, -4.75) {\includegraphics[width=1.9cm]{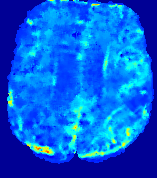}};
	
	\end{tikzpicture}
	} 
	\caption{Predicted concentration time series for the same patient shown in Fig. \ref{M16_param}, where (i)-(vi) correspond to slices \#1-6 respectively. Each grouped row displays (a) the measured concentration image sequences and (b) the predicted concentrations at corresponding time points.}
	\label{M16_ctc}
\end{figure}

%% file: sub/exp_results/box_plots_tmi2.tex
\begin{figure}[t]
	\noindent\resizebox{0.9\textwidth}{!}{
	\begin{tikzpicture}
	% Lesion
	\node at (0, 2.25) {\includegraphics[width=4cm]{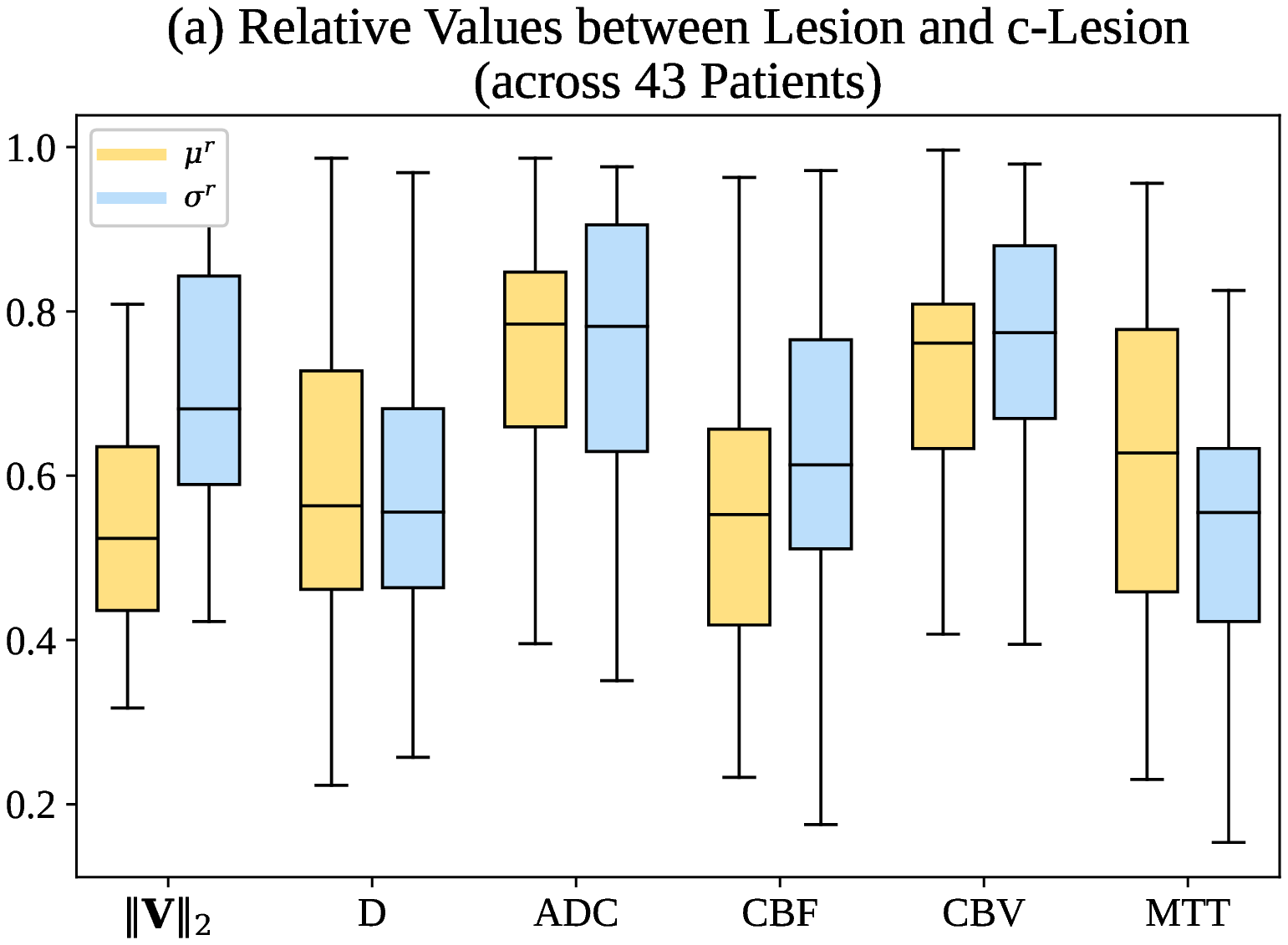}};
	
	%\node at (4.2, 2.25) {\includegraphics[width=4.25cm]{fig/box_plots/t.eps}};
	\node at (0, -0.75) {\includegraphics[width=4cm]{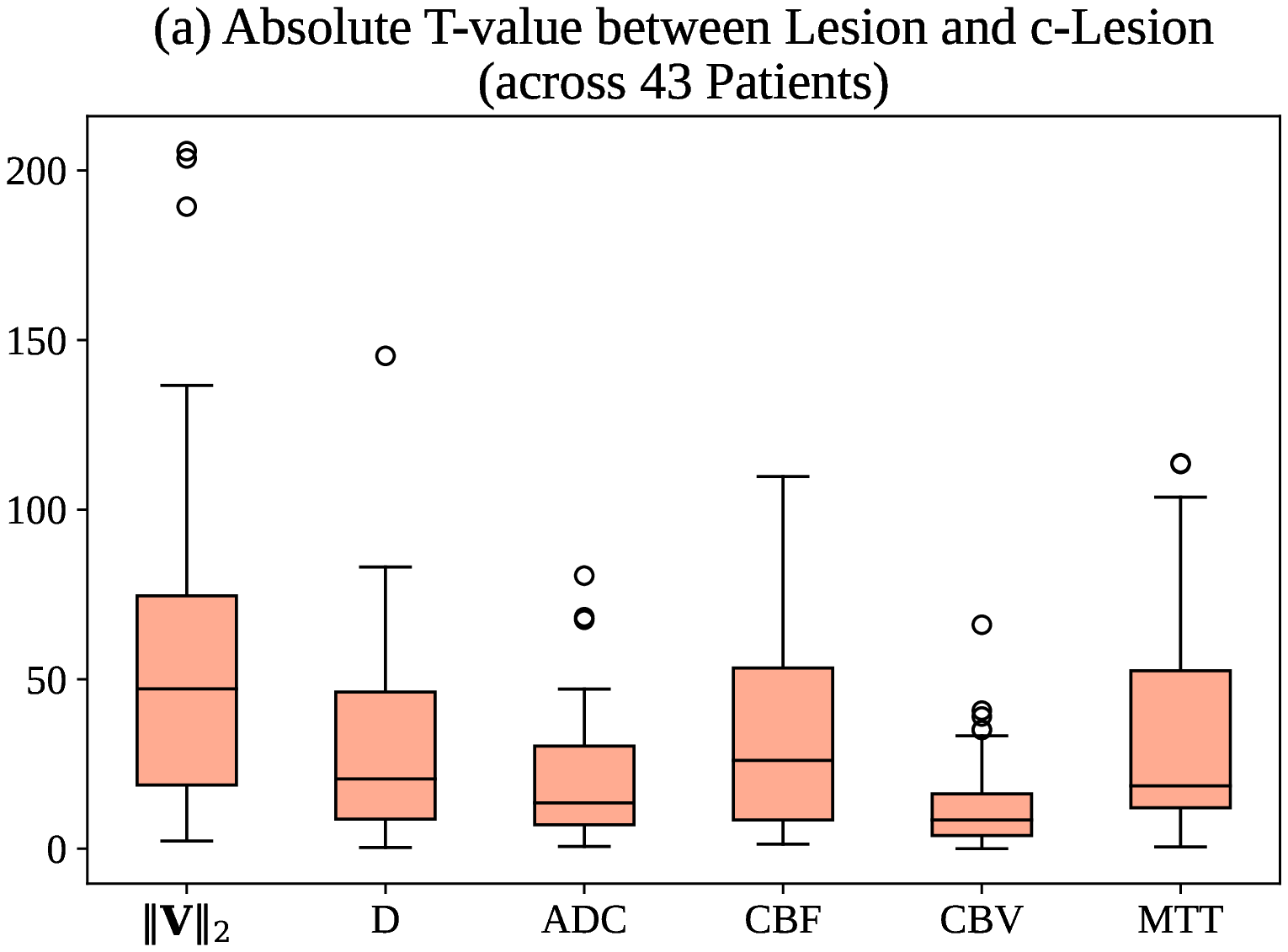}};
	\draw (-1.52 , -2.25) node[draw,scale=0.2,circle,line width=0.075mm]{}; 
	\draw (-1.37 , -2.25) node[draw,scale=0.2,diamond,line width=0.075mm]{}; 
	\draw (-1.22, -2.25) node[draw,scale=0.01,star,star point ratio=5, fill = black]{}; 
	
	\draw (-0.85 , -2.25) node[draw,scale=0.2,circle,line width=0.075mm]{}; 
	\draw (-0.7 , -2.25) node[draw,scale=0.2,diamond,line width=0.075mm]{}; 
	
	\end{tikzpicture}
	}
	\caption{Box plots of (a) relative mean values ($\mu^r$), relative standard deviation ($\sigma^r$) and (b) absolute t-values for PIANO feature maps and ISLES 2017 maps, computed from 43 patients. $\star,\, \diamond,\, \circ$ indicate statistically significant differences between the PIANO feature maps and ADC, CBF, CBV, MTT respectively, based on a paired t-test with Bonferroni correction at a significance level of $0.05$.}
	\label{box_plots}
\end{figure}

%% file: sub/exp_results/table_tmi2_4maps.tex
\begin{table}[t]
\caption{Quantitative comparison between PIANO feature maps and ISLES 2017 maps over 43 subjects, using \emph{Mean}, \emph{Median}, \emph{Standard Deviation (STD)} of relative mean $\mu^r$, relative STD $\sigma^r$ (the lower the better), and absolute t-value (higher absolute value indicates greater difference).}
\resizebox{\linewidth}{!}{
\centering
    \begin{tabular}{P{0.6cm}P{1.5cm}|P{0.8cm}P{0.8cm}|P{0.8cm}P{0.8cm}P{0.8cm}P{0.8cm}} 
       \toprule \\[-2ex]
        \multicolumn{2}{c|}{\bf Maps} & ${\| {\bf{V}} \|_2}$ & $D$ & ADC & CBF & CBV & MTT \\ [0.5ex]
     \midrule\\[-2ex]
       \parbox[t]{5mm}{\multirow{3}{*}{\rotatebox[origin=c]{90}{\thead{Relative\\Mean ($\mu^r$)}}}} & {\emph{Mean}} & {\bf{0.55}}  & 0.60 & 0.76 & 0.57 & 0.89 & 1.83  \\ [1ex]
       &  {\emph{Median}} & {\bf{0.52}}  & 0.56 & 0.78 & 0.55 & 0.80 & 1.59\\ [1ex]
       &  {\emph{STD}} & {\bf{0.13}}  & 0.19 & 0.14 & 0.19 & 0.36 & 0.75 \\ [0.75ex]
     \midrule\\[-2ex]
       \parbox[t]{5mm}{\multirow{3}{*}{\rotatebox[origin=c]{90}{\thead{Relative\\STD ($\sigma^r$)}}}}  & {\emph{Mean}} & 0.72 & {\bf{0.56}} & 0.80 & 0.66 & 0.93 & 2.09  \\ [1ex]
       &  {\emph{Median}} & 0.68 & {\bf{0.56}} & 0.82 & 0.61 & 0.87 & 1.80 \\ [1ex]
       &  {\emph{STD}} & 0.19 & {\bf{0.18}} & 0.27 & 0.23 & 0.33 & 0.97 \\ [0.75ex]
     \midrule\\[-2ex]
       \parbox[t]{5mm}{\multirow{3}{*}{\rotatebox[origin=c]{90}{\thead{Absolute~~\\t-value~~}}}} & {\emph{Mean}} & {\bf{57.76}}  & 29.51 & 20.55 & 32.61 & 13.53 & 33.56 \\ [1ex]
       &  {\emph{Median}} & {\bf{47.13}}  & 20.58 & 13.50 & 26.08 & 8.48 & 18.52 \\ [1ex]
       &  {\emph{STD}} & {\bf{51.83}}  & 27.67 & 19.53 & 27.47 & 14.21 & 31.70  \\ [0.75ex] % [2ex]
          \bottomrule
    \end{tabular}
    \label{table}
}
\end{table}

%% file: sub/exp_peclet.tex
\section{Further Evaluations}

This section provides more detailed experimental results for PIANO. Specifically, Sec.~\ref{sec: peclet} discusses considerations regarding the relationship of advection and diffusion to vessel diameter. Sec.~\ref{sec: effect_robust} and Sec.~\ref{sec: ident} further explore the effectiveness, robustness and identifiability of PIANO.

%%%%%%%%%%%%%%%%%%%%%%%%%%%%%%%%%
%%%%%%%%%%%%%%%%%%%%%%%%%%%%%%%%%
\subsection{Cerebral Blood Velocity and P\'eclet Number}
\label{sec: peclet}

\input{sub/exp_results/box_plots_V_Pe}

As described in Sec.~\ref{sec:piano_feature_maps}, $\| {\bf{V}} \|_2$, is the $2$ norm of the estimated velocity field ${\bf{V}}$ governing the advection process, which describes the transport of CA driven by the cerebral blood flow within the blood vessels. Ivanov et al.~\cite{Ivanov1981} provide an in-depth discussion about blood flow velocities in cerebral capillaries. They report a typical range of blood flow velocities between $0.5$ to $1.5~mm/s$ in cerebral capillaries, precapillaries, and arterioles that are not more than $5~\mu m$ in luminal diameter. Maximum blood flow velocities in humans can reach up to $289~cm/s$ in major cerebral arteries such as the middle cerebral arteries (MCAs)~\cite{Brass1991}. However, such velocities are not observable based on our imaging. Specifically, the ${\bf{V}}$ estimated by PIANO, via observing the transport of CA recorded in PI, should be considered as the velocity field {\it averaged} over space (with voxel spacing of $\approx 1~mm$) and time (with PI temporal resolution of $\approx 1~s$). Estimated velocities are therefore significantly lower than the maximum velocities. In fact, mean velocities across a cardiac cycle ($V_{\text{mean}}$) for cerebral perforating arteries are measured in~\cite{Bouvy2016}, where the authors report $V_{\text{mean}}$ in the semioval centre (CSO) in the range $0.5-1.0~cm/s$, and  in the range of $3.9-5.1~cm/s$ for $V_{\text{mean}}$ in the basal ganglia (BG). Fig.~\ref{box_plots_v} (a) displays the histogram of $\| {\bf{V}} \|_2$, in the unaffected hemispheres (in which we assume blood flow velocities are in the normal range) of the 43 ISLES 2017 stroke patients. In general, $\| {\bf{V}} \|_2$ mainly falls within the range of $0-6~mm/s$ with a mean value of $1.875~mm/s$, which is consistent with the cerebral blood flow velocities reported in the above literature. Fig. \ref{box_plots_v} (b) shows detailed distributions of $\| {\bf{V}} \|_2$ for each patient. We observe a similar range of $\| {\bf{V}} \|_2$ for the different patients.

To assess the relation between the estimated advection and diffusion, we resort to the P\'eclet number (Pe). Pe is a dimensionless number that represents the ratio of the contributions to mass transport by advection to those by diffusion \cite{Leopoldo1992fd}. For mass transfer (i.e., CA in this paper), it is formed as 
\begin{equation}
{\text{Pe}} = \frac{L\| {\bf{V}} \|_2}{D},
\label{eq: peclet}
\end{equation}
where $\| {\bf{V}} \|_2,\, D$ are already defined based on our PIANO feature maps (Sec. \ref{sec:piano_feature_maps}), $L$ is the characteristic length (which we set to $1$ for simplicity). By definition, Pe values range from $0$ to $\infty$, indicating different process behavior, i.e., varying from pure diffusion, to diffusion-dominant transport, to advection-dominant transport, and lastly to pure advection. For structures larger than the micrometer scale, Pe is normally greater than $1$ \cite{Beaudoin2007}, meaning the effects of advection exceed those of diffusion in determining the overall mass flux. %We calculate Pe in the unaffected hemispheres of the 43 patients. 
To achieve better visualizations for all kinds of mass transport, we compute both Pe and the inverse of Pe, with larger Pe (smaller inverse Pe) indicating greater advection and less diffusion (and vice versa). 
%Fig.~\ref{M14_inv_Pe} shows an example of inverse Pe map, where low values (blue) indicate advection-dominant and high values (red) indicate diffusion-dominant.
Fig.~\ref{box_plots_pe} (a) and Fig.~\ref{box_plots_inv_pe} (a) show the histograms of Pe and inverse Pe in the unaffected hemispheres of all patients. Upper outliers exist in both Pe (up to $5.88\times 10^{11}$) and inverse Pe (up to $3.23\times 10^{3}$) histograms, referring to voxels that are dominated by advection and diffusion, respectively. Fig.~\ref{box_plots_pe} (b) and Fig.~\ref{box_plots_inv_pe} (b) show the distributions of Pe and inverse Pe for each patient. Note that there is little across-patient variability with respect to the median of the inverse Pe.

Based on the above discussion about cerebral blood velocity and the P\'eclet number, the velocity and diffusion fields estimated by PIANO fall within reasonable value ranges, and are consistent with value ranges reported in literature as well.

%% file: sub/exp_results/box_plots_V_Pe.tex
\begin{figure}[t]
	\noindent\resizebox{0.9\textwidth}{!}{
	\begin{tikzpicture}
	\node at (0, 0.75) {\includegraphics[width=4cm]{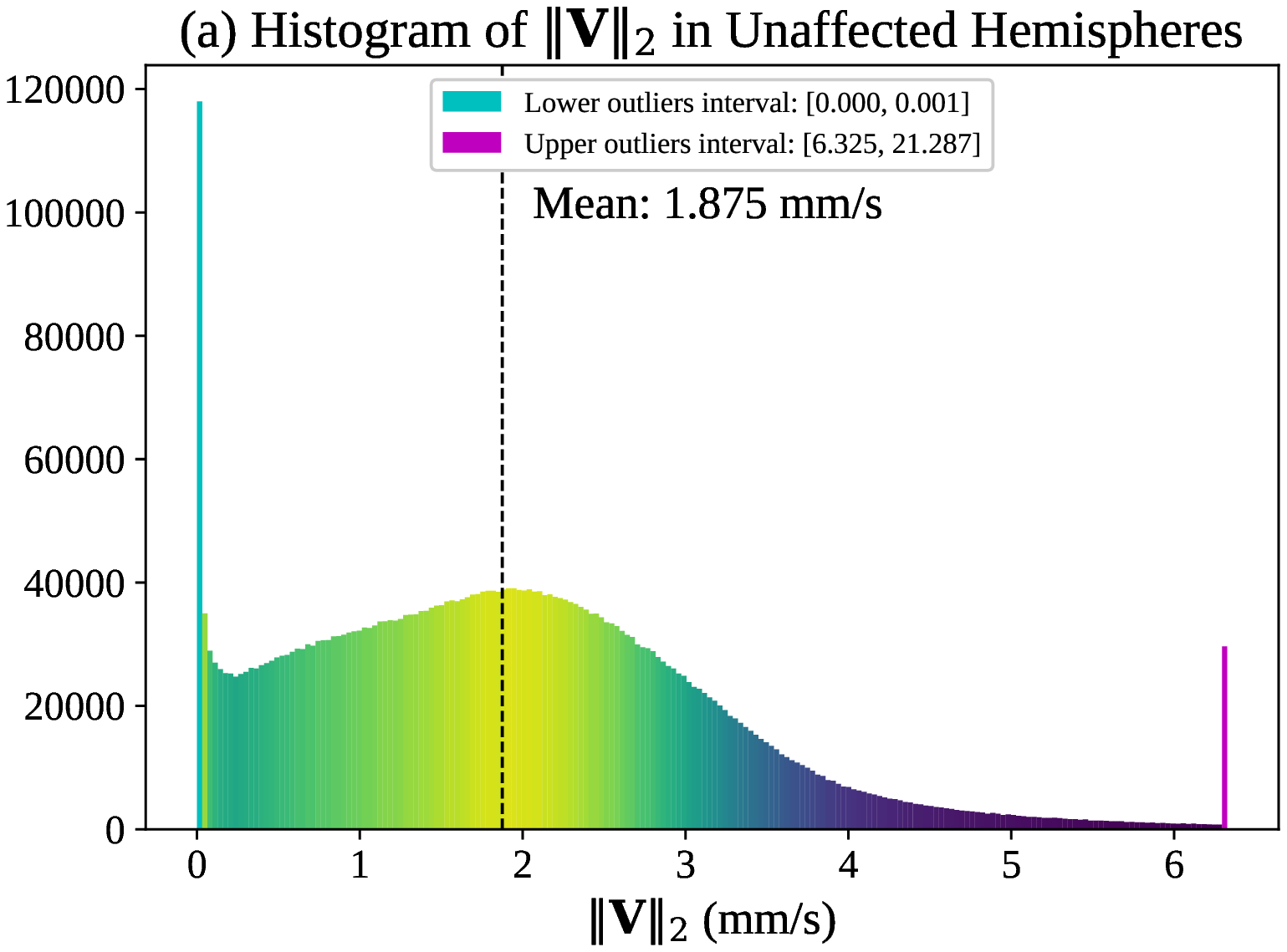}};
	\node at (0, -2.25) {\includegraphics[width=4cm]{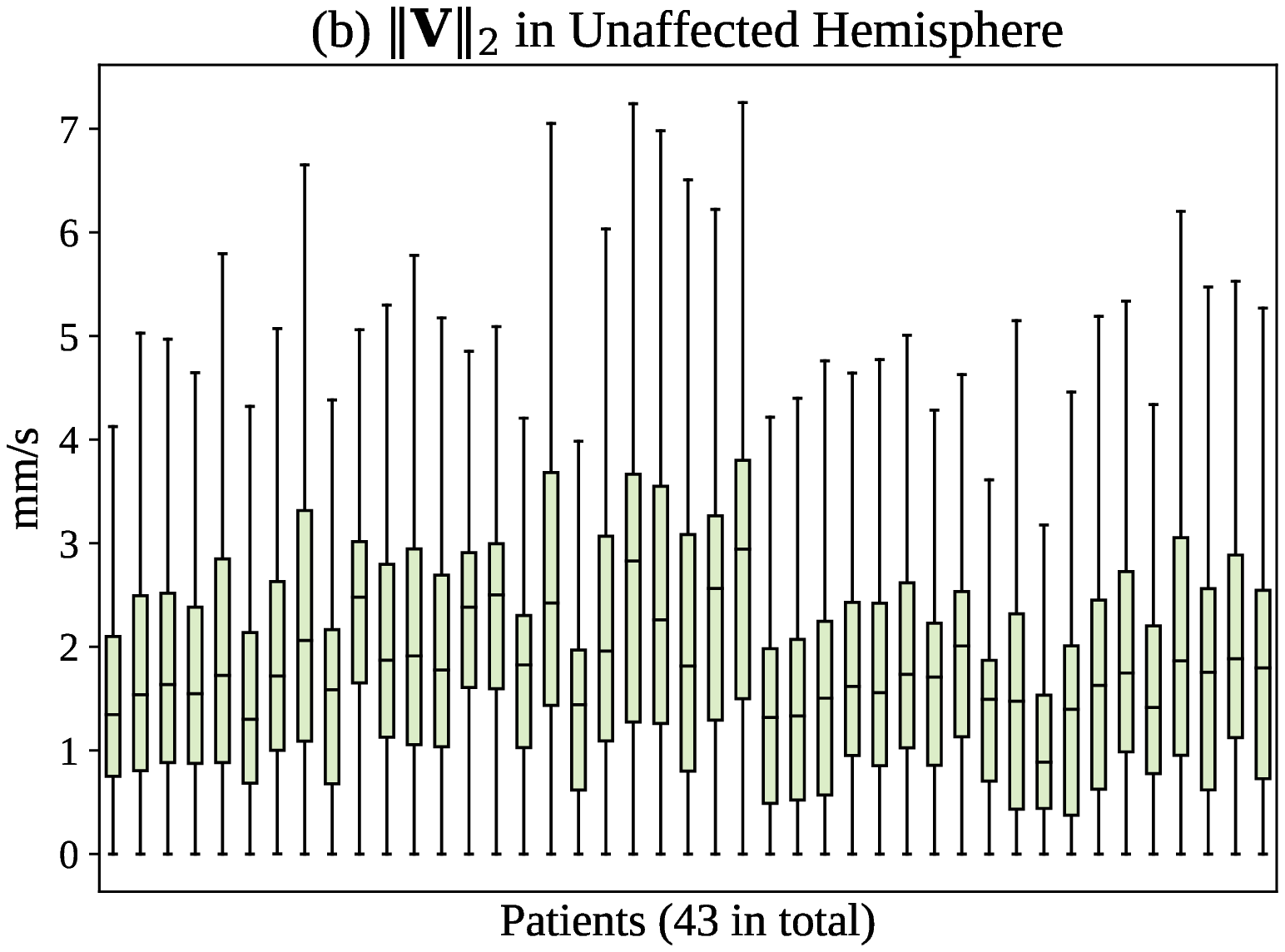}};
	\end{tikzpicture}
	}
	\caption{(a) Histogram of $\| {\bf{V}} \|_2$ in the unaffected hemispheres of 43 ISLES 2017 patients, and (b) corresponding box plots of distribution for individual patients.}
	\label{box_plots_v}
\end{figure}

\begin{figure}[t]
	\noindent\resizebox{0.9\textwidth}{!}{
	\begin{tikzpicture}
	\node at (0, 0) {\includegraphics[width=4cm]{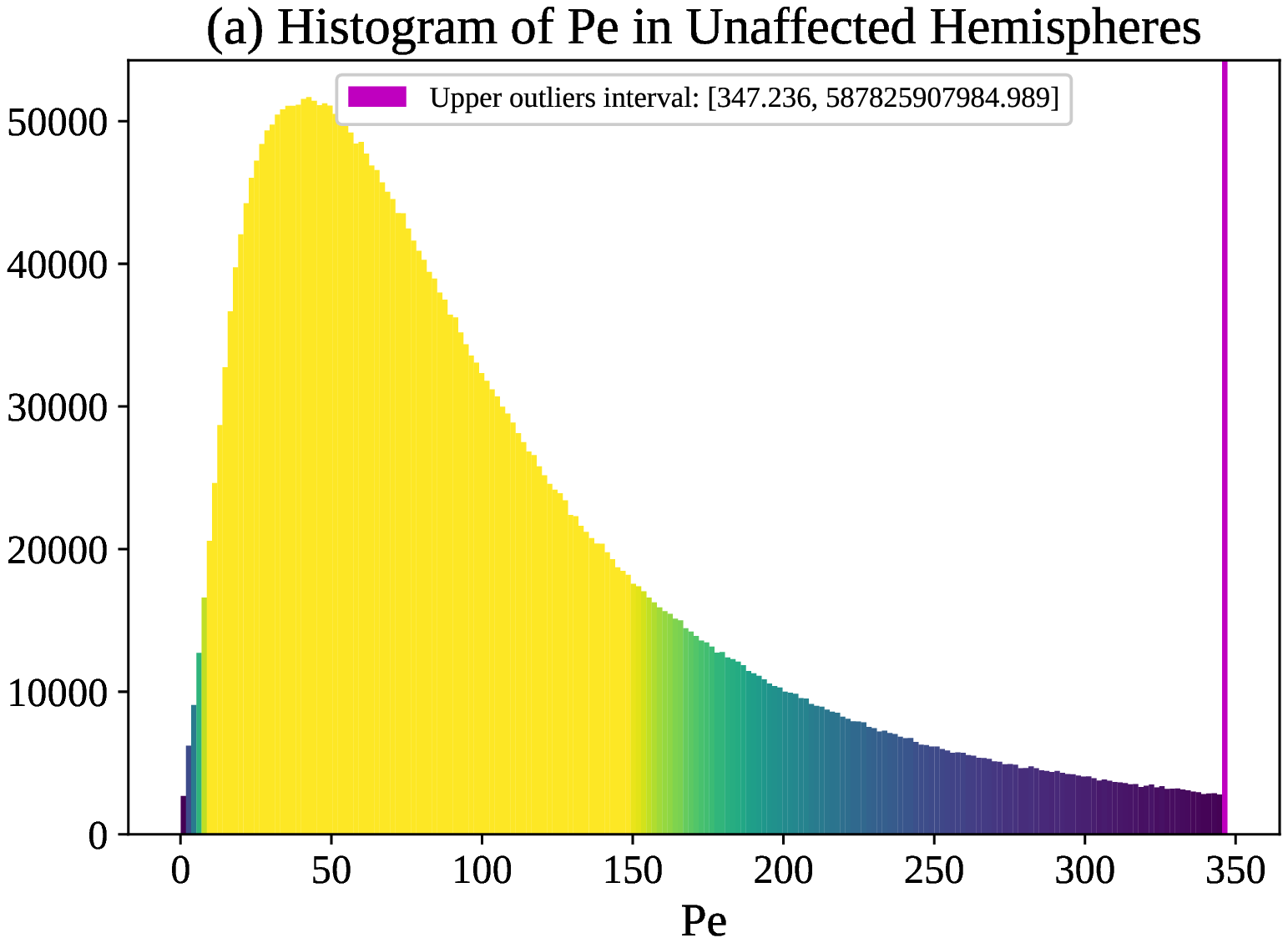}};
	\node at (0, -3) {\includegraphics[width=4cm]{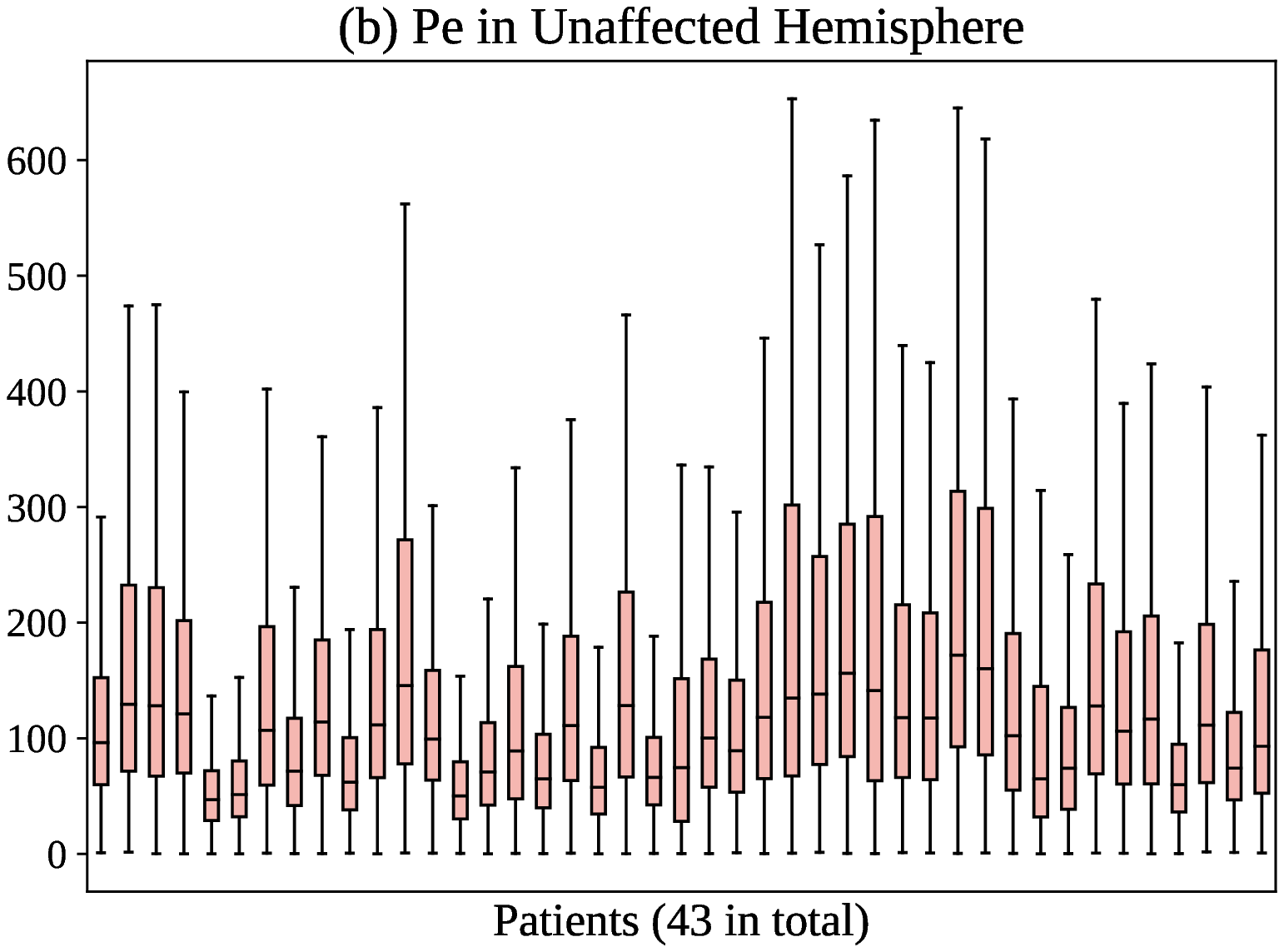}};
	\end{tikzpicture}
	}
	\caption{(a) Histogram of Pe in the unaffected hemispheres of 43 ISLES 2017 patients, and (b) corresponding box plots of distribution for individual patients.}
	\label{box_plots_pe}
\end{figure}

\begin{figure}[t]
	\noindent\resizebox{0.9\textwidth}{!}{
	\begin{tikzpicture}
	\node at (0, 0) {\includegraphics[width=4cm]{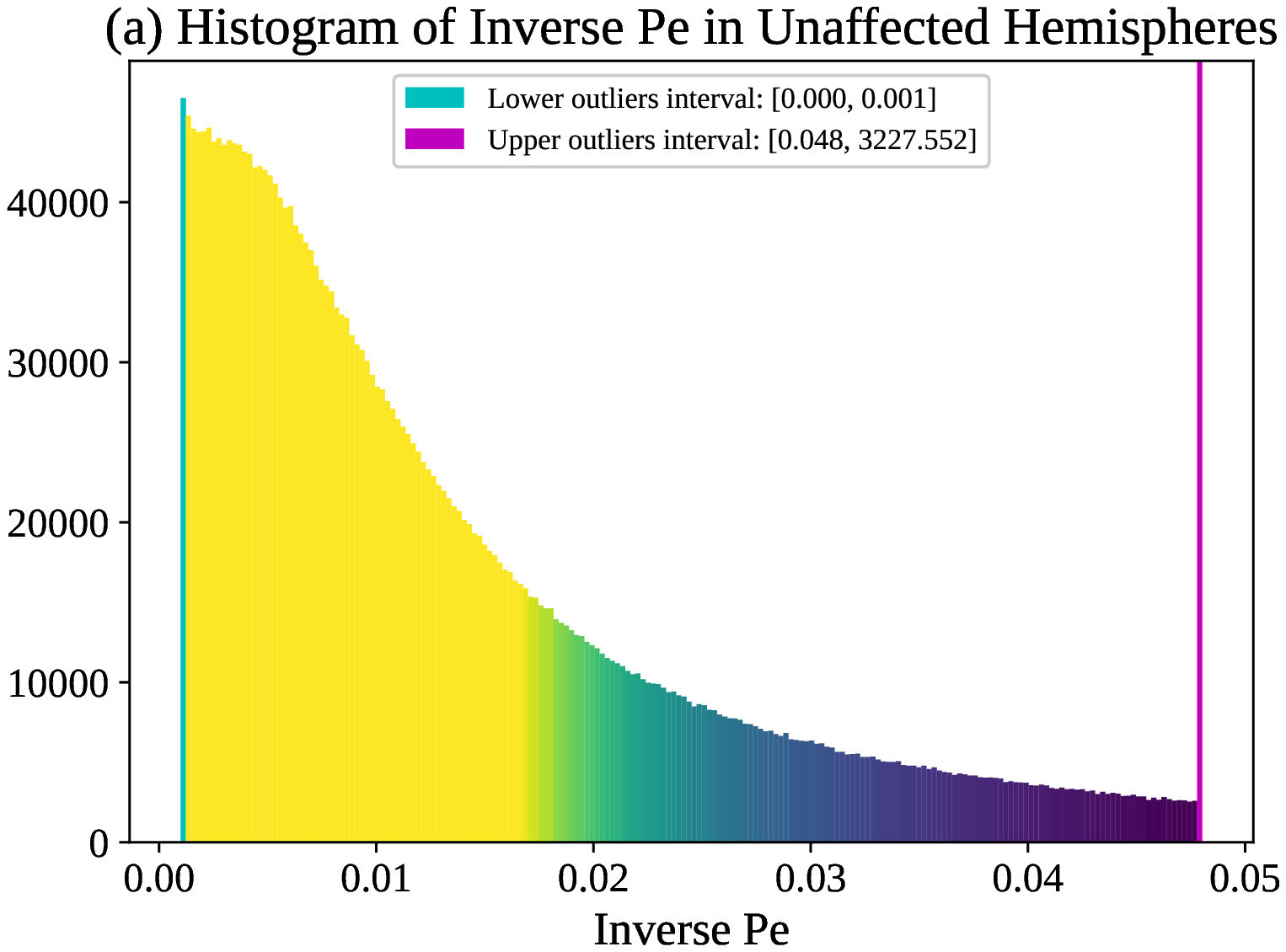}};
	\node at (0, -3) {\includegraphics[width=4cm]{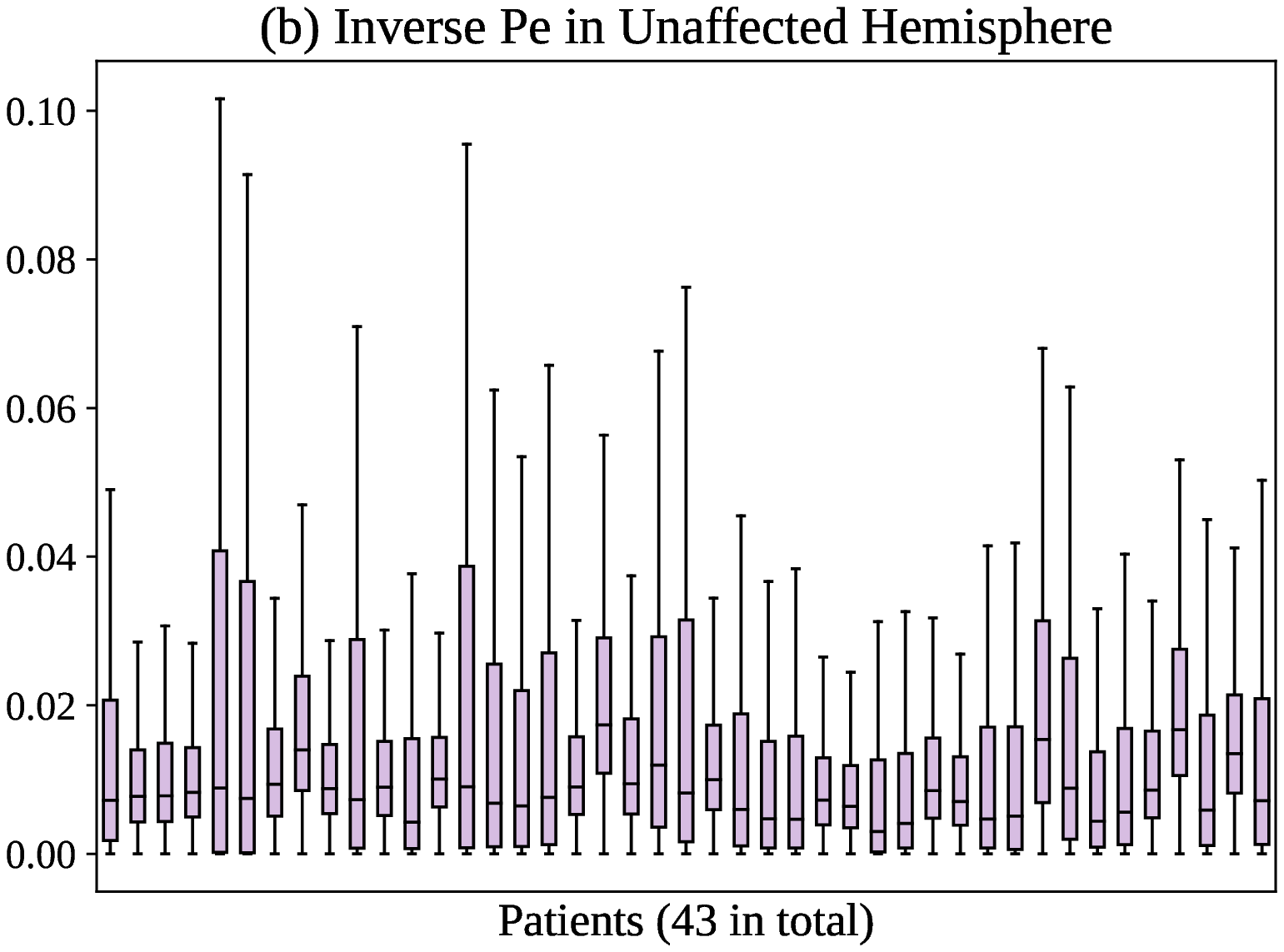}};
	\end{tikzpicture}
	}
	\caption{(a) Histogram of inverse Pe in the unaffected hemispheres of 43 ISLES 2017 patients, and (b) corresponding box plots of distribution for individual patients.}
	\label{box_plots_inv_pe}
\end{figure}

%% file: sub/exp_effect_robust.tex
\subsection{PIANO Effectiveness and Robustness Testing}
\label{sec: effect_robust}

\input{sub/exp_results/effect_robust/box_plots}

Mathematically, PIANO aims at recovering the velocity and diffusion fields of an advection-diffusion process following \Cref{eq: div_free}. In Sec. \ref{sec:quantiative_comparison}, PIANO feature maps showed  greater sensitivity for assessing the lesion compared to conventional perfusion parameter maps. Taking one step back, in this section, we check two essential properties of PIANO: (1) Accuracy of the estimated velocity and diffusion fields, i.e., given a time-series of images capturing an advection process, driven by a certain velocity field, is PIANO capable of recovering the underlying velocity field? Similarly, can PIANO successfully recover a diffusion field governing a diffusion process. (2) Robustness of the estimation to noise. I.e., if measuring errors exist in the given time-series of concentration images, can PIANO still achieve reasonable estimates?

\subsubsection{Advection Imaging via Advection}
\label{sec: adv-adv}
\input{sub/exp_results/effect_robust/adv}

Our goal here is to determine if PIANO can estimate a known velocity field from a concentration time-series consistent with this velocity field.
To this end, for each patient in the ISLES 2017 training set, we treat the velocity field estimated by PIANO as the ground truth velocity field (${\bf{V^{\text{gt}}}}$), and the first image in the concentration time-series dataset $\{C^{t_i}\}$ (as described in Sec. \ref{sec:experimental_results}) is used as the initial condition ($C^0$). We then simulate `advection imaging' $\{C^{t_i}\in\mathbb{R}(\Omega)\vert i = 0,\, 1,\, \ldots,\, 40\}$, i.e., we create a time-series of concentration images driven by the velocity ${\bf{V}}:={\bf{V^{\text{gt}}}}$ only via the advection PDE:
\begin{equation}
\frac{\partial C({\bf{x}}, t)}{\partial t} = -  {\bf{V}}({\bf{x}})\cdot\nabla C({\bf{x}}, t).
\label{eq: adv}
\end{equation}

We further simulate noisy concentrations by adding Rician noise \cite{niethammer2008rice} with variances equalling 2\%, 4\%, 6\%, 8\%, 10\% of the originally simulated concentrations $\{C^{t_i}\}$. We then test how well PIANO can estimate the underlying velocity field via \Cref{eq: adv} with the same model settings (except without estimating the diffusion) as in Sec. \ref{sec:experimental_results} given the original and the noisy concentration time-series. Fig. \ref{effect_adv} shows the estimated $\Vert {\bf{V}}^{\text{est}} \Vert_2$ for one patient. Fig. \ref{sec: effct_box_plots} (a) shows the summarized estimation results for all 43 patients. PIANO almost perfectly captures the underlying velocity field, and maintains excellent performance even when estimating from concentrations with varying noise levels.

%%%%%%%%%%%%%%%%%%%%%%%%%%%%%%%%

\subsubsection{Diffusion Imaging via Diffusion}
\label{sec: diff-diff}
\input{sub/exp_results/effect_robust/diff}

Similarly, starting from the same initial condition $C^0$ as in the `Advection Imaging' experiment for each patient, we simulate concentration time-series $\{C^{t_i}\in\mathbb{R}(\Omega)\vert i = 0,\, 1,\, \ldots,\, 40\}$ via a diffusion PDE, where we define the ground truth diffusivity $D:=D^{\text{gt}}$ via the ADC map of the ISLES 2017 training set (ADC values are scaled by $0.00001$ to ensure numerical stability):
\begin{equation}
\frac{\partial C({\bf{x}}, t)}{\partial t} =  \nabla \cdot \left(D({\bf{x}})\, \nabla C({\bf{x}}, t)\right).
\label{eq: diff}
\end{equation}
Note this is likely not a spatially representative ground-truth for perfusion imaging, as it measures different effects from diffusion imaging. However, we still use it as a quasi-realistic pattern of diffusivity in the brain.
We also added 2\%, 4\%, 6\%, 8\%, 10\% levels of Rician noise to obtain simulations of `Diffusion Imaging'. The estimated $D^{\text{est}}$ given concentrations of all noise levels for one patient are shown in Fig. \ref{effect_diff}, PIANO estimation results for all patients are summarized in Fig. \ref{sec: effct_box_plots} (b). Again, PIANO demonstrates its capability to recover the underlying diffusion field. In Fig. \ref{effect_diff}, when the noise level is increasing, some noisy patterns indeed appear in the associated $D^{\text{est}}$. Note that the ground truth diffusivity applied in this simulation experiment is about ten times larger than the diffusivity estimated in reality (Fig. \ref{M14_param}, Fig. \ref{M16_param}).

%% file: sub/exp_results/effect_robust/box_plots.tex
\begin{figure}[t]
	\noindent\resizebox{0.9\textwidth}{!}{
	\begin{tikzpicture}
	\node at (0, 0) {\includegraphics[width=4cm]{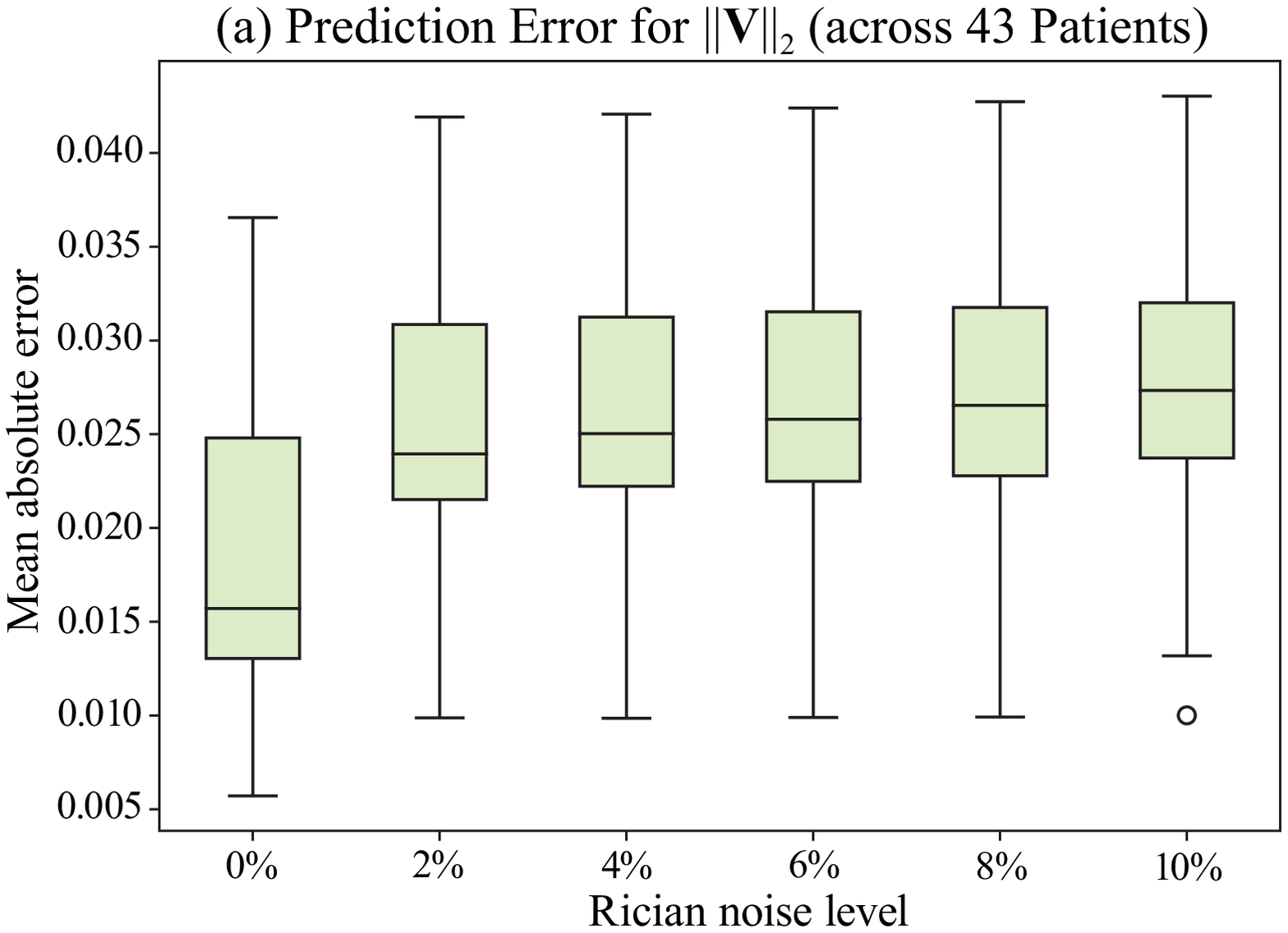}};
	\node at (0, -3) {\includegraphics[width=4cm]{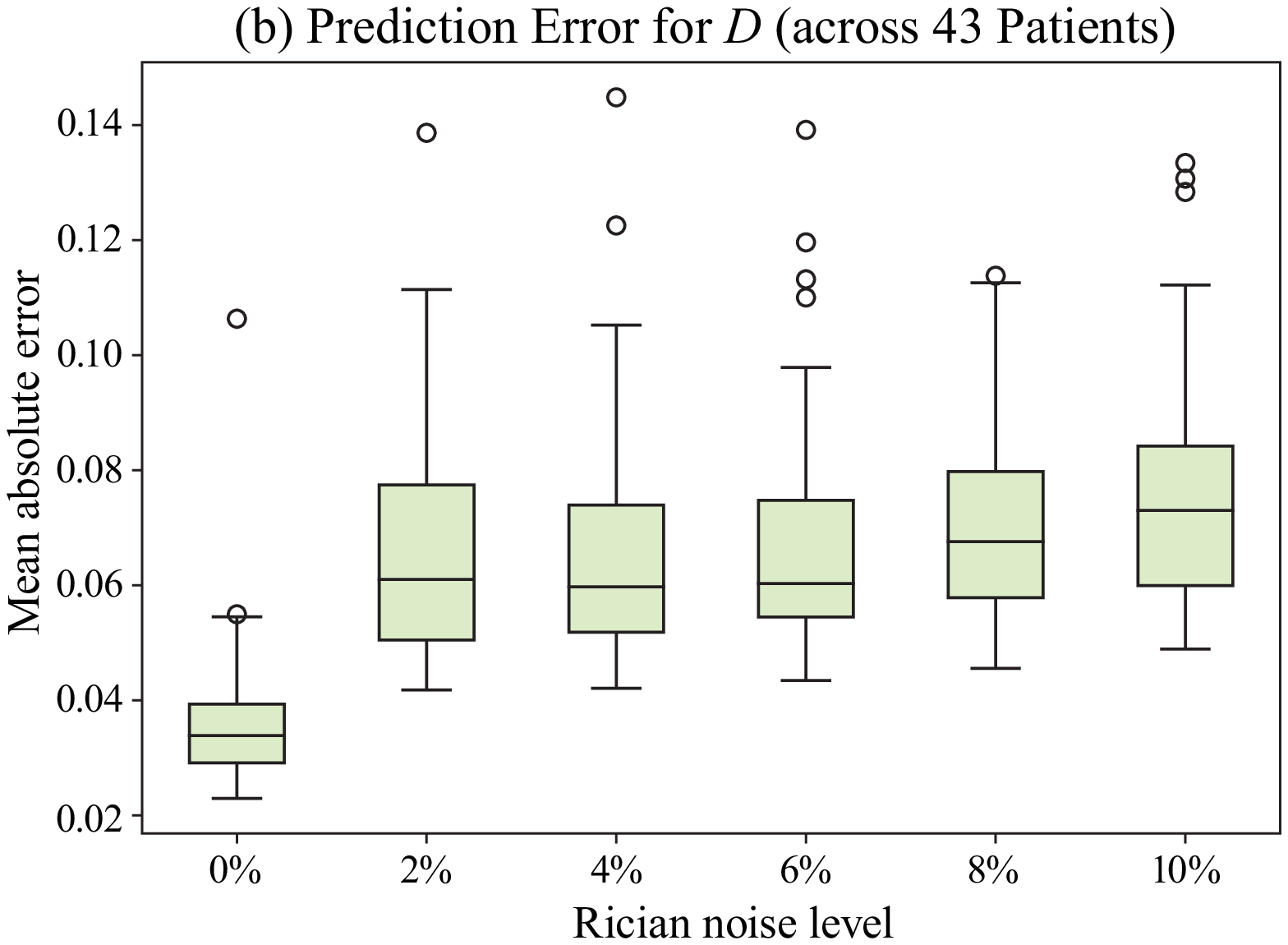}};
	\end{tikzpicture}
	}
	\caption{PIANO effectiveness and robustness testing: box plots of mean absolute error (MAE). (a) Advection Imaging via Advection: MAE of estimated $\| {\bf{V}} \|_2$; (b) Diffusion Imaging via Diffusion: MAE of estimated $D$. To ensure that estimation errors can be compared across different patients, we scaled all estimated feature maps by the maximum value of the corresponding ground truth feature maps.}
	\label{sec: effct_box_plots}
\end{figure}

%% file: sub/exp_results/effect_robust/adv.tex
\begin{figure}[t]
	\noindent\resizebox{\textwidth}{!}{
	\begin{tikzpicture}
	
	% x axis legend
	\node at (0, 1.7)  {{\footnotesize{Slice \#1}}};
	\node at (2, 1.7)  {{\footnotesize{Slice \#2}}};
	\node at (4, 1.7)  {{\footnotesize{Slice \#3}}};
	\node at (6, 1.7)  {{\footnotesize{Slice \#4}}};
	\node at (8, 1.7)  {{\footnotesize{Slice \#5}}};
	\node at (10, 1.7)  {{\footnotesize{Slice \#6}}};
	
	% GT
	\node at (-1.5, 0.25) {\footnotesize{$\Vert {\bf{V}}^{\text{gt}} \Vert_2$}};
	\node at (0, 0.25) {\includegraphics[width=1.9cm]{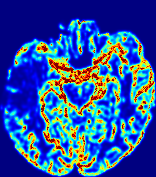}};
	\node at (2, 0.25) {\includegraphics[width=1.9cm]{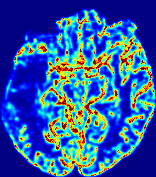}};
	\node at (4, 0.25) {\includegraphics[width=1.9cm]{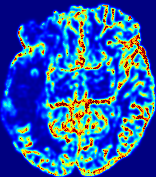}};
	\node at (6, 0.25) {\includegraphics[width=1.9cm]{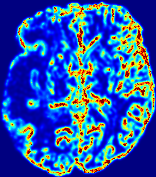}};
	\node at (8, 0.25) {\includegraphics[width=1.9cm]{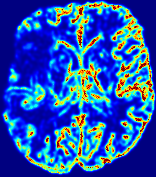}};
	\node at (10, 0.25) {\includegraphics[width=1.9cm]{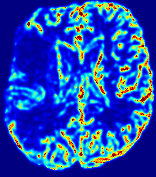}};
	
	% PD
	\node at (-1.5, -2.25) {\footnotesize{(a)}};
	\node at (0, -2.25) {\includegraphics[width=1.9cm]{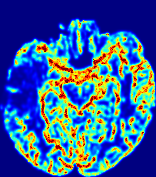}};
	\node at (2, -2.25) {\includegraphics[width=1.9cm]{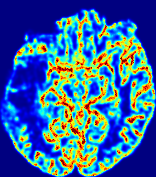}};
	\node at (4, -2.25) {\includegraphics[width=1.9cm]{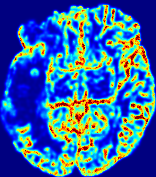}};
	\node at (6, -2.25) {\includegraphics[width=1.9cm]{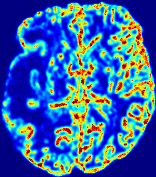}};
	\node at (8, -2.25) {\includegraphics[width=1.9cm]{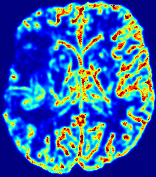}};
	\node at (10, -2.25) {\includegraphics[width=1.9cm]{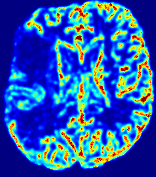}};

	\node at (-1.5, -4.5) {\footnotesize{(b)}};
	\node at (0, -4.5) {\includegraphics[width=1.9cm]{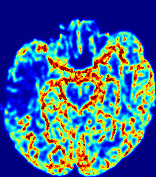}};
	\node at (2, -4.5) {\includegraphics[width=1.9cm]{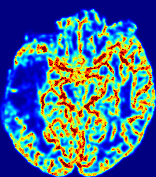}};
	\node at (4, -4.5) {\includegraphics[width=1.9cm]{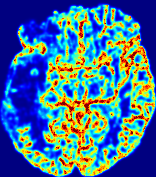}};
	\node at (6, -4.5) {\includegraphics[width=1.9cm]{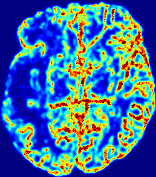}};
	\node at (8, -4.5) {\includegraphics[width=1.9cm]{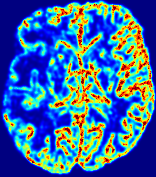}};
	\node at (10, -4.5) {\includegraphics[width=1.9cm]{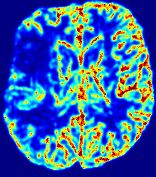}};

	\node at (-1.5, -6.75) {\footnotesize{(c)}};
	\node at (0, -6.75) {\includegraphics[width=1.9cm]{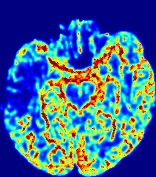}};
	\node at (2, -6.75) {\includegraphics[width=1.9cm]{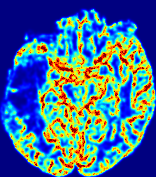}};
	\node at (4, -6.75) {\includegraphics[width=1.9cm]{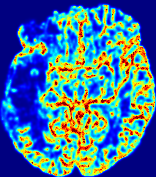}};
	\node at (6, -6.75) {\includegraphics[width=1.9cm]{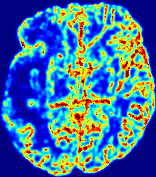}};
	\node at (8, -6.75) {\includegraphics[width=1.9cm]{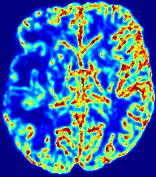}};
	\node at (10, -6.75) {\includegraphics[width=1.9cm]{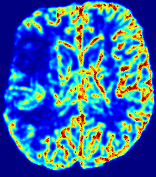}};

	\node at (-1.5, -9) {\footnotesize{(d)}};
	\node at (0, -9) {\includegraphics[width=1.9cm]{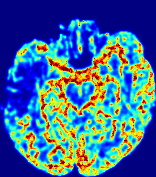}};
	\node at (2, -9) {\includegraphics[width=1.9cm]{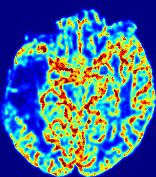}};
	\node at (4, -9) {\includegraphics[width=1.9cm]{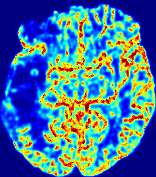}};
	\node at (6, -9) {\includegraphics[width=1.9cm]{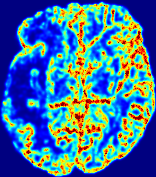}};
	\node at (8, -9) {\includegraphics[width=1.9cm]{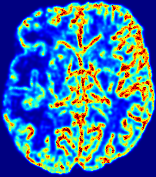}};
	\node at (10, -9) {\includegraphics[width=1.9cm]{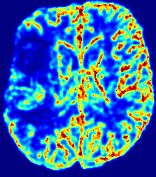}};

	\node at (-1.5, -11.25) {\footnotesize{(e)}};
	\node at (0, -11.25) {\includegraphics[width=1.9cm]{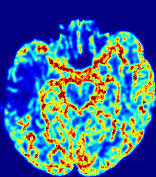}};
	\node at (2, -11.25) {\includegraphics[width=1.9cm]{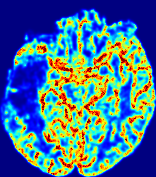}};
	\node at (4, -11.25) {\includegraphics[width=1.9cm]{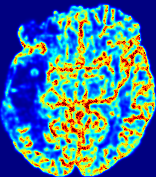}};
	\node at (6, -11.25) {\includegraphics[width=1.9cm]{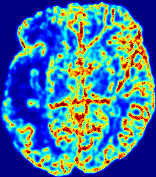}};
	\node at (8, -11.25) {\includegraphics[width=1.9cm]{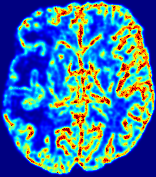}};
	\node at (10, -11.25) {\includegraphics[width=1.9cm]{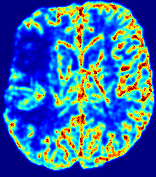}};

	\node at (-1.5, -13.5) {\footnotesize{(f)}};
	\node at (0, -13.5) {\includegraphics[width=1.9cm]{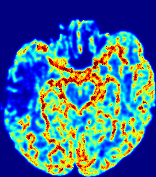}};
	\node at (2, -13.5) {\includegraphics[width=1.9cm]{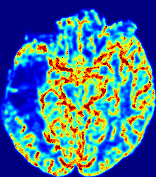}};
	\node at (4, -13.5) {\includegraphics[width=1.9cm]{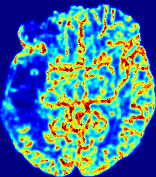}};
	\node at (6, -13.5) {\includegraphics[width=1.9cm]{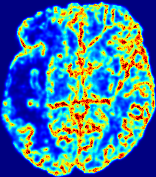}};
	\node at (8, -13.5) {\includegraphics[width=1.9cm]{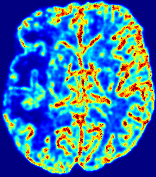}};
	\node at (10, -13.5) {\includegraphics[width=1.9cm]{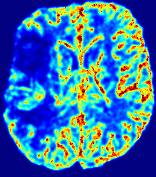}};
	
	% Color bar
	\node at (5, -15.15) {\includegraphics[width=6cm]{fig/cb2.png}};
	\node at (2.3, -15.7)  {$0$};
	\node at (3.4, -15.7)  {$0.3$};
	\node at (4.4775, -15.7)  {$0.6$};
	\node at (5.55, -15.7)  {$0.9$};
	\node at (6.65, -15.7)  {$1.2$};
	\node at (7.75, -15.7)  {$1.5$};
	\node at (8.75, -15.7) {$(mm/s)$};

	\end{tikzpicture}
	}
	\caption{PIANO effectiveness and robustness testing: advection imaging via advection. Top row shows the ground truth $\Vert {\bf{V}}^{\text{gt}} \Vert_2$ used for simulating pure advection. (a)-(f) refer to the results for $\Vert {\bf{V}} \Vert_2$ estimated by PIANO, with simulated advection imaging series where Rician noise at levels 0\%, 2\%, 4\%, 6\%, 8\%, 10\% was added respectively.}
	\label{effect_adv}
\end{figure}

%% file: sub/exp_results/effect_robust/diff.tex
\begin{figure}[t]
	\noindent\resizebox{0.98\textwidth}{!}{
	\begin{tikzpicture}
	
	% x axis legend
	\node at (0, 1.7)  {{\footnotesize{Slice \#1}}};
	\node at (2, 1.7)  {{\footnotesize{Slice \#2}}};
	\node at (4, 1.7)  {{\footnotesize{Slice \#3}}};
	\node at (6, 1.7)  {{\footnotesize{Slice \#4}}};
	\node at (8, 1.7)  {{\footnotesize{Slice \#5}}};
	\node at (10, 1.7)  {{\footnotesize{Slice \#6}}};
	
	% GT
	
	\node at (-1.5, 0.25) {\footnotesize{$D^{\text{gt}}$}};
	\node at (0, 0.25) {\includegraphics[width=1.9cm]{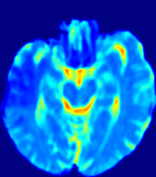}};
	\node at (2, 0.25) {\includegraphics[width=1.9cm]{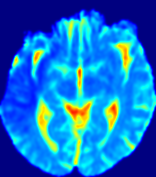}};
	\node at (4, 0.25) {\includegraphics[width=1.9cm]{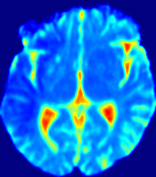}};
	\node at (6, 0.25) {\includegraphics[width=1.9cm]{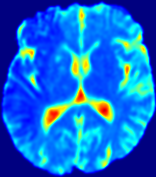}};
	\node at (8, 0.25) {\includegraphics[width=1.9cm]{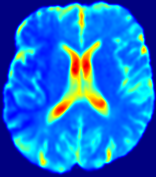}};
	\node at (10, 0.25) {\includegraphics[width=1.9cm]{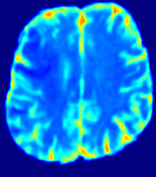}};
	
	% PD
	
	\node at (-1.5, -2.25) {\footnotesize{(a)}};
	\node at (0, -2.25) {\includegraphics[width=1.9cm]{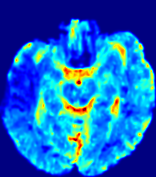}};
	\node at (2, -2.25) {\includegraphics[width=1.9cm]{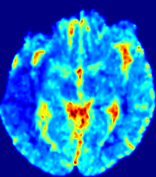}};
	\node at (4, -2.25) {\includegraphics[width=1.9cm]{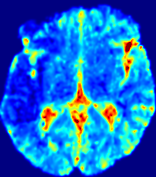}};
	\node at (6, -2.25) {\includegraphics[width=1.9cm]{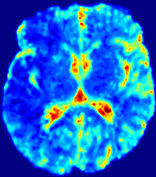}};
	\node at (8, -2.25) {\includegraphics[width=1.9cm]{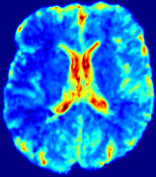}};
	\node at (10, -2.25) {\includegraphics[width=1.9cm]{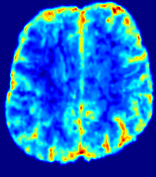}};

	\node at (-1.5, -4.5) {\footnotesize{(b)}};
	\node at (0, -4.5) {\includegraphics[width=1.9cm]{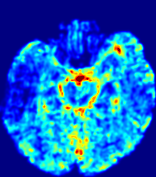}};
	\node at (2, -4.5) {\includegraphics[width=1.9cm]{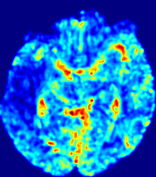}};
	\node at (4, -4.5) {\includegraphics[width=1.9cm]{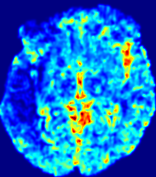}};
	\node at (6, -4.5) {\includegraphics[width=1.9cm]{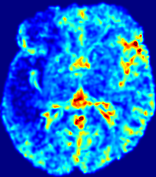}};
	\node at (8, -4.5) {\includegraphics[width=1.9cm]{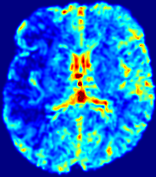}};
	\node at (10, -4.5) {\includegraphics[width=1.9cm]{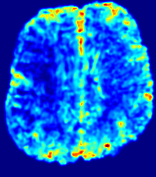}};

	\node at (-1.5, -6.75) {\footnotesize{(c)}};
	\node at (0, -6.75) {\includegraphics[width=1.9cm]{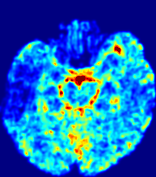}};
	\node at (2, -6.75) {\includegraphics[width=1.9cm]{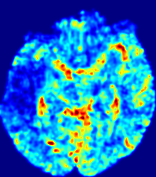}};
	\node at (4, -6.75) {\includegraphics[width=1.9cm]{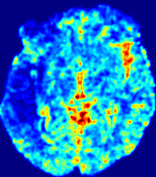}};
	\node at (6, -6.75) {\includegraphics[width=1.9cm]{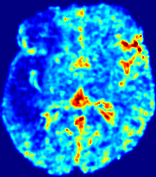}};
	\node at (8, -6.75) {\includegraphics[width=1.9cm]{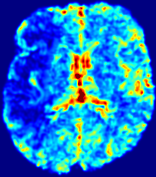}};
	\node at (10, -6.75) {\includegraphics[width=1.9cm]{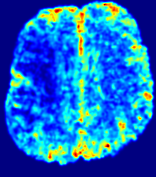}};

	\node at (-1.5, -9) {\footnotesize{(d)}};
	\node at (0, -9) {\includegraphics[width=1.9cm]{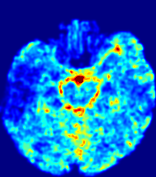}};
	\node at (2, -9) {\includegraphics[width=1.9cm]{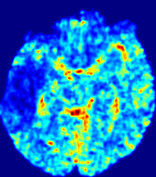}};
	\node at (4, -9) {\includegraphics[width=1.9cm]{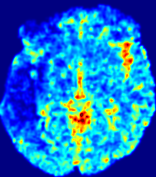}};
	\node at (6, -9) {\includegraphics[width=1.9cm]{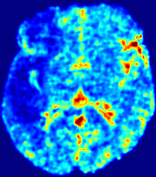}};
	\node at (8, -9) {\includegraphics[width=1.9cm]{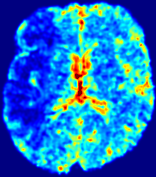}};
	\node at (10, -9) {\includegraphics[width=1.9cm]{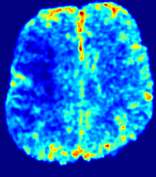}};

	\node at (-1.5, -11.25) {\footnotesize{(e)}};
	\node at (0, -11.25) {\includegraphics[width=1.9cm]{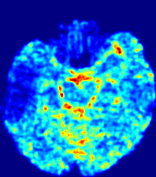}};
	\node at (2, -11.25) {\includegraphics[width=1.9cm]{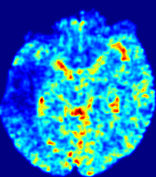}};
	\node at (4, -11.25) {\includegraphics[width=1.9cm]{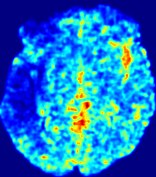}};
	\node at (6, -11.25) {\includegraphics[width=1.9cm]{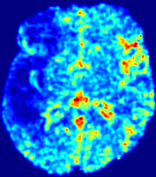}};
	\node at (8, -11.25) {\includegraphics[width=1.9cm]{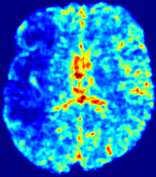}};
	\node at (10, -11.25) {\includegraphics[width=1.9cm]{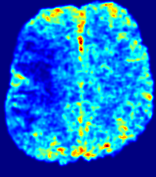}};

	\node at (-1.5, -13.5) {\footnotesize{(f)}};
	\node at (0, -13.5) {\includegraphics[width=1.9cm]{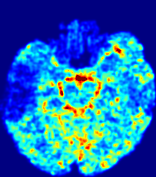}};
	\node at (2, -13.5) {\includegraphics[width=1.9cm]{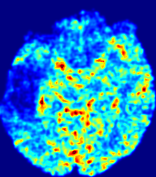}};
	\node at (4, -13.5) {\includegraphics[width=1.9cm]{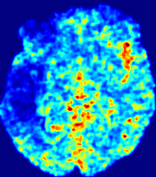}};
	\node at (6, -13.5) {\includegraphics[width=1.9cm]{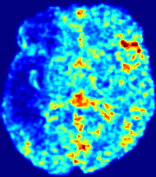}};
	\node at (8, -13.5) {\includegraphics[width=1.9cm]{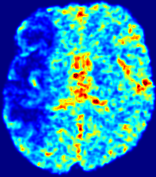}};
	\node at (10, -13.5) {\includegraphics[width=1.9cm]{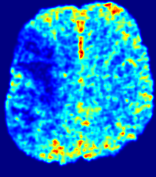}};
	
	% Color bar
	\node at (5, -15.15) {\includegraphics[width=6cm]{fig/cb2.png}};
	\node at (2.3, -15.7)  {$0$};
	\node at (3.4, -15.7)  {$0.06$};
	\node at (4.4775, -15.7)  {$0.12$};
	\node at (5.55, -15.7)  {$0.18$};
	\node at (6.65, -15.7)  {$0.24$};
	\node at (7.75, -15.7)  {$0.30$};
	\node at (9, -15.7) {$(mm^2/s)$};

	\end{tikzpicture}
	}
	\caption{PIANO effectiveness and robustness testing: diffusion imaging via diffusion. Top row shows $D^{\text{gt}}$ used for simulating the ground truth pure diffusion. (a)-(f) refer to the results for $D$ estimated from the ground truth pure diffusion image time-series where Rician noise at levels 0\%, 2\%, 4\%, 6\%, 8\%, 10\% was added respectively.}
	\label{effect_diff}
\end{figure}

%% file: sub/exp_ident.tex
\subsection{PIANO Identifiability Testing}
\label{sec: ident}

We verified in Sec. \ref{sec: effect_robust} the capability of PIANO to estimate the underlying velocity field governing an advection process (`Advection Imaging via Advection'), and the original diffusion field given a diffusion process (`Diffusion Imaging via Diffusion'), respectively. In this section, we further test the identifiability properties of PIANO. Specifically, we let PIANO simultaneously estimate both velocity and diffusion fields given a pure advection (or diffusion) process. The key point for this task is, given a pure advection (or diffusion) process, does PIANO confuse advection with diffusion, resulting in an incorrect estimation for the underlying velocity (or diffusion) field?

\subsubsection{Advection Imaging via Advection-Diffusion}
\label{sec: adv-advdiff}
\input{sub/exp_results/ident/for_adv}

We use the same `Advection Imaging' simulation of Sec. \ref{sec: adv-adv} as the concentration dataset for PIANO. However, instead of modeling pure advection (\Cref{eq: adv}), we let PIANO estimate both velocity ${\bf{V}}^{\text{est}}$ and diffusivity $D^{\text{est}}$ via the advection-diffusion PDE (\Cref{eq: div_free}) underlying the proposed PIANO model. Fig.~\ref{ident_adv} shows the estimated $\Vert {\bf{V}}^{\text{est}} \Vert_2,$ and $D^{\text{est}}$ fields for one patient. Although PIANO has the freedom to estimate both a velocity and a diffusivity field from pure advection, PIANO differentiates well between advection and diffusion: the estimated $\Vert {\bf{V}}^{\text{est}} \Vert_2$ successfully reproduces the ground truth $\Vert {\bf{V}}^{\text{gt}} \Vert_2$ governing the simulated advection process, just as it already did in the `Advection Imaging via Advection' test (Fig. \ref{effect_adv}). More importantly, the estimated diffusivity $D^{\text{est}}$ is orders of magnitudes smaller than $\Vert {\bf{V}}^{\text{est}} \Vert_2$, indicating the estimated diffusion is negligible compared to the estimated advection, which is highly consistent with the underlying pure advection of the simulated data.

%%%%%%%%%%%%%%%%%%%%%%%%%%%%%%%%

\subsubsection{Diffusion Imaging via Advection-Diffusion}
\label{sec: diff-advdiff}
\input{sub/exp_results/ident/for_diff}

Similarly, we test the  behavior of PIANO when estimating both advection and diffusion from a pure diffusion-driven process. The goal is to determine if PIANO is able to recognize that there is only diffusion governing the given concentration time-series. We use the same `Diffusion Imaging' data simulation of Sec. \ref{sec: adv-adv} as the concentration dataset, PIANO estimates both velocity ${\bf{V}}^{\text{est}}$ and diffusivity $D^{\text{est}}$. Estimation results in Fig. \ref{ident_diff} confirm PIANO's identifiability again: the estimated $\Vert {\bf{V}}^{\text{est}} \Vert_2$ is almost invisible compared to $D^{\text{est}}$, even plotted with a $1\%$ value range compared to that for $D^{\text{est}}$. On the other hand, $D^{\text{est}}$ achieves comparable estimation performance as `Diffusion Imaging via Diffusion' in which PIANO predicts $D^{\text{est}}$ alone (shown in Fig. \ref{effect_diff}).

%% file: sub/exp_results/ident/for_adv.tex
\begin{figure}[t]
	\noindent\resizebox{\textwidth}{!}{
	\begin{tikzpicture}
	
	% x axis legend
	\node at (0, -8.15)  {{\footnotesize{Slice \#1}}};
	\node at (2, -8.15)  {{\footnotesize{Slice \#2}}};
	\node at (4, -8.15)  {{\footnotesize{Slice \#3}}};
	\node at (6, -8.15)  {{\footnotesize{Slice \#4}}};
	\node at (8, -8.15)  {{\footnotesize{Slice \#5}}};
	\node at (10, -8.15)  {{\footnotesize{Slice \#6}}};
	
	% Norm V
	% GT
	\node at (-1.6, -2) {\footnotesize{$\Vert \bf{V}^{\text{gt}} \Vert_2$}};
	\node at (0, -2) {\includegraphics[width=1.9cm]{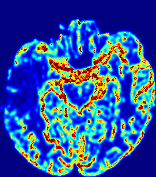}};
	\node at (2, -2) {\includegraphics[width=1.9cm]{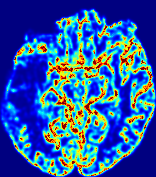}};
	\node at (4, -2) {\includegraphics[width=1.9cm]{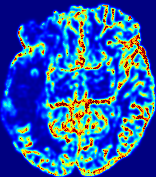}};
	\node at (6, -2) {\includegraphics[width=1.9cm]{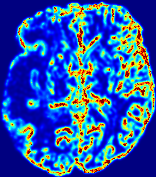}};
	\node at (8, -2) {\includegraphics[width=1.9cm]{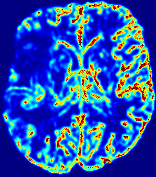}};
	\node at (10, -2) {\includegraphics[width=1.9cm]{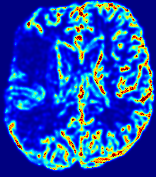}};
	
	% PD
	\node at (-1.6, -4.5) {\footnotesize{$\Vert \bf{V}^{\text{est}} \Vert_2$}};
	\node at (0, -4.5) {\includegraphics[width=1.9cm]{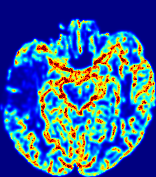}};
	\node at (2, -4.5) {\includegraphics[width=1.9cm]{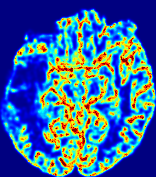}};
	\node at (4, -4.5) {\includegraphics[width=1.9cm]{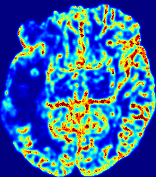}};
	\node at (6, -4.5) {\includegraphics[width=1.9cm]{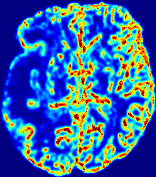}};11.56
	\node at (8, -4.5) {\includegraphics[width=1.9cm]{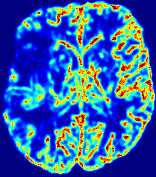}};
	\node at (10, -4.5) {\includegraphics[width=1.9cm]{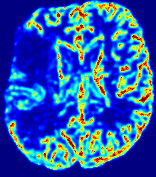}};
	
	% Color bar shift: +1.1; Text shift: +1; Text Gap: -0.575
	\node at (11.2, -3.4) {\includegraphics[width=0.4cm]{fig/cb.png}};
	\node at (11.58, -1.95) {\tiny{$1.5$}};
	\node at (11.58, -2.525) {\tiny{$1.2$}};
	\node at (11.58, -3.1) {\tiny{$0.9$}};
	\node at (11.58, -3.675) {\tiny{$0.6$}};
	\node at (11.58, -4.25) {\tiny{$0.3$}};
	\node at (11.58, -4.825) {\tiny{$0.0$}};
	\node at (11.4, -5.25) {\tiny{$(mm/s)$}};
	
	% D 
	\node at (-1.6, -6.75) {\footnotesize{$D^{\text{est}}$}};
	\node at (0, -6.75) {\includegraphics[width=1.9cm]{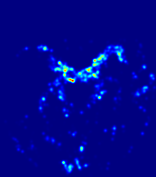}};
	\node at (2, -6.75) {\includegraphics[width=1.9cm]{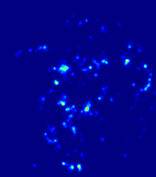}};
	\node at (4, -6.75) {\includegraphics[width=1.9cm]{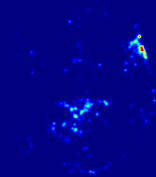}};
	\node at (6, -6.75) {\includegraphics[width=1.9cm]{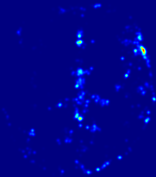}};
	\node at (8, -6.75) {\includegraphics[width=1.9cm]{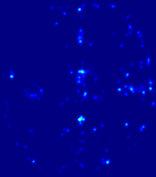}};
	\node at (10, -6.75) {\includegraphics[width=1.9cm]{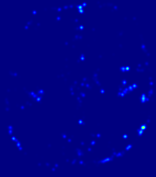}};
	% Color bar shift: +0.1; Unit shift: -0.985;  Text shift: +1; Text Gap: -0.36
	\node at (11.12, -6.6) {\includegraphics[width=0.25cm]{fig/cb.png}};
	\node at (11.56, -5.7) {\tiny{$0.015$}};
	\node at (11.56, -6.06) {\tiny{$0.012$}};
	\node at (11.56, -6.42) {\tiny{$0.009$}};
	\node at (11.56, -6.78) {\tiny{$0.006$}}; % 2.65
	\node at (11.56, -7.12) {\tiny{$0.003$}};
	\node at (11.56, -7.44) {\tiny{$0.000$}};
	\node at (11.45, -7.75) {\tiny{$(mm^2/s)$}}; % 0.2

	\end{tikzpicture}
	}
	\caption{PIANO identifiability testing: advection imaging via advection-diffusion. Top row shows $\Vert {\bf{V}}^{\text{gt}} \Vert_2$ used for simulating ground truth pure advection. Rows below show the estimated $\Vert {\bf{V}}^{\text{est}}  \Vert_2$ and $D^{\text{est}} $ on corresponding slices. Note that the plotted value scale for $D^{\text{est}}$ is 0.01 of that for $\Vert {\bf{V}}^{\text{gt}} \Vert_2$ and  $\Vert {\bf{V}}^{\text{est}} \Vert_2$.}
	\label{ident_adv}
\end{figure}

%% file: sub/exp_results/ident/for_diff.tex
\begin{figure}[t]
	\noindent\resizebox{\textwidth}{!}{
	\begin{tikzpicture}
	
	% x axis legend
	\node at (0, -8.15)  {{\footnotesize{Slice \#1}}};
	\node at (2, -8.15)  {{\footnotesize{Slice \#2}}};
	\node at (4, -8.15)  {{\footnotesize{Slice \#3}}};
	\node at (6, -8.15)  {{\footnotesize{Slice \#4}}};
	\node at (8, -8.15)  {{\footnotesize{Slice \#5}}};
	\node at (10, -8.15)  {{\footnotesize{Slice \#6}}};
	
	% Norm V
	% GT
	\node at (-1.6, -2) {\footnotesize{$D^{\text{gt}}$}};
	\node at (0, -2) {\includegraphics[width=1.9cm]{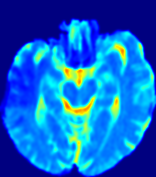}};
	\node at (2, -2) {\includegraphics[width=1.9cm]{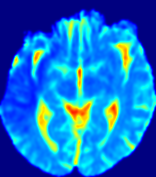}};
	\node at (4, -2) {\includegraphics[width=1.9cm]{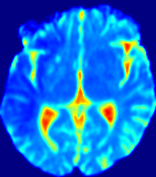}};
	\node at (6, -2) {\includegraphics[width=1.9cm]{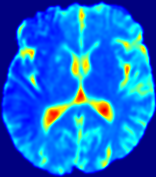}};
	\node at (8, -2) {\includegraphics[width=1.9cm]{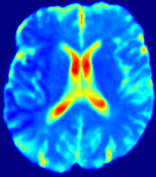}};
	\node at (10, -2) {\includegraphics[width=1.9cm]{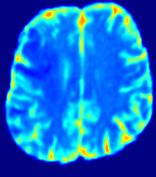}};
	
	% PD
	\node at (-1.6, -4.5) {\footnotesize{$D^{\text{est}}$}};
	\node at (0, -4.5) {\includegraphics[width=1.9cm]{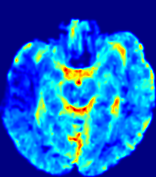}};
	\node at (2, -4.5) {\includegraphics[width=1.9cm]{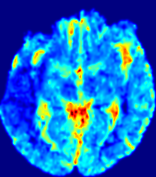}};
	\node at (4, -4.5) {\includegraphics[width=1.9cm]{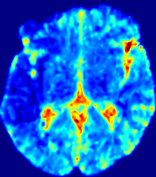}};
	\node at (6, -4.5) {\includegraphics[width=1.9cm]{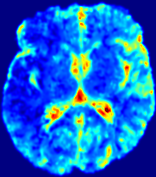}};
	\node at (8, -4.5) {\includegraphics[width=1.9cm]{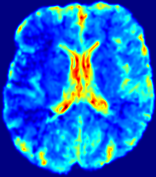}};
	\node at (10, -4.5) {\includegraphics[width=1.9cm]{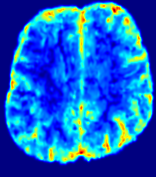}};
	
	% Color bar shift: +1.1; Text shift: +1; Text Gap: -0.575
	\node at (11.2, -3.4) {\includegraphics[width=0.4cm]{fig/cb.png}};
	\node at (11.7, -1.95) {\tiny{$0.30$}};
	\node at (11.7, -2.525) {\tiny{$0.24$}};
	\node at (11.7, -3.1) {\tiny{$0.18$}};
	\node at (11.7, -3.675) {\tiny{$0.12$}};
	\node at (11.7, -4.25) {\tiny{$0.06$}};
	\node at (11.7, -4.825) {\tiny{$0.00$}};
	\node at (11.5, -5.25) {\tiny{$(mm^2/s)$}};
	
	% NV
	\node at (-1.6, -6.75) {\footnotesize{$\Vert \bf{V}^{\text{est}} \Vert_2$}};
	\node at (0, -6.75) {\includegraphics[width=1.9cm]{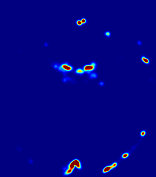}};
	\node at (2, -6.75) {\includegraphics[width=1.9cm]{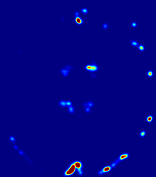}};
	\node at (4, -6.75) {\includegraphics[width=1.9cm]{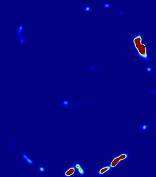}};
	\node at (6, -6.75) {\includegraphics[width=1.9cm]{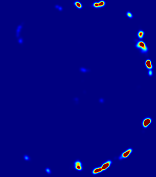}};
	\node at (8, -6.75) {\includegraphics[width=1.9cm]{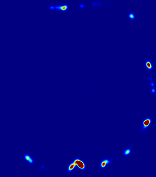}};
	\node at (10, -6.75) {\includegraphics[width=1.9cm]{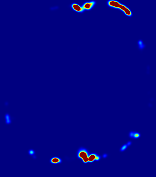}};
	% Color bar shift: +0.1; Unit shift: -0.985;  Text shift: +1; Text Gap: -0.36
	\node at (11.12, -6.6) {\includegraphics[width=0.25cm]{fig/cb.png}};
	\node at (11.6, -5.7) {\tiny{$0.0030$}};
	\node at (11.6, -6.06) {\tiny{$0.0024$}};
	\node at (11.6, -6.42) {\tiny{$0.0018$}};
	\node at (11.6, -6.78) {\tiny{$0.0012$}}; % 2.65
	\node at (11.6, -7.14) {\tiny{$0.0006$}};
	\node at (11.6, -7.5) {\tiny{$0.0000$}};
	\node at (11.5, -7.75) {\tiny{$(mm/s)$}}; % 0.2
	\end{tikzpicture}
	}
	\caption{PIANO identifiability testing: diffusion imaging via advection-diffusion. Top row shows $D^{\text{gt}}$ used for simulating ground truth pure diffusion. Rows below show the estimated $D^{\text{est}}$ and $\Vert {\bf{V}}^{\text{est}} \Vert_2$ on corresponding slices. Note that the plotted value scale for $\Vert {\bf{V}}^{\text{est}} \Vert_2$ is 0.01 of that for $D^{\text{gt}}$ and  $D^{\text{est}}$.}
	\label{ident_diff}
\end{figure}

%% file: sub/con.tex
\section{Conclusions}
We proposed a data-assimilation approach (PIANO) which estimates the velocity and diffusion fields of CA transport via an advection-diffusion PDE. Unlike most postprocessing approaches which treat voxels independently, PIANO considers spatial dependencies and does not require estimating the AIF or deconvolution techniques. We demonstrate that PIANO can successfully resolve velocity and diffusion field ambiguities and results in sensitive measures for the assessment of stroke, comparing favorably to conventional measures of perfusion. Future work will explore clinical applications and thresholds based on statistical atlases.